\documentclass[12pt]{article}
\usepackage{cite}
\usepackage{amsfonts,amsmath,amssymb}
\usepackage{mathtools}
\usepackage{stackengine}\stackMath
\usepackage{bbm}
\usepackage{xspace}
\usepackage{tikz-cd}
\usepackage{cancel}
\usepackage{ulem}
\usepackage[subrefformat=parens,labelformat=parens]{subfig}
\usepackage{graphicx}
\usepackage{longtable}
\usepackage{xcolor}
\usetikzlibrary{calc,shapes}
\usepackage{scalerel}

\usepackage{multirow}
\usepackage[scale=0.8]{geometry}
\numberwithin{equation}{section}
\usepackage[font={small}, labelfont=bf, margin=0pt, width=\textwidth]{caption}
\usetikzlibrary{tikzmark}
\usepackage{hyperref}

 \usepackage[OT2,T1]{fontenc}
\DeclareSymbolFont{cyrletters}{OT2}{wncyr}{m}{n}
\DeclareMathSymbol{\Sha}{\mathalpha}{cyrletters}{"58}

\newcommand\myoverset[2]{\overset{\textstyle #1\mathstrut}{#2}}
\newcommand\myunderset[2]{\underset{\textstyle #1\mathstrut}{#2}}

\newcommand{\fsu}{\mathfrak{su}}
\newcommand{\fso}{\mathfrak{so}}
\newcommand{\fsp}{\mathfrak{sp}}
\newcommand{\fg}{\mathfrak{g}}
\newcommand{\ff}{\mathfrak{f}}
\newcommand{\fe}{\mathfrak{e}}
\newcommand{\Dim}{\text{dim}}

\newcommand{\rk}{\text{rk}}

\begin{document}
~\vspace{2cm}
\begin{center}
{\huge\bfseries A Twist on Heterotic\\[12pt] Little String Duality}\\[10mm]

Hamza Ahmed$^{a,}$\footnote{\href{mailto:ahmed.ha@northeastern.edu}{ahmed.ha@northeastern.edu}}, Paul-Konstantin Oehlmann$^{b,}$\footnote{\href{mailto:oehlmann@ucsb.edu}{oehlmann@ucsb.edu}}, Fabian Ruehle$^{a,c,d}$\footnote{\href{mailto:f.ruehle@northeastern.edu}{f.ruehle@northeastern.edu}} \\[10mm]
\bigskip
{
	{\it ${}^{\text{a}}$ Department of Physics, Northeastern University, Boston, MA 02115}\\[.5em]
 {\it ${}^{\text{b}}$Department of Physics
University of California, Santa Barbara, CA 93106, USA}\\[.5em]
	{\it ${}^{\text{c}}$ Department of Mathematics, Northeastern University, Boston, MA 02115}\\[.5em]
	{\it ${}^{\text{d}}$ NSF Institute for Artificial Intelligence and Fundamental Interactions}\\[.5em]
}
\end{center}
\setcounter{footnote}{0} 
\bigskip\bigskip

\begin{abstract}
\noindent
In this work, we significantly expand the web of T-dualities among heterotic NS5-brane theories with eight supercharges. This is achieved by introducing twists involving outer automorphisms of discrete gauge/flavor factors and tensor multiplet permutations along the compactification circle. We assemble field theory data that we propose as invariants across T-dual theories, comprised of twisted Coulomb branch dimensions, higher group structures and flavor symmetry ranks. Using this data, we establish a detailed field theory correspondence between singularities of the compactification space, the number five-branes in the theory, and the flavor symmetry factors. The twisted theories are realized via M-theory compactifications on non-compact genus-one fibered Calabi-Yau threefolds without section. This approach allows us to prove duality of twisted and (un-)twisted theories by leveraging M/F-theory duality and identifying inequivalent torus fibrations in the same geometry. We construct several new 5D theories, including a novel type of CHL-like twisted theory where the two M9 branes are identified. Using their field theory invariants, we also construct their dual theories.
 
\end{abstract}

\clearpage
\tableofcontents

\section{Introduction} 
\label{sec:intro}
Beyond its role as a proposed UV completion of quantum gravity, string theory has provided deep insights into the structures of strongly coupled theories in various dimensions. One of the most profound lessons learned is the concept of dualities, which means that a theory may have different equivalently valid descriptions with regard to their fundamental degrees freedom.
One of the earliest dualities discovered is T-duality of the $\fe_8^2$ and $\fso_{32}$ heterotic string: The two theories become identical when compactified to 9D on a circle with Wilson lines that break the symmetry group to their common $\fso_{16}^2$ subgroup. This phenomenon appears most prominently in theories coupled to gravity, where lower dimensional theories ``oxidate'' to inequivalent higher dimensional theories upon taking certain (infinite distance) decompactification limits in the moduli space. 

Little String theories (LSTs) are non-trivial, UV complete, interacting theories in six dimensions and, albeit decoupled from gravity, share some of the behavior of gravitational theories, including dualities. This makes LSTs interesting testing grounds to study dualities, while being able to make use of global symmetries (which cannot exist in quantum theories of gravity). Recent advances in the geometric construction of 6D QFTs and generalized symmetries have been used to study the rich network of heterotic~\cite{DelZotto:2022ohj,DelZotto:2022xrh,DelZotto:2023ahf,Ahmed:2023lhj,Bhardwaj:2022ekc} and Type II~\cite{DelZotto:2020sop,Baume:2024oqn} LSTs: The former are obtained from the worldvolume theories of $\fe_8\times \fe_8$ or $\fso_{32}$ NS5 branes probing a transverse singularity $\fg$, together with a choice of flat connection at infinity. Some of the defining data of the theory has to match across dualities, based on field theory arguments~\cite{DelZotto:2020sop}. Among this ``matching data'' are 
\begin{align}
\label{eq:matchingdata}
 \left( \text{dim(CB)}\, , \kappa_R\, , \kappa_P \, , \text{rk}(\ff) \right)\, ,
\end{align}
i.e., the 5D Coulomb branch dimension, the two 2-group structure constants and the rank of the flavor symmetry algebra. These can be used in systematic bottom-up and top-down studies of the T-duality network~\cite{DelZotto:2022ohj,DelZotto:2022xrh,Ahmed:2023lhj,Mansi:2023faa,Baume:2024oqn,Lawrie:2024zon}. The data~\eqref{eq:matchingdata} has been shown to match in all dual theories, using a combination of geometric and field-theoretic arguments. Interestingly, $\kappa_{P}$ is always equal to 2 for Heterotic LSTs, and equal to 0 for Type II LSTs, and hence is conjectured to count the number of M9 branes in the theory~\cite{DelZotto:2020sop}. Also note that -- as already anticipated by Aspinwall and Morrison~\cite{Aspinwall:1997ye} and implied by the matching data \eqref{eq:matchingdata} -- the singularity $\fg$ on either side of the heterotic duality must be the same. 

Gauge theories, however, also admit discrete zero-form symmetries such as charge conjugation and permutation symmetries of tensor multiplets. In fact, such discrete symmetries are often very subtle as for example charge conjugation symmetries that appear on the level of the massless fields may not hold when including additional massive states, such as E-strings and their generalizations. If present, discrete symmetry offers additional degrees of freedom to alter the theory when compactifying the theory on a circle: Analogous to Wilson lines for continuous symmetries, one may switch on discrete Wilson line backgrounds for the discrete zero-form symmetries as well. Switching on a discrete Wilson line modifies the 5D theory substantially and yields, unlike the continuous case, disconnected 5D vacua with typically smaller moduli spaces.
This ``twisted'' circle reduction opens up the possibility of having new twisted T-dualities connecting the theory to other twisted or untwisted LSTs, as initially explored in~\cite{DelZotto:2020sop,Bhardwaj:2019fzv}. 

The goal of this work is to systematically study twisted heterotic LSTs and to discuss how the known network of untwisted duals is extended by discrete symmetry twists. 
Indeed, this question was first considered for LSTs with 16 supercharges in \cite{DelZotto:2020sop}, where it was also established that Coulomb branch and 2-group structure constants 
$\kappa_P$ and $\kappa_R$ remain T-duality invariants for twisted theories. This raises the immediate question whether the full set of invariants~\eqref{eq:matchingdata}, including the recently established flavor rank \cite{Ahmed:2023lhj}, remain invariants in the much richer set of theories with only 8 supercharges. We investigate this question by studying heterotic NS5 brane theories.
These LSTs, unlike their Type II counterparts, always have 8 supercharges, as well as non-trivial flavor symmetries. We show that~\eqref{eq:matchingdata} remain T-duality invariants for these theories. 
 
To systematically study twisted heterotic LSTs and their duals, we start from bottom-up constructions. We consider 6D LSTs with two kinds of discrete symmetries, namely outer automorphisms of gauge/flavor algebras and tensor permutations and discuss how twisting by a particular symmetry leads to distinct 5D theories with different data from the untwisted counterparts. Using the 5D data, we find putative T-duals, both twisted and untwisted, such that the data of~\eqref{eq:matchingdata} matches. 

In order to establish that matching the quantities in~\eqref{eq:matchingdata} is necessary for theories to be T-dual, we construct twisted T-dualities via geometry, more precisely by exploiting F/M-theory duality and two key recent advances. First, we use that circle reduced heterotic LSTs can be engineered in M-theory via a non-compact Calabi-Yau (CY) threefold $X_3$, which is K3 fibered and admits multiple inequivalent torus fibrations. These lift to inequivalent F-theories in 6D. Secondly, we exploit that a circle reduction of F-theory with a discrete twist is encoded in M-theory by a torus fibration that does not possess a section but only an $n$-section.

Such geometries are known to encode the M-theory geometry of twisted circle reduction by a discrete symmetry in F-theory \cite{Mayrhofer:2014opa,Braun:2014oya,Morrison:2014era,Anderson:2014yva,Klevers:2014bqa,Cvetic:2015moa} (also see the review \cite{Cvetic:2018bni}).
Such fibrations are called genus-one fibrations and their twisted fiber structures were recently investigated in great detail in \cite{Bhardwaj:2022ekc,Anderson:2023wkr}. In this work, we want to build on the toric constructions employed in~\cite{DelZotto:2022xrh,Anderson:2023wkr,Bouchard:2003bu,Klevers:2014bqa}. As a by-product, we can use these genus-one fibrations as a geometric consistency check for the discrete symmetries used in the twist of the 6D theory.

This work is structured as follows: In Section~\ref{section:frev}, we review 6D SQFTs and twisted compactifications. In Section~\ref{sec:Twisted}, we discuss the specific case of 6D heterotic LSTs, their possible twists, and the resulting 5D theories. In Section~\ref{sec:MatchingDuals}, we discuss the general structure of twisted heterotic T-duality with gauge algebra twists, which we extend in Section~\ref{sec:basetwists} to include tensor permutation twists as well. Finally, we discuss the geometric construction of twisted heterotic T-duals in Section~\ref{sec:Geometry} and present our conclusions in Section~\ref{sec:conclusion}.
In Appendix~\ref{app:AnTwistDualities}-\ref{app:Mutltitwist}, we describe more twisted LSTs and their duals in detail. Appendix~\ref{app:so32BaseTwistDual} summarizes all $ \mathbbm{Z}_2$ tensor twisted $\fso_{32}$ theories and their $\fe_8$ twisted dual.

\section{Twisted Compactifications in 6D}
\label{section:frev}
In this section, we review the main aspects of twisted 6D compactifications. We will largely follow the general exposition in \cite{Bhardwaj:2019fzv} and summarize details that are scattered throughout the recent literature~\cite{Bhardwaj:2019ngx,Bhardwaj:2020gyu}, which will become important in later sections. We start by summarizing the discrete symmetries that can arise in the context of 6D SUSY gauge theories, and how they are affected by the matter content of the theory. We then introduce the notion of twisting by these discrete symmetries along a compactification circle, to get a twisted 5D KK theory. To illustrate the concepts, we present multiple examples of 6D theories and their twisted 5D KK theories.

\subsection{Discrete symmetries in 6D}
\label{ssec:reviewDiscrete}
Our starting point is a minimal supersymmetric field theory in six dimensions denoted by $\mathcal{L}(\fg,\mathfrak{f}, n_{T})$. The theory comes with a gauge algebra $\mathfrak{g}=\prod_i \mathfrak{g}_i$, which is not necessarily simple, and vector multiplets $V$ in the adjoint representation of $\mathfrak{g}$ as well as hypermultiplets in representations $\mathbf{R}_\fg$. Furthermore, there may be a flavor algebra $\mathfrak{f}$ under which some of the hypermultiplets transform with representation $\mathbf{R}_{\mathfrak{f}}$. Finally, the theory may include $n_{T}$ dynamical tensor multiplets that can participate in 6D gauge anomaly cancellation via the Green-Schwartz mechanism, see~\cite{Weigand:2018rez,Taylor:2011wt,Johnson:2016qar} for reviews.

Among the wide range of symmetries of $\mathcal{L}$, we focus on discrete 0-form symmetries $\mathcal{S}^{(n)}$ \cite{Apruzzi:2017iqe}, with
\begin{enumerate}
 \item $\mathcal{S}^{(n)}=\mathcal{O}^{(n)}$ acting as an outer automorphism of a simple Lie algebra, or as combination of outer automorphisms of multiple simple algebras. 
 \item $\mathcal{S}^{(n)}=\mathcal{B}^{(n)}$ acting as a permutation of the tensor multiplets
 \item $\mathcal{S}^{(n,m)}=\mathcal{B}^{(n)} \circ \mathcal{O}^{(m)}$ acting as a combination of both. 
\end{enumerate}
Here $i$ and $j$ label distinct symmetries of the theory.

Next, we associate a Dirac pairing matrix of the BPS strings that couple to the anti-self-dual 2-form gauge fields, $b_{I}^{(2)}$, of the tensor multiplets. This matrix is a symmetric $(n_{T}+l) \times (n_{T}+l)$ matrix $\eta^{IJ}$ of rank $n_{T}$, where $l$ is the number of \textit{non-dynamical} tensor multiplets in the spectrum, and we assume we are on the \textit{tensor branch} of the theory where the scalars in the tensor multiplet have non-vanishing vacuum expectation values (VEVs). The hypermultiplet content of the theory can then be determined from anomaly cancellation. This data is succinctly encoded in a quiver (where the matter content is not written explicitly),
\begin{align}
\label{eq:quiverF} 
{\overset{\mathfrak{g}_{1}}{n_{1}}} \, \,  
{\overset{\mathfrak{g}_2}{n_{2}}} \, \, 
\cdots \, \,
\underset{\left[\mathfrak{f}_{I}\right]}
{\overset{\mathfrak{g}_k}{n_{k}}} \, \,
\cdots \, \,
{\overset{\mathfrak{g}_{n_{T}+l}}{n_{n_{T}+l}}} \,,
\end{align}
where $n_{I}=\eta^{II}$, and $\eta^{IJ}=-1$ if the $I^\text{th}$ and $J^\text{th}$ node are adjacent and 0 otherwise. Hypermultiplets typically transform in the product of (anti-)fundamental representations of adjacent gauge algebras $\mathfrak{g}_I$ or flavor algebras $\mathfrak{f}_{I}$ (to distinguish the latter from the former, we add brackets around flavor symmetries). From this presentation, one can identify putative discrete symmetries as the outer automorphisms of the gauge algebra $\mathfrak{g}_{i}$ and flavor algebras $\mathfrak{f}_{A}$, in addition to any permutation symmetry of the tensor multiplets, which are present if the quiver can be folded onto itself. Even if matter representations $\mathbf{R}_I$ are not invariant individually under outer automorphism twists $\mathcal{O}^{(n)}_I$ of $\fg_I$, a combined diagonal action $\mathcal{O}=\sum \mathcal{O}_I$ on the gauge and flavor algebras can still be a symmetry of the theory. This means that the matter content of the theory satisfies 
\begin{align}
\label{eq:repcond}
\mathcal{O}^{(n)} \circ\textbf{R}_{\mathfrak{g},\mathfrak{f}}=\textbf{R}_{\mathfrak{g},\mathfrak{f}} \qquad \text{up to complex conjugation} 
\end{align}
where $\circ$ represents the action of the outer automorphism on the representation. A list of outer automorphisms of simple gauge algebras and their action on representations is given in Table~\ref{tab:twistalg}. 

While the massless states can be obtained by demanding anomaly freedom of the quiver gauge theory, massive states are less transparent and need to be deduced from the construction of these theories in string theory~\cite{Heckman:2015bfa, Heckman:2018jxk}. The nodes in \eqref{eq:quiverF} correspond to curves in the geometry of the F-theory background, and $\eta^{IJ}$ is the intersection pairing of these curves. D3 branes wrapping such curves contribute massive states in the 6D theory. For example, an empty node in the quiver with $n_{I}=1$ contributes an E-string theory with a massive state in the adjoint representation of its $\fe_8$ flavor group. Gauge algebras on adjacent nodes gauge a subgroup of the flavor symmetry, and the matter spectrum can be obtained by decomposing the adjoint representation.

Having discussed outer automorphism symmetries, we next turn to discrete symmetries $\mathcal{B}^{(n)}$ associated to permutations of tensor multiplets.
The action on the gauge algebras $\fg_I$, flavor algebras $[\mathfrak{f}_I]$, and the Dirac pairing $\eta_{IJ}$ are
\begin{align}
\mathfrak{g}_{I} \rightarrow \mathfrak{g}_{\mathcal{B}^{(n)}(I)}\,,\qquad 
[\mathfrak{f}_{I}] \rightarrow [\mathfrak{f}_{\mathcal{B}^{(n)}(I)}] \,,\qquad 
\eta_{IJ} \rightarrow \eta_{\mathcal{B}^{(n)}(I)\mathcal{B}^{(n)}(J)} \,,
\end{align}
and these must be invariant for $\mathcal{B}^{(n)}$ to be a symmetry of the theory. 

As a simple example consider the quiver
\begin{align} 
 {\overset{ }{\tikzmarknode[black!70!black]{g}{[\fsu_N]}}} \, \,
{\overset{\mathfrak{su}_{N}}{\tikzmarknode[black!70!black]{e}{2}}} \, \,
{\overset{\mathfrak{su}_{N}}{\tikzmarknode[black!70!black]{c}{2}}} \, \,
{\overset{\mathfrak{su}_{N}}{\tikzmarknode[black!50!black]{d}{2}}} \, \,
{\overset{\mathfrak{su}_{N}}{\tikzmarknode[black!70!black]{f}{2}}} \, \,
{\overset{ }{\tikzmarknode[black!70!black]{h}{[\fsu_N]}}} \, \, \, , \qquad 
\newline
\end{align}
This quiver has a $ \mathbbm{Z}_2$ reflection symmetry acting as 
\begin{align} 
\label{eq:quivquotBase}
{\overset{ }{\tikzmarknode[black!70!black]{g}{[\fsu_N]}}} \, \,
{\overset{\mathfrak{su}_{N}}{\tikzmarknode[black!70!black]{e}{2}}} \, \,
{\overset{\mathfrak{su}_{N}}{\tikzmarknode[black!70!black]{c}{2}}} \, \,
{\overset{\mathfrak{su}_{N}}{\tikzmarknode[black!50!black]{d}{2}}} \, \,
{\overset{\mathfrak{su}_{N}}{\tikzmarknode[black!70!black]{f}{2}}} \, \,
{\overset{ }{\tikzmarknode[black!70!black]{h}{[\fsu_N]}}} \, \, \, , \qquad 
\newline
\begin{tikzpicture}[remember picture,overlay]
\draw[red,<->]([yshift=0.000009ex]c.south) to[bend right]node[below]{\scriptsize} ([yshift=0.000009ex]d.south);
\draw[red,<->]([yshift=0.000009ex]e.south) to[bend right]node[below]{\scriptsize} ([yshift=0.000009ex]f.south);
\draw[red,<->]([yshift=0.000009ex]g.south) to[bend right]node[below]{\scriptsize} ([yshift=0.000009ex]h.south);
\end{tikzpicture}
\end{align}
which leaves the theory invariant. 
 
\subsection{Construction of twisted 5D KK theories}
\label{subsec:bblocks}
Upon circle compactification of a 6D theory $\mathcal{L}(\fg, \mathfrak{f}, n_{T})$ with discrete (0-form) symmetries $\mathcal{S}_{i}$ discussed in the previous section, we can twist by the discrete symmetries by introducing a holonomy $\gamma \in G^{(0)}_{i}$ for the background field associated with $\mathcal{S}_{i}$. We focus on Lie algebra outer automorphisms $\mathcal{O}^{(n)}$ acting on simple algebras $\fg_{I}$ or $\mathfrak{f}_{A}$. The action on gauge fields $A(y)$ along the circle is
\begin{align}
\label{eq:twist}
A(y)=\mathcal{O}^{(n)}(A(y+2\pi)) \, .
\end{align}
The symmetry acts on the roots and weights of the Lie algebra by identifying nodes in the respective Dynkin diagram~(see e.g. \cite{Kim:2004xx,Tachikawa:2011ch,Kim:2019dqn,Bhardwaj:2019fzv,Duan:2021ges,Lee:2022uiq}). The resulting \textit{twisted} algebras are denoted by $\mathfrak{g}^{(n)}$. Identification of roots in the Dynkin diagrams reduces the rank of the Lie algebras, which means that the Coulomb branch dimension and flavor symmetry in the 5D KK theory are reduced compared to the untwisted KK theory. Moreover, the 5D matter representations of $\mathfrak{g}$ and $\mathfrak{f}$ decompose into representations of $\Tilde{\mathfrak{g}}$ and $\Tilde{\mathfrak{f}}$, which are the invariant finite subalgebras of $\mathfrak{g}^{(n)}$ and $\mathfrak{f}^{(m)}$. We summarize the invariant subalgebras of simple Lie algebras and the decomposition of their representations in Table~\ref{tab:twistalginv}. In particular, one obtains several massive representations under this decomposition, with the mass given by the (shifted) KK charges~\cite{Lee:2022uiq, Kim:2020hhh}, which we also summarize in the table.

\subsection{Building blocks for twisting}

\begin{table}[t]
 \centering
 \begin{tabular}{|c|c||c|} \hline
  $\fg $ & $\mathcal{O}^{(n)}$ & $\mathcal{O}^{(n)} \circ \textbf{R}_{\mathfrak{g}}$ \\ \hline\hline
  $\mathfrak{su}_{N} $ & $ \mathbbm{Z}_2$ & $\textbf{F}\leftrightarrow \Bar{\textbf{F}}$, $\mathbf{\Lambda}^{m}\leftrightarrow \Bar{\mathbf{\Lambda}}^{m}$, $\textbf{S}^{k}\leftrightarrow \Bar{\textbf{S}}^{k}$ \\ \hline
  $\mathfrak{so}_{2N} $ & $ \mathbbm{Z}_2$&$\textbf{F}\rightarrow \textbf{F}$, $\textbf{S}\leftrightarrow \textbf{C}$ \\ \hline
  $\mathfrak{e}_{6} $ & $ \mathbbm{Z}_2$& $\textbf{F}\leftrightarrow \Bar{\textbf{F}}$ \\ \hline
  $\mathfrak{so}_{8} $ & $ \mathbbm{Z}_3$&$\textbf{F}\rightarrow \textbf{S}$, $\textbf{S}\rightarrow \textbf{C}$, $\textbf{C}\rightarrow \textbf{F}$ \\ \hline
 \end{tabular}
 \caption{Summary of outer automorphism symmetries $\mathcal{O}^{(n)}$ and their actions on representations $\textbf{R}$ for simple, simply laced Lie algebras $\mathfrak{g}$. Here, \textbf{F} denotes the fundamental representation, $\mathbf{\Lambda}^{m}$ denotes the $m$-index antisymmetric representation, \textbf{S}$^{2}$ denotes the 2-index symmetric representation, and \textbf{S} and \textbf{C} denote spinor and cospinor representations. A bar indicates complex conjugate irrep.}
 \label{tab:twistalg}
\end{table}

\begin{table}[t]
 \centering
 \begin{tabular}{|c|c||c|} \hline
  $\fg^{(n)}$ & $\Tilde{\mathfrak{g}}$ & $ \textbf{R}_{\mathfrak{g}} \rightarrow \textbf{R}_{\Tilde{\mathfrak{g}}}$ \\ \hline\hline
  $\mathfrak{su}_{2N}^{(2)}$ & $\mathfrak{sp}_{N}$ & $\textbf{F} \oplus \Bar{\textbf{F}} \rightarrow \textbf{F}_{0} \oplus \textbf{F}_{1/2}$, $\mathbf{\Lambda}^{2} \oplus \Bar{\mathbf{\Lambda}}^{2}\rightarrow \mathbf{\Lambda}_{0}^{2} \oplus \mathbf{\Lambda}_{1/2}^{2} \oplus \mathbf{1}_{0} \oplus \mathbf{1}_{1/2}$ \\ $\mathfrak{su}_{2N}^{(2')}$ &$\mathfrak{so}_{2N}$ & $\textbf{F} \oplus \Bar{\textbf{F}} \rightarrow \textbf{F}_{0} \oplus \textbf{F}_{1/2}$, \\ \hline
  $\mathfrak{su}_{2N+1}^{(2)}$ & $\mathfrak{sp}_{N}$ & $\textbf{F} \oplus \Bar{\textbf{F}} \rightarrow \textbf{F}_{0} \oplus \textbf{F}_{1/2} \oplus \mathbf{1}_{1/4} \oplus \mathbf{1}_{3/4}$ \\ \hline
  $\mathfrak{so}_{2N}^{(2)}$ & $\mathfrak{so}_{2N-1}$ & $\textbf{F} \rightarrow \textbf{F}_{0} \oplus \mathbf{1}_{1/2}$, $\textbf{S} \oplus \textbf{C}\rightarrow \textbf{S}_{0} \oplus \textbf{S}_{1/2}$ \\ \hline
  $\mathfrak{e}_{6}^{(2)}$ & $\mathfrak{f}_{4}$ & $\textbf{F} \oplus \Bar{\textbf{F}} \rightarrow \textbf{F}_{0} \oplus \textbf{F}_{1/2} \oplus \mathbf{1}_{0} \oplus \mathbf{1}_{1/2}$ \\ \hline
  $\mathfrak{so}_{8}^{(3)}$ & $\mathfrak{g}_{2}$ & $\textbf{F} \oplus \textbf{S} \oplus \textbf{C}\rightarrow \textbf{F}_{0} \oplus \textbf{F}_{1/3} \oplus \textbf{F}_{2/3} \oplus \mathbf{1}_{0} \oplus \mathbf{1}_{1/3} \oplus \mathbf{1}_{2/3}$ \\ \hline
 \end{tabular}
 \caption{List of twisted algebras and their invariant subalgebras. The right column summarizes the matter representations and their twisted KK charges~\cite{Lee:2022uiq}. Note that $\mathfrak{su}_{2N}^{(2)}$ has two different invariant subalgebras.
 }
 \label{tab:twistalginv}
\end{table}

In this section, we discuss several field theory examples and their twisted 5D KK theories. These types of theories will serve as the main building blocks from which the heterotic LSTs and their twisted KK theories are constructed. Since 6D SCFTs are the building blocks of 6D LSTs, their twisted KK theories\footnote{For recent explorations of twisted 6D SCFTs, see \cite{Jefferson:2017ahm,
Bhardwaj:2018yhy,
Bhardwaj:2018vuu,
Jefferson:2018irk,Kim:2019dqn,Bhardwaj:2019fzv,Lee:2022uiq}.} form the building blocks of twisted 5D LSTs. 

We first discuss an $\mathfrak{su}_{N}$ ($N>2$) gauge theory with a single tensor multiplet and self-Dirac pairing $n$ given by the quiver 
\begin{align}
\label{eq:suorg}
{\overset{\mathfrak{su}_{N}}{n}}[\ff] \, \, \, 
\end{align}
where the flavor symmetry $\ff$ depends on $n$. For $\mathfrak{su}_{N}$ with $N>3$, the Dirac pairing values can be $0 \leq n \leq 2$, and for $\mathfrak{su}_{3}$ we have $n=3$ and $\ff=0$. The full
6D flavor symmetry can be inferred from the spectrum, which is fixed by 6D anomaly cancellation. The $ \mathbbm{Z}_2$ outer automorphism symmetry of the $\mathfrak{su}_N$ gauge algebra is a symmetry of the full theory but requires twisting the flavor algebra $\ff$ as well. Since the $ \mathbbm{Z}_2$ acts as complex conjugation, it flips fundamental and anti-fundamental. In general, an $m$-fold anti-symmetrized irrep is exchanged with its conjugate irrep (and likewise for $n$-fold symmetrized irreps) whenever \eqref{eq:repcond} is satisfied. Upon compactifying the theory given by the quiver in~\eqref{eq:suorg} on a circle and twisting by this discrete $ \mathbbm{Z}_{2}$ symmetry, we obtain the 5D theory
\begin{align}
\label{eq:suorg2}
{\overset{\mathfrak{su}_{N}^{(2)}}{n}} [\ff^{(2)}] \, \, \, 
\end{align}
The low energy field content of the theory is readily deduced from the data in Table~\ref{tab:twistalginv}. The twist reduces the gauge group to $\mathfrak{sp}_{\lfloor N/2 \rfloor} \subseteq \fsu_N^{(2)}$, and the 6D matter multiplets are reduced to representations of $\mathfrak{sp}_{\lfloor N/2 \rfloor}$. In particular, since the fundamental and anti-fundamentals of the $\mathfrak{su}_N$ algebras are identified, one gets half the massless degrees of freedom in the 5D twisted theory. For concreteness, let us fix $n=2$ and $N>2$. In this case, there are $2N$ fundamental hypermultiplets in the 6D theory, which means that the flavor algebra is $\ff=\mathfrak{su}_{2N}$. Performing the twisted circle reduction yields an $\mathfrak{sp}_{\lfloor N/2 \rfloor}$ gauge algebra, with $N$ massless hypermultiplets in its fundamental representation. Hence, the flavor symmetry in 5D is $\mathfrak{so}_{2N}$, under which the fundamentals transform as half-hypermultiplets.

However, for $N$ even, there is another choice for the twist corresponding to an inequivalent outer automorphism of the $\mathfrak{su}_{N}$ gauge algebra \cite{Bhardwaj:2020kim, Henning:2021ctv}. In this case, the outer automorphism leads to an $\mathfrak{so}_{N} \subseteq \fsu_N^{(2)}$ gauge algebra in 5D with $N$ fundamentals\footnote{It is customary to abuse notation and talk about (bi-)fundamentals also for algebras other than $\fsu_N$, which then refers to the lowest-dimensional irrep.} transforming with a 5D $\fsp_N$ flavor symmetry.
To distinguish the two twists, we write $\fsu_N^{(2)}$ for $\fsp_{N/2}$ and $\fsu_N^{(2')}$ for $\fso_{N}$. Note that both finite subalgebras are part of the same twisted algebra and are smoothly connected in the 5D CB moduli space. To summarize the $n=2$ example, we can obtain the following 5D twisted theories from the 6D theory~\eqref{eq:suorg}
\begin{equation}
\label{eq:chart1}
\begin{tikzcd}
\text{6D theory}:& & & {\overset{\mathfrak{su}_{N}}{2}} [\mathfrak{su}_{2N}]\arrow[rd,"\text{for even $N$}"] \arrow[dl,swap,"\text{all $N$}"] & \\
\text{Twisted 5D theories}:& &{\overset{\mathfrak{su}_{N}^{(2)}}{2}} [\fsu_{2N}^{(2')}] \, \, \, && {\overset{\mathfrak{su}_{N}^{(2')}}{2}}[\fsu_{2N}^{(2)}] \,.
\end{tikzcd}
\end{equation}
Thus, the 5D twisted theory is obtained by twisting the gauge algebra with one outer automorphism and the flavor algebra with the other.

Next, we consider a more complicated theory, which is the tensor branch of a 6D SCFT known as an orbi-instanton theory of type $\mathfrak{g}=\mathfrak{su}_{N}$, with $N>4$, of rank $k$. A classification of orbi-instanton theories of ADE type was given in \cite{Frey:2018vpw}. We consider the $A_{N-1}$ type theory
\begin{align}
\label{eq:1ex2} 
\lbrack \mathfrak{e}_{8}\rbrack  \, \,
1 \, \,
2 \, \,
{\overset{\mathfrak{su}_{2}}{2}} \, \,
{\overset{\mathfrak{su}_{3}}{2}} \, \,
{\overset{\mathfrak{su}_{4}}{2}} \, \,
\cdots {\overset{\mathfrak{su}_{N-1}}{2}}\, \, 
\underbrace{{\overset{\mathfrak{su}_{N}}{2}} \ldots 
{\overset{\mathfrak{su}_{N}}{2}} \, \,}_{\times (k)} \, \,
\lbrack \mathfrak{su}_{N}\rbrack  \, \, ,
\end{align}
which has $k$ repetitions of the $\fsu_N$ gauge algebra blocks. The $\mathfrak{su}_{N>2}$ algebras in the above quiver all have $ \mathbbm{Z}_{2}$ outer automorphisms. In isolation, the respective matter multiplets are incompatible with the induced $\mathbbm{Z}_2$ conjugation symmetry, but the full orbi-instanton theory admits one diagonally-acting complex conjugation symmetry which can be used to twist upon circle compactification to 5D. However, note that for odd $N$, each of the $k$ copies of twisted $\mathfrak{su}_{N}^{(2)}$ algebras can only be twisted to a maximal subalgebra $\mathfrak{sp}_{\lfloor N/2 \rfloor}$. The $\mathfrak{sp}-\mathfrak{sp}$ bi-fundamental hypermultiplets in the reduction always stay massive, cf.~\cite{Bhardwaj:2019fzv}. For even $N$, one can alternate twists such that $\fso$-$\fsp$ massless bi-fundamentals appear in the 5D theory,
\begin{align}
\label{eq:1ex3} 
\lbrack \mathfrak{e}_{8}\rbrack \, \,
1 \, \,
2 \, \,
{\overset{\mathfrak{su}_{2}}{2}} \, \,
{\overset{\mathfrak{su}_{3}^{(2)}}{2}} \, \,
{\overset{\mathfrak{su}_{4}^{(2')}}{2}} \, \,
{\overset{\mathfrak{su}_{5}^{(2)}}{2}} \, \,
\cdots
\underbrace{
{\overset{\mathfrak{su}_{N}}{2}^{(2/2')}} \, \,}_{\times (k)} \, \,
\lbrack \mathfrak{su}_{N}^{(2/2')} \rbrack \,.
\end{align}
The flavor algebra depends on which twist was used for the last $\mathfrak{su}_{N}$ algebra. This is consistent with results in~\cite{Bhardwaj:2019fzv}, and we will later provide a top-down construction of such a twisted orbi-instanton theory via F-theory, which serves as a cross-check for the field-theoretic analysis. More generally, we can have a chain of $\mathfrak{su}_{2N_i}$ algebras, which has an overall $ \mathbbm{Z}_{2}$ outer automorphism we can use to twist. We will realize such chains later when we consider $\mathfrak{so}_{32}$ heterotic LSTs.

Next, we analyze 6D theories with $\mathfrak{so}$ gauge algebras, and determine the consistent twisted 5D theories. Consider an $\mathfrak{g}=\mathfrak{so}_{2N+8}$ gauge theory with a single tensor multiplet,
\begin{align}
{\overset{\mathfrak{so}_{2N+8}}{4}}[\fsp_{2N}] \,.
\end{align}
Anomaly cancellation requires $2N$ hypermultiplets in the fundamental of $\mathfrak{so}_{2N+8}$, leading to an $\mathfrak{sp}_{2N}$ flavor symmetry. The $ \mathbbm{Z}_{2}$ outer automorphism of $\mathfrak{so}_{2N+8}$ leaves the fundamental irrep invariant, but interchanges spinors and co-spinors (both of which are present in the massless spectrum in the above theory). Hence, twisting by this symmetry satisfies condition~\eqref{eq:repcond}. From Table~\ref{tab:twistalginv}, we see that the 5D gauge theory is given by the invariant subalgebra $\mathfrak{so}_{2N-1}$, and the $2N$ fundamentals of $\mathfrak{so}_{2N}$ descend to $2N$ fundamentals of $\mathfrak{so}_{2N-1}$ in addition to massive states. The 5D twisted theory is then given by
\begin{align}
{\overset{\mathfrak{so}_{2N+8}^{(2)}}{4}}[\fsp_{2N}] \,.
\end{align}
Note that the flavor algebra does not admit an outer automorphism, which is consistent with the fact that it has only real representations.

Next, consider a 6D $\mathfrak{sp}_{N}$ gauge theory with a single tensor multiplet,
\begin{align}
\label{eq:spex1}
{\overset{\mathfrak{sp}_{N}}{1}}[\mathfrak{so}_{4N+16}] \,.
\end{align}
The theory is anomaly free if we include $2N+8$ fundamental hypermultiplets of $\mathfrak{sp}_{N}$, which transform under an $\mathfrak{so}_{4N+16}$ flavor symmetry. The gauge algebra has no outer automorphisms, but the flavor symmetry does. Moreover, since only (bi-)fundamentals of the $\mathfrak{so}$ flavor group appear, one might naively think that the $ \mathbbm{Z}_{2}$ outer automorphism of the $\mathfrak{so}$ algebra is a symmetry of the theory. Note, however, that for an $\fsp_N$ gauge theory on a $1$ curve, there exists a single massive spinor\footnote{This was first observed in the chiral ring of the Higgs branch operator of D-D conformal matter theories.} (but no co-spinor) of the $\fso_{4N+8}$ flavor symmetry. This spinor transforms under the conjugation symmetry and hence breaks it~\cite{Distler:2022yse, Baume:2021chx}.

However, if we gauge a subalgebra of the $\mathfrak{so}_{4N+16}$ flavor symmetry, the resulting theory can have a $ \mathbbm{Z}_{2}$ outer automorphism: consider for example the theory
\begin{align} 
\label{eq:spinex1}
\lbrack \mathfrak{sp}_{N}\rbrack  \, \,
{\overset{\mathfrak{so}_{2N+8}}{4}} \, \,  
{\overset{\mathfrak{sp}_{N}}{1}} \, \, 
\lbrack \mathfrak{so}_{2N+8}\rbrack  \, ,
\end{align}
which is related to \eqref{eq:spex1} by gauging a $\mathfrak{so}_{2N+8}$ subalgebra of the $\mathfrak{so}_{4N+16}$ flavor symmetry coupled to another tensor multiplet, and adding matter to cancel the 6D gauge anomaly. Again, the theory has (bi-)fundamentals of the $\mathfrak{so}$ algebras, such that the outer automorphisms of these algebras could be symmetries of the theory. However, we need to again take the massive spinors into account. Under the branching
\begin{align}
\label{decomp-1}
\mathfrak{so}_{4N+16} \rightarrow \mathfrak{so}_{2N+8} \oplus \mathfrak{so}_{2N+8}
\end{align}
the spinor $\mathbf{S}$ of $\mathfrak{so}_{4N+16}$ decomposes as
\begin{align}
\mathbf{S} \rightarrow (\mathbf{S}_{1},\mathbf{S}_{2}) \oplus (\mathbf{C}_{1},\mathbf{C}_{2})\,,
\end{align}
where the subscripts label the two $\mathfrak{so}_{2N+8}$ factors. The respective bi-spinors do have a complex conjugation symmetry when acting on both $\mathfrak{so}_{2N+8}$ algebras simultaneously, which exchanges spinor and cospinor representations. Hence, we can have a twisted 5D theory only if we twist by both outer automorphisms simultaneously, resulting in
\begin{align} 
\lbrack \mathfrak{sp}_{N}\rbrack  \, \,
{\overset{\mathfrak{so}_{2N+8}^{(2)}}{4}} \, \,  
{\overset{\mathfrak{sp}_{N}}{1}} \, \, 
\lbrack \mathfrak{so}_{2N+8}^{(2)}\rbrack  \,.
\end{align}
The above can be generalized to cases where neighboring $\fso$ gauge or flavor factors have different ranks. 

With this result, we exemplify some consistent twisted theories that will appear later in the F-theory constructions. In particular, consider the tensor branch of the rank $M$ $(\mathfrak{so}_{2N+8},\mathfrak{so}_{2N+8})$ conformal matter theory
\begin{align} 
\lbrack \mathfrak{so}_{2N+8}\rbrack  \, \,
{\overset{\mathfrak{sp}_{N}}{1}}
\underbrace{{\overset{\mathfrak{so}_{2N+8}}{4}} \, \,  
{\overset{\mathfrak{sp}_{N}}{1}}}_{M-1} \, \, 
\lbrack \mathfrak{so}_{2N+8}\rbrack  \,.
\end{align}
For this theory, one has a $ \mathbbm{Z}_{2}$ symmetry corresponding to the combined outer automorphism of all $\mathfrak{so}$ gauge algebra factors in the quiver. The twisted 5D theory is
\begin{align} 
\lbrack \mathfrak{so}_{2N+8}^{(2)}\rbrack  \, \,
{\overset{\mathfrak{sp}_{N}}{1}}
\underbrace{{\overset{\mathfrak{so}_{2N+8}^{(2)}}{4}} \, \,  
{\overset{\mathfrak{sp}_{N}}{1}}}_{M-1} \, \, 
\lbrack \mathfrak{so}_{2N+8}^{(2)}\rbrack  \,.
\end{align} 
A case where we need to decompose the $\fso$ flavor group into three different factors is given by the quiver 
\begin{align} 
\label{weirdso}
\lbrack \mathfrak{sp}_{2N-5}\rbrack  \, \,
{\overset{\mathfrak{so}_{4N+2}}{4}} \, \,  
\underset{{\left[\mathfrak{u}_{1}\right]}}
{\overset{\mathfrak{sp}_{2N-1}}{1}} \, \, 
{\overset{\mathfrak{so}_{4N+8}}{4}} \, \, 
\lbrack \mathfrak{sp}_{2N+1}\rbrack  \, \,
\end{align}
for $N>2$. Naively there are three independent $ \mathbbm{Z}_{2}$ outer automorphisms, one for each $\mathfrak{so}$ algebra and one for $\fso_2 \sim \mathfrak{u}_{1}$,
whose outer automorphism maps a charge $q$ to $-q$. If we decouple all tensors adjacent to the $\mathfrak{sp}_{2N-1}$ gauge algebra, the flavor symmetry becomes $\mathfrak{so}_{8N+12}$. Hence, we have the branching
\begin{align}
\label{decomp-0}
\mathfrak{so}_{8N+12} \rightarrow \mathfrak{so}_{4N+2}\oplus \mathfrak{so}_{4N+8} \oplus \mathfrak{u}_{1} \,.
\end{align}
The spinor of $\mathfrak{so}_{8N+12}$ then decomposes as
\begin{align}
\mathbf{S} \rightarrow (\mathbf{S}_{1},\mathbf{S}_{2})_{-1} \oplus (\mathbf{C}_{1},\mathbf{S}_{2})_{1} \oplus (\mathbf{S}_{1},\mathbf{C}_{2})_{1} \oplus (\mathbf{C}_{1},\mathbf{C}_{2})_{-1}\, ,
\end{align}
where the subscript indicates the $\mathfrak{u}_{1}$ charge. The important observation is that the matter content on the right hand side is only invariant under any two of the three $ \mathbbm{Z}_{2}$ outer automorphisms! This is due to the fact that the spinor and cospinor irreps are complex conjugates of each other for $\mathfrak{so}_{4N+2}$, but are real irreps for $\mathfrak{so}_{4N+8}$. Since any two choices of twists are consistent, we can obtain three twisted compactifications:
\begin{align}
\label{3so}
\text{\MakeUppercase{\romannumeral 1}} : \qquad \qquad 
\lbrack \mathfrak{sp}_{2N-5}\rbrack  \, \,
{\overset{\mathfrak{so}_{4N+2}^{(2)}}{4}} \, \,  
\underset{{\left[\mathfrak{u}_{1}\right]}}
{\overset{\mathfrak{sp}_{2N-1}}{1}} \, \, 
{\overset{\mathfrak{so}_{4N+8}^{(2)}}{4}} \, \, 
\lbrack \mathfrak{sp}_{2N+1}\rbrack  \, \,
 \, , \qquad 
\nonumber
\\ 
\text{\MakeUppercase{\romannumeral 2}} : \qquad \qquad 
\lbrack \mathfrak{sp}_{2N-5}\rbrack  \, \,
{\overset{\mathfrak{so}_{4N+2}^{(2)}}{4}} \, \,  
{\overset{\mathfrak{sp}_{2N-1}}{1}} \, \, 
{\overset{\mathfrak{so}_{4N+8}}{4}} \, \, 
\lbrack \mathfrak{sp}_{2N+1}\rbrack  \, \,
 \, , \qquad 
\nonumber
\\ 
\text{\MakeUppercase{\romannumeral 3}} : \qquad \qquad 
\lbrack \mathfrak{sp}_{2N-5}\rbrack  \, \,
{\overset{\mathfrak{so}_{4N+2}}{4}} \, \,  
{\overset{\mathfrak{sp}_{2N-1}}{1}} \, \, 
{\overset{\mathfrak{so}_{4N+8}^{(2)}}{4}} \, \, 
\lbrack \mathfrak{sp}_{2N+1}\rbrack  \, \,
 \, , \qquad 
\end{align}
where in the last two cases, the $\mathfrak{u}_{1}$ is twisted to ``nothing'' in 5D.

There is another distinct twisting pattern for alternating $\mathfrak{sp}$ and $\mathfrak{su}$ algebras with bi-fundamentals. Consider the quiver 
\begin{align}
\label{ancase}
{\overset{\mathfrak{su}_{N}}{2}}\, \, 
{\overset{\mathfrak{sp}_{N}}{1}}
[\mathfrak{so}_{2N+16}] \, \, \, 
\end{align}
where, naively, the theory has a discrete symmetry associated to the $\mathfrak{su}_{N}$ and $\mathfrak{so}_{2N+16}$ algebras, since the massless hypermultiplets in the theory satisfy \eqref{eq:repcond}. The corresponding twisted 5D theory was already considered in \cite{Bhardwaj:2020kim}, but without taking the $\mathfrak{so}$ type flavor symmetry into account. As we have seen, there is a massive state on the (1)-node which transforms as a spinor under this flavor symmetry. The branching that leads to the above theory is
\begin{align}
\label{decomp2}
\mathfrak{so}_{4N+16} \rightarrow \mathfrak{so}_{2N} \oplus \mathfrak{so}_{2N+16} \rightarrow \mathfrak{su}_{N} \oplus \mathfrak{u}_{1} \oplus \mathfrak{so}_{2N+16}
\end{align}
The $\mathfrak{u}_{1}$ is actually ABJ anomalous~\cite{Apruzzi:2020eqi,Lee:2018ihr,Ahmed:2023lhj}, so we will ignore it. We first take $N$ to be odd. In this case, the spinor of $\mathfrak{so}_{4N+16}$ decomposes as
\begin{align}
\mathbf{S} \rightarrow (\boldsymbol{r},\mathbf{S}_{1}) \oplus (\boldsymbol{\Bar{r}},\mathbf{C}_{1})
\end{align}
where $\boldsymbol{r}$ is some complex representation of $\mathfrak{su}_{N}$. An outer automorphism of any $\mathfrak{su}_{N > 2}$ type algebra always maps (complex) representation to their complex conjugates and so under this action, $\boldsymbol{r}$ and $\boldsymbol{\Bar{r}}$ are exchanged. However, this action is \textit{not} a symmetry of the theory \eqref{ancase}, since the spinor and cospinor of $\mathfrak{so}_{2N+16}$ are complex for $N$ odd. Hence, only the combined $ \mathbbm{Z}_{2}$ outer automorphism of the $\mathfrak{su}_{N}$ and $\mathfrak{so}_{2N+16}$ is a symmetry of the theory. The twisted 5D theory is
\begin{align}
\label{ancase1}
{\overset{\mathfrak{su}_{N}^{(2)}}{2}}
{\overset{\mathfrak{sp}_{N}}{1}}
[\mathfrak{so}_{2N+16}^{(2)}] \, \, \, 
\end{align}
For even $N$, the spinor decomposes as
\begin{align}
\mathbf{S} \rightarrow (\boldsymbol{r},\mathbf{S}_{1}) \oplus (\boldsymbol{r'},\mathbf{C}_{1})
\end{align}
where this time $\boldsymbol{r}$ and $\boldsymbol{r'}$ are (pseudo-)real representations of $\mathfrak{su}_{N}$. The spinor and cospinor irreps are (pseudo-)real as well, and hence the outer automorphism of the $\mathfrak{so}_{2N+16}$ algebra is not a symmetry of the theory. However, the outer automorphism of the $\mathfrak{su}_{N}$ algebra is a symmetry of the theory, and if we pick the (2') twist in Table~\ref{tab:twistalginv}, we also obtain massless matter in 5D in the form of $\mathfrak{so}-\mathfrak{sp}$ half-bi-fundamentals. This results in the twisted 5D theory:
\begin{align}
\label{ancase2}
{\overset{\mathfrak{su}_{N}^{(2')}}{2}}
{\overset{\mathfrak{sp}_{N}}{1}}
[\mathfrak{so}_{2N+16}^{(1)}] \, .
\end{align}
Thus, the presence of the massive spinor dictates the twisting patterns and therefore the possible 5D theories.

We can generalize the above examples to the case where a (1)-node with an $\mathfrak{sp}_{N}$ gauge algebra has $m$ neighboring (gauge or flavor) algebras, each of which has a non-trivial outer automorphism. One can then check that the number of distinct twisted 5D theories is given by
\begin{align}
\label{n5d}
n_{\text{5D}}=\sum_{i \in 2  \mathbbm{Z}}^{m} \frac{m!}{(m-i)!(i)!}
\end{align}
A general 6D theory will have $k$ such (1)-nodes, and some may ``share'' an adjacent algebra, in which case one needs to be careful when determining the number of possible 5D twisted theories. 

E-string states are the final type of sources that constrain the discrete symmetries of a 6D theory. To illustrate this, consider an $(\mathfrak{e}_{6},\mathfrak{e}_{6})$ conformal matter theory with quiver
\begin{align}
\label{eq:1ex1} 
\lbrack \mathfrak{e}_{6}\rbrack  \, \,
1 \, \,
{\overset{\mathfrak{su}_{3}}{3}} \, \,  
1 \, \, 
\lbrack \mathfrak{e}_{6}\rbrack  \,.
\end{align}
Both the $\mathfrak{su}_{3}$ and $\mathfrak{e}_{6}$ algebras admit $ \mathbbm{Z}_{2}$ outer automorphisms, leading to 5D gauge algebras $\mathfrak{f}_{4}$ and $\mathfrak{sp}_{1}$ respectively. As before, we need to study which combinations of outer automorphisms lead to a symmetry of the theory~\eqref{eq:1ex1}. For this, we need to consider the matter content of the theory. Since $\mathfrak{su}_{3}$ gauge algebras on a $(3)$-node do not have any matter (they form a so-called non-higgsable cluster), the quiver has no massless matter. However, there are massive states originating from the massive E-strings. An isolated E-string theory, represented by one empty (1)-node, comes with an $\mathfrak{e}_{8}$ flavor symmetry. In a quiver, a part of (or even the full) flavor symmetry is gauged by the algebras of the adjacent nodes. For the quiver~\eqref{eq:1ex1} above, we have to consider the decomposition
\begin{align}
\label{decomp1}
\mathfrak{e}_{8} \rightarrow \mathfrak{su}_{3} \oplus \mathfrak{e}_{6}\,,
\end{align}
to obtain the $\mathfrak{e}_{6}$ flavor symmetry. Under this breaking, the $\textbf{248}$ of $\mathfrak{e}_{8}$ decomposes as
\begin{align}
\textbf{248} \rightarrow \textbf{(8,1)} \oplus \textbf{(1,78)} \oplus \textbf{(3,27)} \oplus \textbf{(}\overline{\textbf{{3}}},\overline{\textbf{{27}}}\textbf{)} \, ,
\end{align}
providing massive states of the theory that may become light at the origin of the E-string tensor branch. These states need to be considered when determining the symmetries of the full theory. The bi-fundamental massive matter is only invariant under the simultaneous action of both $\fsu_3$ and $\fe_6$ $ \mathbbm{Z}_{2}$ symmetries \cite{Bhardwaj:2019fzv}, and hence the quiver admits only a single diagonally-acting $ \mathbbm{Z}_2$ symmetry we may twist by, which results in the twisted 5D theory
\begin{align}
\label{eq:1ex1t} 
\lbrack \mathfrak{e}_{6}^{(2)}\rbrack  \, \,
1 \, \,
{\overset{\mathfrak{su}_{3}^{(2)}}{3}} \, \,  
1 \, \, 
\lbrack \mathfrak{e}_{6}^{(2)}\rbrack  \, \,.
\end{align}
Conformal matter theories of type $(\mathfrak{e}_{6},\mathfrak{e}_{6})$ with higher rank can be studied analogously.

\section{Heterotic LSTs with outer automorphism twists}
\label{sec:Twisted}
6D LSTs are a special type of 6D QFTs whose Dirac pairing matrix has a one-dimensional kernel, which gives rise to a single non-dynamical tensor multiplet in the spectrum~\cite{Bhardwaj:2015xxa}. Since gravity is decoupled in LSTs, they can have global discrete 0-form symmetries of the type discussed before, which can be used to twist the theory upon circle reduction. Furthermore, LSTs have an intrinsic scale, set by the VEV of the scalar in the non-dynamical tensor multiplet. Little string theories also admit T-dualities that relate two (or more) inequivalent 6D LSTs upon circle compactification to 5D and an adequate choice of Wilson lines~\cite{Intriligator:1997dh}. Moreover, the non-dynamical tensor field generates a 1-form symmetry, which can mix with other 0-form global symmetries resulting in the structure of a 2-group characterized by its structure constants~\cite{DelZotto:2020sop}.

A lot of recent research in this area is focused on identifying necessary criteria (by matching data across theories) for when theories can be T-dual. In this work, we find that the criteria worked out for untwisted theories continue to hold upon considering twisted T-duals. More precisely, the following data is invariant under (twisted) T-duality and hence has to agree among all T-dual theories:
\begin{align}
\label{eq:invdata} \framebox{$
\begin{array}{ll} 
\Dim(\text{CB})=n_{T} + \text{rk}(\fg^{(n)})\,, & \qquad \Dim(\text{WL})=\rk(\fg_{F}^{(n)})\, , \\ 
& \\ \kappa_{R}=-\sum_{I=1}^{n_{T}+1}\ell_{\text{LST},I}{h}^\vee_{\mathfrak{g}^{(n)}_{I}}\,,&\qquad 
\kappa_{P}=-\sum_{I=1}^{n_{T}+1}\ell_{\text{LST},I}(\eta^{II}-2)\,.
\end{array}
$}
\end{align}
Here, dim(CB) is the dimension of the 5D Coulomb branch, dim(WL) is the number of Wilson lines for the $n$-twisted flavor algebra $\ff^{(n)}$, and $\kappa_{R}$ and $\kappa_{P}$ are the 2-group structure constants. Moreover, $\mathfrak{g}^{(n)}$ refers to the $n$-twisted 6D gauge algebra of the LST with dual Coxeter number ${h}^\vee_{\mathfrak{g}^{(n)}_{I}}$. The $\ell_{\text{LST},I}$ are the components of the null vector of $\eta^{IJ}$, and correspond to the \textit{little string charge} under the $\mathfrak{u}_1$ LST 1-form symmetry. The matching data is similar for twists applied to the tensor multiplets, as we will discuss in detail in 
Section~\ref{sec:basetwists}. 

A field theory justification for why the twisted Coulomb branch dimension and the flavor rank have to match is that they correspond to continuous deformations of the theories. The 2-group structure constants have to match because they are related to certain mixed anomalies in the 6D theory. Geometric constructions of untwisted LSTs via F/M-theory were used to establish that dim(CB), $\kappa_{R}$ and $\kappa_{P}$ are T-duality invariants~\cite{DelZotto:2022ohj,DelZotto:2022xrh}, while for the match of rk($\mathfrak{g}_{F}^{(1)}$), the geometry had to be supplemented with field-theoretic arguments~\cite{Ahmed:2023lhj}. We focus on heterotic LSTs, whose T-duality descends from the usual T-duality between the $\mathfrak{e}_{8} \times \mathfrak{e}_{8}$ and $\mathfrak{so}_{32}$ heterotic string. We first review the construction of these theories, and then move to their twisted versions. We end the section by matching the relevant data across twisted T-duals.

\subsection{Review: Heterotic LSTs and untwisted T-duality}
Heterotic little string theories arise as the worldvolume theories of NS5 branes in heterotic string theory. In particular, a rank $M$ LST with minimal SUSY arises as the worldvolume theory of $M$ NS5 branes probing an ALE singularity $\mathbbm{C}^{2}/\Gamma_{\mathfrak{g}}$ in the heterotic string context, where $\Gamma_{\mathfrak{g}} \subset SU(2)$ and $\mathfrak{g}$ labels the Lie algebra according to the McKay correspondence \cite{Intriligator:1997dh,Blum:1997mm,Blum:1997fw}. From the point of view of the worldvolume theory, the $\mathfrak{e}_{8} \times \mathfrak{e}_{8}$ or $\mathfrak{so}_{32}$ gauge bundle corresponds to the flavor symmetry of the LST.
Since $\pi_{1}(S^{3}/\Gamma_{\mathfrak{g}}) \cong \Gamma_{\mathfrak{g}}$ (where the lens space $S^{3}/\Gamma_{\fg}$ is the boundary of the ALE space), one also has to specify a choice of flat connection at infinity. This choice is encoded in the morphisms $\mu_{a}$ and $\lambda$:
\begin{align}
\label{eq:flatcons}
\mu_{a}&: \pi_{1}(S^{3}/\Gamma_{\mathfrak{g}}) \cong \Gamma_{\mathfrak{g}} \hookrightarrow \text{E}_{8}\,. \nonumber \\
\lambda&: \pi_{1}(S^{3}/\Gamma_{\mathfrak{g}}) \cong \Gamma_{\mathfrak{g}} \hookrightarrow \text{Spin}(32)/\mathbbm{Z}_{2}\,,
\end{align}
where $a=1,2$ labels the two $E_{8}$ factors. Non-trivial values of these morphisms break the respective flavor groups to their commutants. The $\fe_8\times\fe_8$ and $\fso_{32}$ heterotic LSTs are denoted by $\mathcal{K}_{M}(\mu_{1},\mu_{2},\mathfrak{g})$ and $\phantom{}_\pm\mathcal{\Tilde{K}}_{M}(\lambda;\mathfrak{g})$, respectively, where the $+$ and $-$ indicate $\mathfrak{so}_{32}$ theories with and without vector structure, which we discuss in more detail in Section~\ref{subsec:twso32}.

NS5-brane theories in the $\mathfrak{e}_{8} \times \mathfrak{e}_{8}$ LST are inherently strongly coupled and are therefore best described in the M-theory Ho\v{r}ava-Witten (HW) setup~\cite{Horava:1995qa}. For our purposes, a rank $M$ $\mathfrak{e}_{8} \times \mathfrak{e}_{8}$ LST, of ADE type $\mathfrak{g}$, is given by the quiver
\begin{align}
\lbrack \mathfrak{e}_{8}\rbrack  \, \,
{\overset{\mathfrak{g}}{1}}\, \,
\underbrace{{\overset{\mathfrak{g}}{2}}\, \, \, \,
{\overset{\mathfrak{g}}{2}}\, \,\, \, 
. \, \,
. \, \,
. \, \,
{\overset{\mathfrak{g}}{2}}\, \, \, \,
{\overset{\mathfrak{g}}{2}} \, \,}_{M-1}
{\overset{\mathfrak{g}}{1}}\, \,
\lbrack \mathfrak{e}_{8}\rbrack  \,.
\end{align}
Alternatively, one may obtain such an LST by fusing two rank-one orbi-instanton SCFTs $\mathcal{T}(\mu_{a},\mathfrak{g})$, along with insertions of superconformal matter SCFTs $\mathcal{T}_{M-2}(\mathfrak{g},\mathfrak{g})$, for $M>2$. The fusion of these theories can be thought of as gauging the diagonal global symmetry algebras $\mathfrak{g}$ of the two SCFTs, which we denote by $\mathrel{\stackon[-1pt]{{-}\mkern-5mu{-}\mkern-5mu{-}\mkern-5mu{-}}{\mathfrak{g}}}$, and then coupling the gauge algebra $\mathfrak{g}$ to a tensor multiplet. This process can be written as
\begin{align}
\label{eq:orbiconforbi}
\mathcal{K}_{N}(\mu_{1},\mu_{2},\mathfrak{g})=\mathcal{T}(\mu_{1},\mathfrak{g}) \mathrel{\stackon[-1pt]{{-}\mkern-5mu{-}\mkern-5mu{-}\mkern-5mu{-}}{\mathfrak{g}}} \mathcal{T}_{M-2}(\mathfrak{g},\mathfrak{g}) \mathrel{\stackon[-1pt]{{-}\mkern-5mu{-}\mkern-5mu{-}\mkern-5mu{-}}{\mathfrak{g}}} \mathcal{T}(\mu_{2},\mathfrak{g})\,.
\end{align}
The above presentation is beneficial in the discussion of twisted LSTs, since it reduces the discussion to twists of the individual SCFT building blocks. These orbi-instanton building blocks and several Higgs branches were classified in \cite{Frey:2018vpw}, and the structure of the conformal matter theories was worked out in \cite{DelZotto:2014fia}.

For the $\mathfrak{so}_{32}$ case, one has a description in terms of Type~I string theory via strong-weak duality \cite{Blum:1997fw,Intriligator:1997dh}. The rank $M$ of this LST is given by the number of D5 branes in the Type~I theory, and the $\mathfrak{so}_{32}$ flavor symmetry arises from 16 D9 branes on top of an O9 orientifold plane. The quiver of an $\mathfrak{so}_{32}$ LST of type $\mathfrak{g}$ has the shape of the corresponding affine Dynkin diagram. For example for $\fg=\fe_8$, the quiver is
\begin{align*}
 [\fso_{32}] \, {\overset{\mathfrak{\fsp}_{M+9}}{1}} \, {\overset{\fso_{4M+20}}{4}}\, {\overset{\mathfrak{\fsp}_{3M+3}}{1}} \, {\overset{\fso_{8M+8}}{4}}\, {\overset{\mathfrak{\fsp}_{5M-3}}{1}} \, \overset{{\overset{\mathfrak{\fsp}_{3M-5}}{1^*}}}{\overset{\fso_{12M-4}}{4}} \, {\overset{\mathfrak{\fsp}_{4M-4}}{1}} \, {\overset{\fso_{4M+4}}{4}}\,.
\end{align*}
We summarize all quiver shapes in Table~\ref{tab:so32top}. Note that for some flavor configurations of $\fso_{32}$, one can fold the base quiver~\cite{Intriligator:1997dh}, which can be used to describe theories with O$7^+$ planes~\cite{Oehlmann:2024cyn}.

It has been proposed~\cite{DelZotto:2020sop} that the 2-group structure constant $\kappa_P$ counts the number of M9 branes, and hence $\kappa_P=2$ for the heterotic string. Moreover, since $\kappa_P$ is invariant under T-duality, it follows that 
heterotic LSTs close under T-duality. The 5D Coulomb branch and $\kappa_R$ are additive quantities in the quiver and can be computed from the constituents $\mathcal{T}(\mu_i,\fg)$ and $\mathcal{T}_{M}(\fg,\fg)$ of the quiver in the $\fe_8$ case, and similarly for the $\fso_{32}$ case. Therefore, the 5D Coulomb branch dimension and $\kappa_R$ are linear functions in the number of five-branes $M$, 
\begin{align}
 \label{eq:LinearRelation}
 (CB,\kappa_R)(\mathcal{K}_{M}(\fg))= (c_1 + c_2 M,~r_1+r_2 M) \, .
\end{align}
The coefficients $c_2$ and $r_2$ in these linear relations determine what we call the \textit{fractionalization coefficients} of the five-branes probing a singularity $\fg$ in either theory, and are given by~\cite{Baume:2024oqn}
\begin{align}
c_2 = h^\vee_{\fg^{(1)}} \, , \qquad r_2 =| \Gamma_\fg | \, .
\end{align}
From this, we see that the type of singularity on either side of the (untwisted) T-duality
\begin{align}
 \mathcal{K}_{M}(\mu_1, \mu_2; \fg)\,
 \xleftrightarrow{\text{T-dual }} \, \mathcal{\phantom{}_\pm\Tilde{K}}_{M}(\lambda;\mathfrak{g})
\end{align}
must be identical, and the holonomies need to be chosen such that the flavor rank and as well as the constant coefficients $c_1$ and $r_1$ in~\eqref{eq:LinearRelation} match.

\subsection{Twisted heterotic LSTs}
\label{section:twisthet} 
The main focus of this work is to explore the landscape of T-dual LSTs when we use discrete symmetries of the 6D theory to twist the 5D theory along the circle direction. In this section, we will focus on twists by outer automorphisms $\mathcal{O}$ of the gauge and flavor symmetries of an LST $\mathcal{K}$. As explained in Section~\ref{section:frev}, these twists reduce the rank of gauge and flavor algebras and therefore affect the data in \eqref{eq:invdata}, such that
\begin{align}
 \text{dim(CB}( \mathcal{K}^{(n)} )) \leq \text{dim(CB}( \mathcal{K}^{(1)} )) \,, \qquad \text{dim(WL}( \mathcal{K}^{(n)} )) \leq \text{dim(WL}( \mathcal{K}^{(1)} )) \, .
\end{align}
Since the action on tensor multiplets is trivial, the LS charges are unchanged and hence $\kappa_P$ is invariant under twisting. Since $\kappa_R$ is given in terms of the dual Coxeter number of the gauge algebra $\fg$, it also stays invariant under fibral outer automorphism twists $\fg^{(1)} \rightarrow \fg^{(n)}$, since
\begin{align}
 h^\vee_{\fg^{(1)}} =h^\vee_{\fg^{(n)}} \,.
\end{align}
Hence, the 2-group structure constants are unchanged under twisting by outer automorphisms and they match those obtained in 6D, which we may also describe as the untwisted 5D KK-theory 
\begin{align}
\mathfrak{\kappa}_{R}( \mathcal{K}^{(n)} ) = \mathfrak{\kappa}_{R}( \mathcal{K}^{(1)} )\,,\qquad
\mathfrak{\kappa}_{P}( \mathcal{K}^{(n)} ) = \mathfrak{\kappa}_{P}( \mathcal{K}^{(1)} )\,.
\end{align}
However, note that the twisted/untwisted affine extension of an algebra that appears in the 5D theory does have different dual Coxeter numbers. For example both $\fe_6^{(2)}$ and $\ff_4^{(1)}$ have the same $\mathfrak{f}_4$ finite subalgebra, but their dual Coxeter numbers are different. Thus, this does not hold for twists $\mathcal{B}^{(n)}$ that act on the tensor multiplets, as we will discuss in detail in Section~\ref{sec:basetwists}.

\subsubsection{Twisted \texorpdfstring{$\fe_8\times\fe_8$}{e8xe8} heterotic LSTs}
We start by discussing the general structure of twisted $\mathfrak{e}_{8} \times \mathfrak{e}_{8}$ heterotic LSTs. First, recall that such theories can be viewed as the fusion of two rank-one orbi-instanton theories $\mathcal{T}(\mu_i, \fg)$ along a rank $M-2$ conformal matter chain $\mathcal{T}_{M-2}(\fg,\fg)$. From the 6D perspective, we may twist any component of this chain individually as long as the flavor algebras $\fg$ along which we fuse are the same (twisted) algebras. This results in a twisted quiver 
\begin{align}
\label{eq:orbicmorbi}
\mathcal{K}_{M}(\mu_{1},\mu_{2},\mathfrak{g})&\xrightarrow{} \mathcal{K}^{(n_{\text{CM}}, n_{\alpha}, m_{\beta})}_{N}(\mu_{1},\mu_{2},\mathfrak{g})\nonumber \\[4pt] 
&=\mathcal{T}(\mu_{1}^{(n_{\alpha})},\mathfrak{g}^{(n_{\text{CM}})}) \mathrel{\stackon[-1pt]{{-}\mkern-5mu{-}\mkern-5mu{-}\mkern-5mu{-}}{\mathfrak{g}^{(n_{\text{CM}})}}} \mathcal{T}_{N-2}(\mathfrak{g}^{(n_{\text{CM}})},\mathfrak{g}^{(n_{\text{CM}})}) \mathrel{\stackon[-1pt]{{-}\mkern-5mu{-}\mkern-5mu{-}\mkern-5mu{-}}{\mathfrak{g}^{(n_{\text{CM}})}}} \mathcal{T}(\mu_{2}^{(m_{\beta})},\mathfrak{g}^{(n_{\text{CM}})})\,,
\end{align}
which we use to label the twisted 5D KK-theory. We will suppress the flat connections $\mu_{1}, \mu_{2}$ from now on to not clutter the notation too much.

We denote twists along the conformal matter chain by $\mathcal{O}^{(n_{\text{CM}})}$, and twists of the two orbi-instanton pieces by $\mathcal{O}^{(n_{\alpha})}$ and $\mathcal{O}^{(m_{\beta})}$, where $\alpha$ and $\beta$ label different discrete symmetries, and hence different twists. When there is only one possible twist, we drop the subscripts. As discussed in Section~\ref{section:frev}, twisting a single conformal matter piece requires the twisting of the entire rank $M-2$ conformal matter chain. As for the orbi-instanton pieces, we can have multiple choices for twisting. This is because, as argued in Section~\ref{section:frev}, one can have multiple choices in the case where a (1)-node has more than two neighboring algebras. The fusion of these pieces to an LST can then lead to inequivalent twisted 5D theories. Note that $n_{\alpha}, n_{\beta} \neq 1$ if $n_{\text{CM}} \neq 1$, since in order to fuse the orbi-instanton piece to the rank $M-2$ twisted conformal matter theory, we need to twist at least the ``endpoints'' of the orbi-instanton theories. 

In \cite{DelZotto:2022ohj}, it was shown how the SCFT pieces contribute to~\eqref{eq:LinearRelation}, with $r_1,c_1$ originating from the orbi-instanton pieces and $r_2,c_2$ originating from the $M-2$ conformal matter theories $\mathcal{T}_{M-2}(\fg,\fg)$. The latter depend only on $M$ and the singularity $\fg$, so we can discuss their contributions to the twisted theories systematically. On a partial tensor branch, a rank $M$ conformal matter chain is given by 
\begin{align}
\label{eq:CMchain}
 \mathcal{T}_N(\fg,\fg): \quad [\fg] \,\, \underbrace{ \overset{\fg}{2} \, \, \overset{\fg}{2} \, \, \ldots \overset{\fg}{2} }_{\times M}\, \, [\fg] \, .
\end{align}
We view the above pieces individually as
\begin{align} 
\label{eq:CMLink}
 [\fg] \, \, \overset{\fg}{2} \, ,
\end{align}
which must be replaced by the conformal matter contribution according to~\cite{DelZotto:2014hpa}. If $\fg$ admits an order $n_{\text{CM}}$ outer automorphism, we can twist the respective conformal matter block and compute $r_2$ and $c_2$. The twisting conditions reviewed in Section~\ref{ssec:reviewDiscrete} can be used to deduce how to twist the conformal matter block and the entire rank $M-2$ chain. From the action of the twist, we can compute the twisted fractionalization coefficients $c_2$ (as argued before, $r_2$ is invariant).
\begin{table}[t!]
 \centering
 \begin{tabular}{|c|c|c|}\hline 
 $\fg^{(n)}$  & Quiver & $(c_2,r_2) $\\ \hline
  $\fso_{2N+8}^{(1)} $ & $[\fso_{2N+8}^{(1)}] \, \overset{\fsp_N^{(1)}}{1} \overset{\fso_{2N+8}^{(1)}}{4}$ & (6+2N,8+4N) \\
  $\fso_{2N+8}^{(2)} $ & $[\fso_{2N+8}^{(2)}] \, \overset{\fsp_N^{(1)}}{1} \overset{\fso_{2N+8}^{(2)}}{4}$ & (5+2N,8+4N) \\

  $\fso_{8}^{(3)} $ & $[\fso_{8}^{(3)}] \, 1 \, \overset{\fso_{8}^{(3)}}{4}$ & (4,8) \\ \hline 
  $\fe_{6}^{(1)} $ & $[\fe_6^{(1)}] \, 1 \, \overset{\fsu_3^{(1)}}{3} \, 1 \overset{\fe_6^{(1)}}{6}$ & (12,24) \\ 
   $\fe_{6}^{(2)} $ & $[\fe_6^{(2)}] \, 1 \, \overset{\fsu_3^{(2)}}{3} \, 1 \overset{\fe_6^{(2)}}{6}$ & (9,24) \\ \hline
$\fe_{7}^{(1)} $ & $[\fe_7^{(1)}] \, 1 \, \overset{\fsu_2^{(1)}}{2}\, \overset{\fso_7^{(1)}}{3}\, \overset{\fsu_2^{(1)}}{2} \, 1 \overset{\fe_7^{(1)}}{8}$ & (18,48) \\ \hline
$\fe_{8}^{(1)} $ & $[\fe_8^{(1)}] \, 1 \, 
2 \, \overset{\fsu_2^{(1)}}{2} \, \overset{\fg_2^{(1)}}{3}\, 1 \, \overset{\ff_4^{(1)} }{5} \,1 \,\overset{\fg_2^{(1)}}{3} \,\overset{\fsu_2^{(1)}}{2} \, 2 \,1 \, \overset{\fe_8}{12}$ & (30,120) \\ \hline
 \end{tabular}
 \caption{Summary of twisted conformal matter blocks, Coulomb branch dimensions $c_2$, and 2-group contributions $r_2$. }
 \label{tab:TCMSummary}
\end{table} 
We have summarized the relevant conformal matter blocks, as well as their Coulomb branch dimensions and 2-group contributions in Table~\ref{tab:TCMSummary}. 
Notably, for the twisted conformal blocks, we are still able to express $c_2$ in terms of the dual Coxeter number of its untwisted Langlands dual $L(\fg^{(n)})=\widetilde{\fg}^{(1)}$,
\begin{align}
 c_2 (\fg^{(n)})= h^\vee_{\widetilde{\fg}^{(1)}} \, , \qquad r_2 = |\Gamma_{\fg}|\,,
\end{align}
see also in \cite{DelZotto:2014hpa,Baume:2024oqn}. Concretely, the duals are
\begin{align}
    L(\fso_{2N+8}^{(2)})= \fsp^{(1)}_{2N+6}\,,\qquad L(\fso_{8}^{(3)})= \fg_2^{(1)}\,,\qquad L(\fe_{6}^{(2)})= \ff_4^{(1)}\,.
\end{align}
In addition, we have a choice for the rank $1$ orbi-instanton theory $\mathcal{T}(\mu,\fg)$ and its twist. While the conformal matter theories only have ADE singularities, the orbi-instanton theories also require specification of a non-trivial holonomy that breaks the $\fe_8$ flavor symmetry. This leads to a large number of possibilities, with typically tens of choices for each $\fg$. Furthermore, when the tensor branch of the orbi-instanton theory has (1)-nodes, there can be multiple ways to twist the theory, depending on the neighboring algebras. We list the orbi-instanton theories and their twists for $\fg=\fso_8$ and flavor ranks 7 and 8 in Table~\ref{tab:so8OrbiInstantonsso8}, and for $\fg=\fe_6$ in Table~\ref{tab:e6OrbiInstantons}. Parts of this classification have previously appeared in~\cite{Frey:2018vpw}. 
\begin{table}[t!]
 \centering
\renewcommand{\arraystretch}{2}
 \begin{tabular}{|c c|c|c|} 
 \hline
&&$(\mathbbm{Z}_2, \mathbbm{Z}_3)$ twist&\# of distinct twists\\[-12pt]
 \multicolumn{2}{|c|}{$\fg=\fso_8$}& possible? & $( \mathbbm{Z}_2,  \mathbbm{Z}_3,  \mathbbm{Z}_2 \times  \mathbbm{Z}_{2}, \mathbbm{Z}_2 \times  \mathbbm{Z}_{3}) $\\ \hline
 $ \mathcal{T}(\fe_8,\fso_8)$ & $ [ \fe_8] \, 1 \,2 \,\overset{\fsu_2}{2}\, \overset{\fg_2}{3}\, 1\, [\fso_8]\ldots $& (\checkmark,\checkmark) & (1,1,0,0) \\
   $ \mathcal{T}(\fe_7\times \fsu_2,\fso_8)$ & $ [ \fe_7] \,1\, \overset{\fsu_2}{2}\, \overset{\fso_7}{\underset{\displaystyle [\fsu_2]}{3}} \, 1 \,[\fso_8]\ldots $ & (\checkmark,X) & (1,0,0,0) \\
 $ \mathcal{T}(\fso_{16},\fso_8)$ & $ [ \fso_{16}]\, \overset{\fsp_2}{1} \, \overset{\fso_7}{3} \,1\, [\fso_8]\ldots $& $(\checkmark, $X$ )$ & (2,0,1,0) \\
$ \mathcal{T}(\fso_{8}
\times \fso_8,\fso_8)$ & $ [ \fso_{8}]\, 1 \, \overset{\displaystyle \overset{\displaystyle [\fso_8]}{1}}{\overset{\fso_8}{4}} \, 1\, [\fso_8]\ldots $ & $(\checkmark,\checkmark)$ & (1,1,0,0) \\ 
$ \mathcal{T}(\fso_{12}
\times \fso_4,\fso_8)$ & $ [ \fso_{12}]\, \overset{\fsp_1}{1}\, \overset{\fso_8}{\displaystyle \underset{\displaystyle [\fsu_2^2]}{3}}\, 1\, [\fso_8]\ldots $ & $(\checkmark,$X$)$ & (1,0,0,0) \\ 

$ \mathcal{T}(\fsu_{8}
\times \mathfrak{u}_1,\fso_8)$ & $ [ \fsu_{8}] \,\overset{\fsu_4}{2} \,\underset{\displaystyle [\mathfrak{u}_1]}{1} \,\overset{\fso_8}{4} \, 1 \,[\fso_8]\ldots $ & $(\checkmark,$X$)$ & (3,0,0,0) \\ \hline
$ \mathcal{T}(\fe_7
 ,\fso_8)$ & $ [ \fe_{7}]\, 1 \,\overset{\fsu_2}{2} \, \overset{\fg_2}{3}\, 1\, [\fso_8]\ldots $ & $(\checkmark,\checkmark)$ & (1,1,0,0)\\ 

$ \mathcal{T}(\fe_6 \times \mathfrak{u}_1
 ,\fso_8)$ & $ [ \fe_{6}]\, 1 \,\overset{\fsu_3}{3} \, \underset{\displaystyle [\mathfrak{u}_{1}^{2}]}{1} \, [\fso_8]\ldots $ & ($\checkmark,\checkmark$) & (2,1,1,1) \\ 
 
$\mathcal{T}( \fso_{12} \times \fsu_2,\fso_8)$ & $ [ \fso_{12}] \,\overset{\fsp_1}{1} \, \overset{\fso_7}{\displaystyle \underset{\displaystyle [\fsu_2]}{3}}\, 1 \,[\fso_8]\ldots $ & $(\checkmark,$X$)$ & (2,0,1,0) \\ 

$\mathcal{T}( \fso_{8} \times \fsu_2^3,\fso_8)$ & $ [ \fso_{8}] \, 1 \, \overset{\fso_8}{\underset{\displaystyle [\fsu_2^3]}{3}} \, 1 \,[\fso_8]\ldots $ &$(\checkmark,\checkmark)$ & (1,1,0,0) \\ 
$\mathcal{T}( \fsu_{6} \times \mathfrak{u}_1^2,\fso_8)$ & $ [ \fsu_{6}] \,\overset{\fsu_3}{2} \, \underset{\displaystyle [\mathfrak{u}_1^2]}{1} [\fso_8]\ldots $ & $(\checkmark,\checkmark)$ & (2,1,1,1) \\ \hline
 \end{tabular}
 \caption{ $\fg=\fso_8$ orbi-instanton theories with flavor ranks 8 and 7 according to \cite{Frey:2018vpw}, and their possible twists.}
 \label{tab:so8OrbiInstantonsso8}
\end{table}
Any two combinations of the orbi-instanton theories can be fused into a 6d LST, and then twisted upon compactification to 5D. As highlighted in Table~\ref{tab:so8OrbiInstantonsso8}, most of the orbi-instanton theories admit $\mathbbm{Z}_2$ or $\mathbbm{Z}_3$ twists, since one can almost always twist $\mathfrak{so}_{8}$ flavor factor.

\begin{table}[t!]
 \centering
\renewcommand{\arraystretch}{2} 
 \begin{tabular}{|c c|c|} 
 \hline
&&\# of distinct twists\\[-12pt]
 \multicolumn{2}{|c|}{$\fg=\fe_6$} & $( \mathbbm{Z}_2,  \mathbbm{Z}_2 \times  \mathbbm{Z}_{2}) $\\ \hline
 $ \mathcal{T}(\fe_8,\fe_6)$ & $ [ \fe_8] 1 2 \overset{\fsu_2}{2} \overset{\fg_2}{3} 1 \overset{\ff_4}{5} 1 \overset{\fsu_3}{3}1 [\fe_6]\ldots $ & (1,0) \\
   $ \mathcal{T}(\fe_6 \times \fsu_3,\fe_6)$ & 
   $ [ \fe_6] 1 \overset{\fsu_3}{3} 1  \overset{\overset{\overset{[\fsu_3]}{1}}{ \fe_6}}{6} 1 \overset{\fsu_3}{3}1 [\fe_6]\ldots $ 
  & (1,0) \\
 $ \mathcal{T}(\fsu_{9},\fe_6)$ & $ [ \fsu_{9}] \overset{\fsu_6}{2} \overset{\fsu_3}{2} 1 [\fe_6]\ldots $ & (1,0) \\ 

  $ \mathcal{T}(\fso_{14}\times \mathfrak{u}_1,\fe_6)$ & $ [ \fso_{14}] 
 \overset{\fsp_2}{1} \overset{\fso_{10}}{4} \underset{[\mathfrak{u}_1]}{1} \overset{\fsu_3}{3} 1 [\fe_6]\ldots $ & (2,0) \\ \hline
 
 $ \mathcal{T}(\fe_{7},\fe_6)$ & $ [ \fe_{7}] 1 \overset{\fsu_2}{2} \overset{\fg_2}{3} 1 \overset{\ff_4}{5} 1 \overset{\fsu_3}{3} 1 [\fe_6]\ldots $ & (1,0) \\ 
  $ \mathcal{T}(\fsu_6 \times \fsu_3,\fe_6)$ & $ [ \fsu_6] \overset{\fsu_3}{2} 1 \overset{\overset{[\fsu_3]}{1}}{\overset{\fe_6}{6}} 1\overset{\fsu_3}{3}1 [\fe_6]\ldots $ & (1,0) \\ 

$ \mathcal{T}(\fe_{6}\times \mathfrak{u}_1,\fe_6)$ & $ [ \fe_6] 1 \overset{\fsu_3}{3} 1 \overset{\fe_6}{\underset{[\mathfrak{u}_1]}{5}} 1 \overset{\fsu_3}{3} 1 [\fe_6] \ldots $ & (1,0) \\ 

$ \mathcal{T}(\fso_{12}\times \mathfrak{u}_1,\fe_6)$ & $ [ \fso_{12}] \overset{\fsp_1}{1} \overset{\fso_7}{3} \overset{\fsu_2}{2} \underset{[\mathfrak{u}_1]}{1} [\fe_6] \ldots$ & (2,1)\\ 

  $ \mathcal{T}(\fsu_{7}\times \mathfrak{u}_1,\fe_6)$ & $ [ \fsu_{7}] \overset{\fsu_5}{2} \overset{\fsu_3}{\underset{[\mathfrak{u}_1]}{2}} 1 [\fe_6]\ldots $ & (1,0) \\ 

  $ \mathcal{T}(\fso_{10}\times \fsu_2 \times \mathfrak{u}_1,\fe_6)$ & $ [ \fso_{10}] \overset{\fsp_1}{1} \overset{\fso_{10}}{\underset{[\fsu_2]}{4}} \underset{[\mathfrak{u}_1]}{1} \overset{\fsu_3}{3} 1 [\fe_6] \ldots$ & (2,0) \\ 
  \hline
 \end{tabular}
 \caption{ $\fg=\fe_6$ orbi-instanton theories with flavor ranks 8 and 7 according to \cite{Frey:2018vpw}, and their possible twists.}
 \label{tab:e6OrbiInstantons}
\end{table}

In the following we discuss a particular rich example, where we have eight inequivalent twisted theories. 
This LST arises from fusing rank $N$ $\mathfrak{so}_{8}-\mathfrak{so}_{8}$ conformal matter with two copies of the orbi-instanton theory
\begin{align} 
\label{eq:orbitun1}
\lbrack \mathfrak{e}_{6}\rbrack  \, \, 
1 \, \,
{\overset{\mathfrak{su}_{3}}{3}} \, \,
\underset{\left[\mathfrak{u}_{1} \times \mathfrak{u}_{1}\right]}
1 \, \,
\overset{\mathfrak{so}_{8}}{4} \, \,
\cdots \lbrack \mathfrak{so}_{8}\rbrack\,,
\end{align}
cf.\ Table~\ref{tab:so8OrbiInstantonsso8}. The 6D LST is
\begin{align}
\label{exagain}
\mathcal{K}^{(1,1,1)}_{N}(\mathfrak{so}_{8}): \qquad 
\lbrack \mathfrak{e}_{6}\rbrack  \, \, 
1 \, \,
{\overset{\mathfrak{su}_{3}}{3}} \, \,
\underset{\left[\mathfrak{u}_{1} \times \mathfrak{u}_{1}\right]}
1 \, \,
\underbrace{
\overset{\mathfrak{so}_{8}}{4} \, \,
\underset{\left[\mathfrak{u}_{1} \times \mathfrak{u}_{1}\right]}
1 \, \,}_{\times (N)} \, \,
{\overset{\mathfrak{su}_{3}}{3}} \, \,
1 \, \,
\lbrack \mathfrak{e}_{6}\rbrack \, ,
\end{align}
and the invariant 6D data is
\begin{align}
\text{rk}(\ff)=16\, , \qquad  \text{dim(CB)}=6N +8 \, , \qquad \kappa_{R}=8N+10\, .
\end{align}
From Table \ref{tab:so8OrbiInstantonsso8}, one can deduce that this theory has eight distinct discrete symmetries, due to the various allowed twists of the orbi-instanton pieces. More explicitly, these inequivalent 5D twisted theories are given as:
\begin{align}
\label{ex4t1am}
&\mathcal{K}^{(1,2_{1},1)}_{N}(\mathfrak{so}_{8}): \qquad 
\lbrack \mathfrak{e}_{6}^{(2)}\rbrack  \, \, 
1 \, \,
{\overset{\mathfrak{su}_{3}^{(2)}}{3}} \, \,
1 \, \,
\underbrace{
\overset{\mathfrak{so}_{8}}{4} \, \,
{\overset{{\left[\mathfrak{u}_{1} \times \mathfrak{u}_{1}\right]}}{1}} \, \,}_{\times (N)} \, \,
{\overset{\mathfrak{su}_{3}}{3}} \, \,
1 \, \,
\lbrack \mathfrak{e}_{6}\rbrack  \, ,  
\qquad & \begin{array}{l}\text{dim(CB)}= 6N+7 \\
 \text{rk}(\ff)=12 \end{array} \\
\label{ex4t2am}
&\mathcal{K}^{(1,2_{1},2_{1})}_{N}(\mathfrak{so}_{8}): \qquad 
\lbrack \mathfrak{e}_{6}^{(2)}\rbrack  \, \, 
1 \, \,
{\overset{\mathfrak{su}_{3}^{(2)}}{3}} \, \,
1 \, \,
\underbrace{
\overset{\mathfrak{so}_{8}}{4} \, \,
1 \, \,}_{\times (N)} \, \,
{\overset{\mathfrak{su}_{3}^{(2)}}{3}} \, \,
1 \, \,
\lbrack \mathfrak{e}_{6}^{(2)}\rbrack  \, , 
\qquad &\begin{array}{l}\text{dim(CB)}= 6N+6 \\
   \text{rk}(\ff)=8
\end{array}\\
\label{ex4t3am}
&\mathcal{K}^{(2,2_{2},2_{2})}_{N}(\mathfrak{so}_{8}): \qquad
\lbrack \mathfrak{e}_{6}\rbrack  \, \, 
1 \, \,
{\overset{\mathfrak{su}_{3}}{3}} \, \,
{\overset{{\left[\mathfrak{u}_{1}\right]}}{1}}\, \,
\underbrace{
\overset{\mathfrak{so}_{8}^{(2)}}{4} \, \,
{\overset{{\left[\mathfrak{u}_{1}\right]}}{1}} \, \,}_{\times (N)} \, \,
{\overset{\mathfrak{su}_{3}}{3}} \, \,
1 \, \,
\lbrack \mathfrak{e}_{6}\rbrack  \, , 
\qquad &\begin{array}{l}\text{dim(CB)}= 5N+8 \\
   \text{rk}(\ff)=14
\end{array} \\
\label{ex4t4am}
&\mathcal{K}^{(2,2_{3},2_{2})}_{N}(\mathfrak{so}_{8}): \qquad 
\lbrack \mathfrak{e}_{6}^{(2)}\rbrack  \, \, 
1 \, \,
{\overset{\mathfrak{su}_{3}^{(2)}}{3}} \, \,
{\overset{{\left[\mathfrak{u}_{1}\right]}}{1}} \, \,
\underbrace{
\overset{\mathfrak{so}_{8}^{(2)}}{4} \, \,
{\overset{{\left[\mathfrak{u}_{1}\right]}}{1}} \, \,}_{\times (N)} \, \,
{\overset{\mathfrak{su}_{3}}{3}} \, \,
1 \, \,
\lbrack \mathfrak{e}_{6}\rbrack  \, , 
\qquad &\begin{array}{l} \text{dim(CB)}= 5N+7 \\ 
  \text{rk}(\ff)=12
\end{array}\\ 
\label{ex4t5am}
&\mathcal{K}^{(2,2_{3},2_{3})}_{N}(\mathfrak{so}_{8}): \qquad 
\lbrack \mathfrak{e}_{6}^{(2)}\rbrack  \, \, 
1 \, \,
{\overset{\mathfrak{su}_{3}^{(2)}}{3}} \, \,
{\overset{{\left[\mathfrak{u}_{1}\right]}}{1}}\, \,
\underbrace{
\overset{\mathfrak{so}_{8}^{(2)}}{4} \, \,
{\overset{{\left[\mathfrak{u}_{1}\right]}}{1}} \, \,}_{\times (N)} \, \,
{\overset{\mathfrak{su}_{3}^{(2)}}{3}} \, \,
1 \, \,
\lbrack \mathfrak{e}_{6}^{(2)}\rbrack  \, , 
\qquad &\begin{array}{l}\text{dim(CB)}= 5N+6 \\
  \text{rk}(\ff)=10
\end{array}\\
\label{ex4t6am}
&\mathcal{K}^{(3,3,3)}_{N}(\mathfrak{so}_{8}): \qquad 
\lbrack \mathfrak{e}_{6}\rbrack  \, \, 
1 \, \,
{\overset{\mathfrak{su}_{3}}{3}} \, \,
1 \, \,
\underbrace{
\overset{\mathfrak{so}_{8}^{(3)}}{4} \, \,
1 \, \,}_{\times (N)} \, \,
{\overset{\mathfrak{su}_{3}}{3}} \, \,
1 \, \,
\lbrack \mathfrak{e}_{6}\rbrack  \, , 
\qquad &\begin{array}{l}\text{dim(CB)}= 4N+8 \\ 
  \text{rk}(\ff)=12
\end{array} \\ 
\label{ex4t7am}
&\mathcal{K}^{(3,6,3)}_{N}(\mathfrak{so}_{8}): \qquad 
\lbrack \mathfrak{e}_{6}^{(2)}\rbrack  \, \, 
1 \, \,
{\overset{\mathfrak{su}_{3}^{(2)}}{3}} \, \,
1 \, \,
\underbrace{
\overset{\mathfrak{so}_{8}^{(3)}}{4} \, \,
1 \, \,}_{\times (N)} \, \,
{\overset{\mathfrak{su}_{3}}{3}} \, \,
1 \, \,
\lbrack \mathfrak{e}_{6}\rbrack  \, , 
\qquad & \begin{array}{l} \text{dim(CB)}= 4N+7 \\
  \text{rk}(\ff)=10
\end{array} \\
\label{ex4t8am}
&\mathcal{K}^{(3,6,6)}_{N}(\mathfrak{so}_{8}): \qquad 
\lbrack \mathfrak{e}_{6}^{(2)}\rbrack  \, \, 
1 \, \,
{\overset{\mathfrak{su}_{3}^{(2)}}{3}} \, \,
1 \, \,
\underbrace{
\overset{\mathfrak{so}_{8}^{(3)}}{4} \, \,
1 \, \,}_{\times (N)} \, \,
{\overset{\mathfrak{su}_{3}^{(2)}}{3}} \, \,
1 \, \,
\lbrack \mathfrak{e}_{6}^{(2)}\rbrack  \, , 
\qquad &\begin{array}{l}\text{dim(CB)}= 4N+6 \\
  \text{rk}(\ff)=8
\end{array}
\end{align} 
where we have also shown the data that changes under the twisting. All theories have different Coulomb branch dimensions and are therefore in disconnected components of the 5D moduli space. In Section~\ref{sec:Geometry}, we discuss this specific example again in relation to the Tate-Shafarevich and Weil-Ch$\hat{\text{a}}$talet group.

\subsubsection{Twisted \texorpdfstring{$\fso_{32}$}{so32} LSTs}
\label{subsec:twso32}
Next, we consider twisted $\fso_{32}$ heterotic LSTs. Similar to the $\fe_8$ case, their general structure is specified by the number $M$ of NS5 branes, a singularity $\fg \in$ ADE and a flavor holonomy.\footnote{For special choices of flavor holonomies $\lambda$, quivers can have a similar topology to quivers without vector structure, as discussed in~\cite{Oehlmann:2024cyn}.} The gauge algebras that can appear are typically of $\fso,\fsp$ and $\fsu$ type, and one can use the results from Section~\ref{ssec:reviewDiscrete} to determine how the different quivers can be twisted. In Table~\ref{tab:so32top}, we summarize the gauge algebras and base quiver topologies of all $\mathfrak{so}_{32}$ LSTs, with and without vector structure. Such a theory may then be twisted by an $ \mathbbm{Z}_2$ outer automorphism of the $\mathfrak{so}$ and $\mathfrak{su}$ algebras:
\begin{align}
\mathcal{\phantom{}_\pm\Tilde{K}}_{M}(\lambda;\mathfrak{g}) \xrightarrow{} \mathcal{\phantom{}_\pm\Tilde{K}}_{M}^{(k_{i})}(\lambda;\mathfrak{g})
\end{align}
where $k_{i}$ is the order of the twist; since sometimes multiple twists are possible, we add an index $i$. We again suppress the flat connection $\lambda$ in the notation from this point on, unless we need to distinguish the theories in the first and second column of Table~\ref{tab:so32top}. Such theories have simple twisting patterns: For instance, the theories in the second column in Table~\ref{tab:so32top} (except the $\mathfrak{e}_{7}$ type) have at most a single possible twist. More specifically, we have the following twisted theories with maximal flavor rank, arising from the $\mathfrak{so}_{32}$ theories without vector structure:
\begin{align}
\label{exsot1het}
\mathcal{\phantom{}_-\Tilde{K}}_{M}^{(2)}(\mathfrak{su}_{2N}): \qquad 
\lbrack \mathfrak{su}_{16}^{(2)}\rbrack  \, \,
\overset{\fsu^{(2')}_{2M+8N}}{1} \overset{\fsu^{(2)}_{2M+8(N-1)}}{2} \overset{\fsu^{(2')}_{2M+8(N-2)}}{2} ... \overset{\fsu^{(2/2')}_{2M+8}}{2} \, \, \overset{\fsu^{(2'/2)}_{2M}}{1} \, , 
\end{align}
\begin{align}
\label{exsot2het}
\mathcal{\phantom{}_-\Tilde{K}}_{M}^{(2)}(\mathfrak{so}_{4N+8}): \qquad 
\overset{ \displaystyle \overset{ \lbrack \mathfrak{su}_{16}^{(2)}\rbrack  \, \,\fsu^{(2')}_{2M+8+4N}}{ 2} }{ \underset{ \displaystyle \, \, \overset{ \fsu^{(2')}_{2M+4N}}{2}}{\, \, \overset{\fsu^{(2)}_{4M+8N}}{2}}} \overset{\fsu^{(2')}_{4M+8(N-1)}}{2} \, \overset{\fsu^{(2)}_{4M+8(N-2)}}{ 2 } ... \overset{\fsu^{(2')}_{4M+8}}{ 2\ } \,\overset{\fsp_{2M }}{1}\, ,
\end{align} 
\begin{align}
\label{exsot3het}
\mathcal{\phantom{}_-\Tilde{K}}_{M}^{(2)}(\mathfrak{so}_{4N+10}): \qquad
\overset{ \displaystyle \overset{ \lbrack \mathfrak{su}_{16}^{(2)}\rbrack  \, \,\fsu^{(2')}_{2M+8+4N}}{ 2} }{ \underset{ \displaystyle \, \overset{\fsu^{(2')}_{2M+4N}}{2}}{\, \, \overset{\fsu^{(2)}_{4M+8N}}{2}}} \, \overset{\fsu^{(2')}_{4M+8(N-1)}}{2} \, \overset{\fsu^{(2)}_{4M+8(N-2)}}{ 2} \hspace{-0.1cm}... \hspace{-0.1cm}\overset{\fsu^{(2/2')}_{4M+8}}{ 2} \,\overset{\fsu^{(2'/2)}_{4M}}{1}\, .
\end{align}
These theories are obtained from the building blocks discussed in Section~\ref{subsec:bblocks}. Similar arguments apply to the $\fsu_{N}$ heterotic $\mathfrak{so}_{32}$ theories with vector structure, which then also have a single possible $ \mathbbm{Z}_{2}$ discrete symmetry. Note that the $\mathbbm{Z}_2$ quotient changes the Coulomb branch dimension and the fractionalization coefficient $c_2$ for each such twist. We list the universal (twisted) fractionalization coefficients $c_2,r_2$ in Table~\ref{tab:so32TwistSummary}.

Let us discuss the remaining theories with vector structure in Table~\ref{tab:so32top}, and the $\mathfrak{e}_{7}$ theory without vector structure. A new feature is that we can have $\mathfrak{sp}$ algebras adjacent to multiple $\mathfrak{so}$ gauge or flavor algebras. To see which twists are allowed, we have to take the massive bi-spinor states into account, as reviewed in Section~\ref{ssec:reviewDiscrete}. Indeed, having an outer automorphism symmetry requires an $\mathfrak{sp}$ algebra be adjacent to at least two $\mathfrak{so}$ gauge or flavor algebras. To illustrate this, consider the family of $\fso_{4N+8}$ theories $\, \mathcal{\phantom{}_+\Tilde{K}}_{M}(\mathfrak{so}_{4N+8})$,
\begin{align}
\label{eq:6DSO32D4n8}
\, \mathcal{\phantom{}_+\Tilde{K}}_{M}^{(1)}(\mathfrak{so}_{4N+8}): \qquad  \,\lbrack \mathfrak{so}_{K}\rbrack  \, \, \overset{\fsp_{M+a_{1}(N)}}{1 } \myoverset{\overset{\overset{{\left[ \mathfrak{so}_{P} \right]}}{\mathfrak{sp}_{M+a_{2}(N)}}}{1}}{ \overset{\fso_{4M+a_{3}(N)}}{4 } }\,\overset{\fsp_{2M+a_{4}(N)}}{1 } ...\myoverset{\overset{\overset{{\left[ \mathfrak{so}_{Q} \right]}}{\mathfrak{sp}_{M+a_{5}(N)}}}{1}}{ \overset{\fso_{4M+a_{6}(N)}}{4 } } \overset{\fsp_{M+a_{7}(N)}}{1 }
\lbrack \mathfrak{so}_{L}\rbrack \,,
\end{align}
where $K,L,P,Q \geq 2$, the parameters $a_{i}$ are fixed by anomaly cancellation, and the number of gauge nodes in the quiver is $N+2$. The $ \mathbbm{Z}_2$-twisted 5D quivers are then
\begin{align}
\, \mathcal{\phantom{}_+\Tilde{K}}_{M}^{(2)}(\mathfrak{so}_{4N+8}): \qquad \,\lbrack \mathfrak{so}_{K}^{(2)}\rbrack  \, \, \overset{\fsp_{M+a_{1}(N)}}{1 } \myoverset{\overset{\overset{{\left[ \mathfrak{so}_{P}^{(2)} \right]}}{\mathfrak{sp}_{M+a_{2}(N)}}}{1}}{ \overset{\fso^{(2)}_{4M+a_{3}(N)}}{4 } }\,\overset{\fsp_{2M+a_{4}(N)}}{1 } ...\myoverset{\overset{\overset{{\left[ \mathfrak{so}_{Q}^{(2)} \right]}}{\mathfrak{sp}_{M+a_{5}(N)}}}{1}}{ \overset{\fso^{(2)}_{4M+a_{6}(N)}}{4 } } \overset{\fsp_{M+a_{7}(N)}}{1 }
\lbrack \mathfrak{so}_{L}^{(2)}\rbrack \,.
\end{align}
Since the $ \mathbbm{Z}_{2}$ symmetry is unique, these are the only 5D twisted theories that can arise from the 6D theory~\eqref{eq:6DSO32D4n8}, and we will find that they are T-duals of $\mathfrak{e}_{8} \times \mathfrak{e}_{8}$ LSTs in Section~\ref{sec:MatchingDuals}. However, as explained in around.~\eqref{weirdso} and~\eqref{3so}, when more than two $\mathfrak{so}$ algebras are adjacent to an $\mathfrak{sp}$ algebra, multiple inequivalent 5D twisted theories are possible, corresponding to distinct $ \mathbbm{Z}_{2}$ symmetries. In particular, the 6D theory $\,\mathcal{\phantom{}_+\Tilde{K}}_{M}(\mathfrak{so}_{4N+8})$ 
\begin{align}
\label{6dsoex2}
\, &\mathcal{\phantom{}_+\Tilde{K}}_{M}^{(1)}(\mathfrak{so}_{4N+8}):\nonumber\\ & \,\lbrack \mathfrak{so}_{k}\rbrack  \, \, \overset{\fsp_{M+a_{1}(N)}}{1 } \myoverset{\overset{\overset{{\left[ \mathfrak{so}_{m} \right]}}
{\mathfrak{sp}_{M+a_{2}(N)}}}{1}}{ \overset{\fso_{4M+a_{3}(N)}}{4 } }
\,\overset{\fsp_{2M+a_{4}(N)}}{\underset{[\fso_{2}]}{1}} 
\,\overset{\fso_{4M+a_{5}(N)}}{4 }
...\,\overset{\fso_{4M+a_{6}(N)}}{4 } 
\,\overset{\fsp_{2M+a_{7}(N)}}{\underset{[\fso_{2}]}{1}}
\myoverset{\overset{\overset{{\left[ \mathfrak{so}_{p} \right]}}{\mathfrak{sp}_{M+a_{8}(N)}}}{1}}{ \overset{\fso_{4M+a_{9}(N)}}{4 } } \overset{\fsp_{M+a_{10}(N)}}{1 }
\lbrack \mathfrak{so}_{l}\rbrack
\end{align}
has seven possible 5D twisted quivers 
\begin{align}
\label{appsogen1}
\, &\mathcal{\phantom{}_+\Tilde{K}}_{M}^{(2_{1})}(\mathfrak{so}_{4N+8}):\nonumber\\ & \,\lbrack \mathfrak{so}_{K}^{(2)}\rbrack  \, \, \overset{\fsp_{M+a_{1}(N)}}{1 } \myoverset{\overset{\overset{{\left[ \mathfrak{so}_{P}^{(2)} \right]}}
{\mathfrak{sp}_{M+a_{2}(N)}}}{1}}{ \overset{\fso^{(2)}_{4M+a_{3}(N)}}{4 } }
\,\overset{\fsp_{2M+a_{4}(N)}}{\underset{[\fso^{(2)}_{2}]}{1}} 
\,\overset{\fso_{4M+a_{5}(N)}}{4 }
...\,\overset{\fso_{4M+a_{6}(N)}}{4 } 
\,\overset{\fsp_{2M+a_{7}(N)}}{\underset{[\fso^{(2)}_{2}]}{1}}
\myoverset{\overset{\overset{{\left[ \mathfrak{so}_{p}^{(2)} \right]}}{\mathfrak{sp}_{M+a_{8}(N)}}}{1}}{ \overset{\fso^{(2)}_{4M+a_{9}(N)}}{4 } } \overset{\fsp_{M+a_{10}(N)}}{1 }
\lbrack \mathfrak{so}_{L}^{(2)}\rbrack \,,
\end{align}
\begin{align}
\label{appsogen2}
\, &\mathcal{\phantom{}_+\Tilde{K}}_{M}^{(2_{2})}(\mathfrak{so}_{4N+8}):\nonumber\\ &  \,\lbrack \mathfrak{so}_{K}^{(2)}\rbrack  \, \, \overset{\fsp_{M+a_{1}(N)}}{1 } \myoverset{\overset{\overset{{\left[ \mathfrak{so}_{P}^{(2)} \right]}}
{\mathfrak{sp}_{M+a_{2}(N)}}}{1}}{ \overset{\fso^{(2)}_{4M+a_{3}(N)}}{4 } }
\,\overset{\fsp_{2M+a_{4}(N)}}{\underset{[\fso^{(2)}_{2}]}{1}} 
\,\overset{\fso_{4M+a_{5}(N)}}{4 }
...\,\overset{\fso_{4M+a_{6}(N)}}{4 } 
\,\overset{\fsp_{2M+a_{7}(N)}}{\underset{[\fso_{2}]}{1}}
\myoverset{\overset{\overset{{\left[ \mathfrak{so}_{p} \right]}}{\mathfrak{sp}_{M+a_{8}(N)}}}{1}}{ \overset{\fso_{4M+a_{9}(N)}}{4 } } \overset{\fsp_{M+a_{10}(N)}}{1 }
\lbrack \mathfrak{so}_{L}\rbrack \,,
\end{align}
\begin{align}
\label{appsogen3}
\, &\mathcal{\phantom{}_+\Tilde{K}}_{M}^{(2_{3})}(\mathfrak{so}_{4N+8}):\nonumber\\ & \,\lbrack \mathfrak{so}_{K}\rbrack  \, \, \overset{\fsp_{M+a_{1}(N)}}{1 } \myoverset{\overset{\overset{{\left[ \mathfrak{so}_{P} \right]}}
{\mathfrak{sp}_{M+a_{2}(N)}}}{1}}{ \overset{\fso_{4M+a_{3}(N)}}{4 } }
\,\overset{\fsp_{2M+a_{4}(N)}}{\underset{[\fso_{2}]}{1}} 
\,\overset{\fso_{4M+a_{5}(N)}}{4 }
...\,\overset{\fso_{4M+a_{6}(N)}}{4 } 
\,\overset{\fsp_{2M+a_{7}(N)}}{\underset{[\fso^{(2)}_{2}]}{1}}
\myoverset{\overset{\overset{{\left[ \mathfrak{so}_{p}^{(2)} \right]}}{\mathfrak{sp}_{M+a_{8}(N)}}}{1}}{ \overset{\fso^{(2)}_{4M+a_{9}(N)}}{4 } } \overset{\fsp_{M+a_{10}(N)}}{1 }
\lbrack \mathfrak{so}_{L}^{(2)}\rbrack \,,
\end{align}
\begin{align}
\label{appsogen4}
\, &\mathcal{\phantom{}_+\Tilde{K}}_{M}^{(2_{4})}(\mathfrak{so}_{4N+8}):\nonumber\\ & \,\lbrack \mathfrak{so}_{K}^{(2)}\rbrack  \, \, \overset{\fsp_{M+a_{1}(N)}}{1 } \myoverset{\overset{\overset{{\left[ \mathfrak{so}_{P}^{(2)} \right]}}
{\mathfrak{sp}_{M+a_{2}(N)}}}{1}}{ \overset{\fso^{(2)}_{4M+a_{3}(N)}}{4 } }
\,\overset{\fsp_{2M+a_{4}(N)}}{\underset{[\fso_{2}]}{1}} 
\,\overset{\fso^{(2)}_{4M+a_{5}(N)}}{4 }
...\,\overset{\fso^{(2)}_{4M+a_{6}(N)}}{4 } 
\,\overset{\fsp_{2M+a_{7}(N)}}{\underset{[\fso_{2}]}{1}}
\myoverset{\overset{\overset{{\left[ \mathfrak{so}^{(2)}_{p} \right]}}{\mathfrak{sp}^{(2)}_{M+a_{8}(N)}}}{1}}{ \overset{\fso^{(2)}_{4M+a_{9}(N)}}{4 } } \overset{\fsp_{M+a_{10}(N)}}{1 }
\lbrack \mathfrak{so}_{L}^{(2)}\rbrack \,,
\end{align}
\begin{align}
\label{appsogen5}
\, &\mathcal{\phantom{}_+\Tilde{K}}_{M}^{(2_{5})}(\mathfrak{so}_{4N+8}):\nonumber\\ & \,\lbrack \mathfrak{so}_{K}\rbrack  \, \, \overset{\fsp_{M+a_{1}(N)}}{1 } \myoverset{\overset{\overset{{\left[ \mathfrak{so}_{P} \right]}}
{\mathfrak{sp}_{M+a_{2}(N)}}}{1}}{ \overset{\fso_{4M+a_{3}(N)}}{4 } }
\,\overset{\fsp_{2M+a_{4}(N)}}{\underset{[\fso^{(2)}_{2}]}{1}} 
\,\overset{\fso^{(2)}_{4M+a_{5}(N)}}{4 }
...\,\overset{\fso^{(2)}_{4M+a_{6}(N)}}{4 } 
\,\overset{\fsp_{2M+a_{7}(N)}}{\underset{[\fso^{(2)}_{2}]}{1}}
\myoverset{\overset{\overset{{\left[ \mathfrak{so}_{p} \right]}}{\mathfrak{sp}_{M+a_{8}(N)}}}{1}}{ \overset{\fso_{4M+a_{9}(N)}}{4 } } \overset{\fsp_{M+a_{10}(N)}}{1 }
\lbrack \mathfrak{so}_{L}\rbrack \,,
\end{align}
\begin{align}
\label{appsogen6}
\, &\mathcal{\phantom{}_+\Tilde{K}}_{M}^{(2_{6})}(\mathfrak{so}_{4N+8}):\nonumber\\ & \,\lbrack \mathfrak{so}_{K}\rbrack  \, \, \overset{\fsp_{M+a_{1}(N)}}{1 } \myoverset{\overset{\overset{{\left[ \mathfrak{so}_{P} \right]}}
{\mathfrak{sp}_{M+a_{2}(N)}}}{1}}{ \overset{\fso_{4M+a_{3}(N)}}{4 } }
\,\overset{\fsp_{2M+a_{4}(N)}}{\underset{[\fso^{(2)}_{2}]}{1}} 
\,\overset{\fso^{(2)}_{4M+a_{5}(N)}}{4 }
...\,\overset{\fso^{(2)}_{4M+a_{6}(N)}}{4 } 
\,\overset{\fsp_{2M+a_{7}(N)}}{\underset{[\fso_{2}]}{1}}
\myoverset{\overset{\overset{{\left[ \mathfrak{so}^{(2)}_{p} \right]}}{\mathfrak{sp}_{M+a_{8}(N)}}}{1}}{ \overset{\fso^{(2)}_{4M+a_{9}(N)}}{4 } } \overset{\fsp_{M+a_{10}(N)}}{1 }
\lbrack \mathfrak{so}^{(2)}_{L}\rbrack \,,
\end{align}
\begin{align}
\label{appsogen7}
\, &\mathcal{\phantom{}_+\Tilde{K}}_{M}^{(2_{7})}(\mathfrak{so}_{4N+8}):\nonumber\\ & \,\lbrack \mathfrak{so}^{(2)}_{K}\rbrack  \, \, \overset{\fsp_{M+a_{1}(N)}}{1 } \myoverset{\overset{\overset{{\left[ \mathfrak{so}^{(2)}_{P} \right]}}
{\mathfrak{sp}_{M+a_{2}(N)}}}{1}}{ \overset{\fso^{(2)}_{4M+a_{3}(N)}}{4 } }
\,\overset{\fsp_{2M+a_{4}(N)}}{\underset{[\fso_{2}]}{1}} 
\,\overset{\fso^{(2)}_{4M+a_{5}(N)}}{4 }
...\,\overset{\fso^{(2)}_{4M+a_{6}(N)}}{4 } 
\,\overset{\fsp_{2M+a_{7}(N)}}{\underset{[\fso^{(2)}_{2}]}{1}}
\myoverset{\overset{\overset{{\left[ \mathfrak{so}_{p} \right]}}{\mathfrak{sp}_{M+a_{8}(N)}}}{1}}{ \overset{\fso_{4M+a_{9}(N)}}{4 } } \overset{\fsp_{M+a_{10}(N)}}{1 }
\lbrack \mathfrak{so}_{L}\rbrack \,,
\end{align}
\normalsize
where $\mathfrak{so}_{2} \simeq \mathfrak{u}_{1}$, such that $\mathfrak{so}_{2}^{(2)}$ indicates that the $\mathfrak{u}_{1}$ flavor algebra has been removed by the twist. 
Note that the pairs in \{\eqref{appsogen2}, \eqref{appsogen3}\} and \{\eqref{appsogen6}, \eqref{appsogen7}\} have the exact same 5D matching data~\eqref{eq:invdata}. As we discuss in more detail in Section~\ref{sec:MatchingDuals}, the theories in either pair satisfy the criterion to be twisted T-dual LSTs. The other twisted 5D theories have inequivalent data and hence cannot be twisted T-dual to each other. 

\begin{table}[t]
 \centering
 \begin{tabular}{|c|c|}\hline 
  $ \mathcal{\phantom{}_{+/-}\Tilde{K}}_{M}^{(k)}(\mathfrak{g}) $ & $(c_2 , r_2 )$ \\ \hline \hline
  $ \mathcal{\phantom{}_\pm\Tilde{K}}_{M}^{(1)}(\mathfrak{su}_{N})$ & $(N,N)$\\ \hline
  $ \mathcal{\phantom{}_+\Tilde{K}}_{M}^{(2)}(\mathfrak{su}_{2N})$ & $(1+N,2N)$\\ 
  $ \mathcal{\phantom{}_-\Tilde{K}}_{M}^{(2)}(\mathfrak{su}_{2N}) $ & $(N,2N)$\\ \hline
$\mathcal{\phantom{}_-\Tilde{K}}_{M}^{(2)}(\mathfrak{su}_{2N+1})$ & $(1+N,1+2N)$\\ \hline 
$ \mathcal{\phantom{}_\pm\Tilde{K}}_{M}^{(1)}(\mathfrak{so}_{2N+8})$ & $(6+ 4N,8+ 4N ) $ \\
$ \mathcal{\phantom{}_+\Tilde{K}}_{M}^{(2)}(\mathfrak{so}_{4N+6})$ & $( 3 + 4N, 4 + 8 N) $ \\
$ \mathcal{\phantom{}_-\Tilde{K}}_{M}^{(2)}(\mathfrak{so}_{4N+6})$ & $( 2+2N , 4+8N ) $ \\
$ \mathcal{\phantom{}_-\Tilde{K}}_{M}^{(2)}(\mathfrak{so}_{4N+8})$ & $( 4+2N, 8+8N ) $ \\ \hline 
  $ \mathcal{\phantom{}_+\Tilde{K}}_{M}^{(2)}(\mathfrak{e}_{6})$ & $( 9 , 24 )$ \\ \hline
 $ \mathcal{\phantom{}_-\Tilde{K}}_{M}^{(2)}(\mathfrak{e}_{7})$ & $( 12 , 48 )$ \\ \hline
 \end{tabular}
 \caption{Summary of the universal 
 $\fso_{32}$ NS5 brane fractionalization coefficients $c_2$ and $r_2$, for each type
 of twisted singularity with and without vector structure.
 }
 \label{tab:so32TwistSummary}
\end{table}

Similar considerations apply to the $\fso_{N}$, $\fe_{6}$ and $\fe_{8}$ type of $\mathfrak{so}_{32}$ theories with and without vector structure. As before, the flat connection $\mathfrak{\lambda}$ dictates the resulting 6D flavor algebra and hence how many 5D outer-automorphism-twisted theories a 6D theory can have. Importantly, the twist of $\fso_{2N}$ gauge algebras lowers the Coulomb branch dimension by one, but does not affect the universal fractionalization coefficient $c_2$. In contrast, twisting $\fsu_n$ gauge algebras reduces the Coulomb branch dimension by a factor of two. This becomes important when matching different singularity types across dualities. All this only scratches the surface of the rich structure of possible $\fso_{32}$ twisted theories, since their specific symmetry is also dictated by the choice of holonomy $\mathfrak{\lambda}$. We leave studying the details of this relation for future work, and focus instead on twisted theories that are T-duals of $\mathfrak{e}_{8} \times \mathfrak{e}_{8}$ (twisted or untwisted) theories, both from the field theory and the F-theory geometry point of view. 

\begin{table}[t]
\begin{center}
\scriptsize
\begin{tabular}{|c|c|c|}
\hline
$\fg$ & $\mathcal{\phantom{}_+\Tilde{K}}_{M}(\fg)$ & $\mathcal{\phantom{}_-\Tilde{K}}_{M}(\fg)$ \\ \hline 

\hspace{-0.1cm}$\fsu_{2n-1}$ & 
$\overset{\fsp_{2M+4n+a_0}}{1} \overset{\fsu_{4M+8(n-1)+a_1}}{2} \,\overset{\fsu_{4M+8(n-2)+a_2}}{2} ... \overset{\fsu_{4M+8+a_k}}{1} $
& - \\ \hline

\hspace{-0.1cm}$\fsu_{2n-2}$ & 
$\overset{\fsp_{2M+4n+a_0}}{1} \overset{\fsu_{4M+8(n-1)+a_1}}{2} \,\overset{\fsu_{4M+8(n-2)+a_2}}{2} ... \overset{\fsu_{4M+8}}{2} \, \, \overset{\fsp_{2M}+a_k}{1} $
& \hspace{-0.3cm}
$ \hspace{-0.1cm}\overset{\fsu_{2M+8n+a_0}}{1} \overset{\fsu_{2M+8(n-1)+a_1}}{2} \overset{\fsu_{2M+8(n-2)+a_2}}{2} ... 
\overset{\fsu_{2M+a_r}}{1} $ \hspace{-0.3cm}
 \\ \hline 

\hspace{-0.1cm}$\fso_{4n+8}$ & 
$ \,  \overset{\fsp_{4n+M+a_1}}{1 } \,\overset{ \displaystyle \overset{\fsp_{4N+M+a_0}}{1} }{ \overset{\fso_{16N+4M+a_1}}{4 } }\,\overset{\fsp_{8n+2M+a_2 }}{1 } ...\overset{ \displaystyle \overset{\fsp_{M+a_t}}{1} }{\overset{\fso_{4M+16+a_{t-1}}}{4 } } \,  \overset{\fsp_{M+a_{t+1}}}{1 } $
 & 
 $ \hspace{-0.5cm}
 \overset{ \displaystyle \overset{ \fsu_{2M+4n+a_0}}{ 2} }{ \underset{ \displaystyle \, \, \overset{ \fsu_{2M+4n+a_2}}{2}}{\, \, \overset{\fsu_{4M+8n+a_1}}{2}}} \overset{\fsu_{4M+8(n-1)+a_3}}{2} \, \overset{\fsu_{4M+8(n-2)+a_4}}{ 2 } ... 
 \,\overset{\fsp_{2M +a_r}}{1} $\hspace{-0.4cm}
 
 \\ \hline 
 \hspace{-0.1cm}$\fso_{4n+10} $&
 $ \, \, \overset{\fsp_{M+4n+a_1}}{1} \, \, \overset{ \displaystyle \overset{\fsp_{ M+4n+a_0}}{ 1}}{
\overset{\fso_{4M+16n+a_2}}{4} 
} ... \overset{\fsp_{2M +a_{t-1} }}{1} \, \overset{\fsu_{2M+a_t }}{2} \, $
 &
 $
  \overset{ \displaystyle \overset{ \fsu_{2M+4n+a_0}}{ 2} }{ \underset{ \displaystyle \, \overset{\fsu_{2M+4n+a_2}}{2}}{\, \, \overset{\fsu_{4M+8n+a_1}}{2}}} \, \overset{\fsu_{4M+8(n-1)+a_3}}{2} \, \overset{\fsu_{4M+8(n-2)+a_4}}{ 2} \hspace{-0.1cm}...
  \,\overset{\fsu_{4M+a_r}}{1} $ \hspace{-0.2cm}
 \\ \hline 
 $\fe_6$ & $ 
 \, {\overset{\mathfrak{\fsp}_{M+a_0}}{1}} \, {\overset{\fso_{4M+a_1}}{4}}\, {\overset{\mathfrak{\fsp}_{3M +a_2}}{1}} \, {\overset{\mathfrak{\fsu}_{4M +a_3}}{2}} \, {\overset{\mathfrak{\fsu}_{2M+a_4}}{2}}
 $ & \\ \hline 
 
 $\fe_7$ & $ \hspace{-0.2cm} 
 \hspace{-0.2cm} {\overset{\mathfrak{\fsp}_{M+a_0}}{1}} \, {\overset{\fso_{4M+a_1}}{4}}\, {\overset{\mathfrak{\fsp}_{3M+a_2}}{1}} \, \overset{\displaystyle \overset{\fsp_{2M+a_7}}{1}}{{\overset{\fso_{8M+a_3}}{4}}} \, {\overset{\mathfrak{\fsp}_{3M+a_4}}{1}} \, {\overset{\fso_{4M+a_5}}{4}} \, \overset{\fsp_{M+a_6}}{1} \hspace{-0.2cm}$ 
 & $
 \, {\overset{\mathfrak{\fsu}_{2M+a_0}}{2}} \, {\overset{\mathfrak{\fsu}_{4M+a_1}}{2}} \, {\overset{\mathfrak{\fsu}_{6M+a_2}}{2}} \, 
 {\overset{\mathfrak{\fsp}_{4M+a_3}}{1}}\, {\overset{\fso_{4M+a_4}}{4}}
$ 
 \\ \hline 
 $\fe_8$ & $  
 \hspace{-0.2cm} {\overset{\mathfrak{\fsp}_{M+a_0}}{1}} \, {\overset{\fso_{4M+a_1}}{4}}\, {\overset{\mathfrak{\fsp}_{3M+a_2}}{1}} \, {{\overset{\fso_{8M+a_3}}{4}}} \, {\overset{\mathfrak{\fsp}_{5M+a_4}}{1}} \, \overset{\displaystyle \overset{\fsp_{3M+a_8}}{1}}{\overset{\fso_{12M+a_5}}{4}} \, \overset{\fsp_{4M+a_6}}{1}
 {\overset{\fso_{4M+a_7}}{4}}
 \hspace{-0.2cm}$ 
 &-
 \\ \hline 
 
\end{tabular}
\caption{Quivers for $M$ $\fso_{32}$ instantons probing a $\fg$-singularity. The integers $a_i$ are fixed by the flavor symmetry.}
\label{tab:so32top}
\end{center}
\end{table}

\section{Matching twisted T-duals} 
\label{sec:MatchingDuals}
In this section, we show how twisted LSTs enlarge the network of T-dual theories. For this, we rely on the matching data~\eqref{eq:invdata} of twisted T-duals. We split the discussion of twisted T-dual LSTs into three parts: First we discuss matching twisted singularities $\fg$ in Section~\ref{subsec:mtwist}, then matching twisted flavor symmetries in Section~\ref{sec:flavmatch}, and finally the possibility of multi-twisted T-duals in Section~\ref{subsec:mtduals}.

\begin{table}[t]
 \centering
 \begin{tabular}{|c|c|c|} \hline
  $\fg^{(n_{\text{CM}})}$ & $\fg'$ \\ \hline \hline
  $\mathfrak{su}_{2N}^{(2)}$ $(N\geq3)$ & $\mathfrak{so}_{2N+4}^{(1)}$ \\ \hline
  $\mathfrak{so}_{2N+6}^{(2)}$ & $\mathfrak{so}_{4N+8}^{(1)}$ \\ \hline
  $\mathfrak{e}_{6}^{(2)}$ & $\mathfrak{e}_{7}^{(1)}$ \\ \hline
  $\mathfrak{so}_{8}^{(3)}$ & $\mathfrak{e}_{6}^{(1)}$ \\ \hline
 \end{tabular}
 \caption{Twisted T-dual algebras $\mathfrak{g}^{(n_{\text{CM}})}$ to $n_{\text{CM}}$ five branes on $\fg'$ untwisted, such that~\eqref{eq:identifySing} holds.
 }
 \label{tab:twistuntwist}
\end{table}

\subsection{Matching twisted singularities}
\label{subsec:mtwist} 
Recall that the specification of a singularity $\fg$ on top of $M$ NS5 branes in a heterotic LST fixes the fractionalization coefficients $(c_2,r_2)$ that contribute to the Coulomb branch and 2-group structure coefficients as described in~\eqref{eq:LinearRelation}. For a $k$-twisted LST to be dual to an $\ell$-twisted LST, we require
\begin{align}
 (\text{dim}(\text{CB}),~\text{dim}(\text{WL}),~\kappa_R,~\kappa_P)_{\mathcal{K}^{(k)}(\fg,\mu_i)}= (\text{dim}(\text{CB}),~\text{dim}(\text{WL}),~\kappa_R,~\kappa_P)_{\mathcal{K}^{(\ell)}(\hat{\fg},\hat{\mu}_i)} 
\end{align}
By Equation~\eqref{eq:invdata}, $\kappa_{R}$ and $\kappa_{P}$ do not depend on the outer automorphism twist and are hence invariant, as long as the twist acts trivially on the tensor multiplets. In contrast, the fractionalization coefficients do change. While $c_1$ and $r_1$ correspond to a finite piece that can depend on a relative shift in the number of five-branes $M$ or the flavor symmetries, the coefficients $(c_2,r_2)$ need to match across any (twisted) singularity $\fg$. However, since the number $M$ of five-branes could change in the duality, it is enough to demand 
\begin{align}
\label{eq:identifySing}
 e_1 (c_2 , r_2) = e_2 (\tilde{c}_2,\tilde{r}_2) \qquad\text{ with }\qquad e_i \in \mathbbm{N}^*\, .
\end{align}
Since a singularity $\fg$ (almost always) has a unique pair of fractionalization coefficients $(c_2,r_2)$, with $r_2$ being invariant upon twisting as described in~\eqref{eq:identifySing},
we can identify potential dual theories by scanning over all combinations subject to~\eqref{eq:identifySing}. While all data in~\eqref{eq:invdata} has to match, this provides a simple and strong necessary condition for two (or more) LSTs to match, independent of many of the details such as their flavor symmetries. 

We first discuss matching T-dual $\mathfrak{e}_{8} \times \mathfrak{e}_{8}$ heterotic LSTs
\begin{align}
\mathcal{K}^{(n_{\text{CM}}, n_{\alpha}, m_{\beta})}_{n_{\text{CM}}M}(\mu_{1},\mu_{2},\mathfrak{g}) \quad \xleftrightarrow{\text{T-dual~}} \quad \mathcal{K}_{M}^{(1,n'_{\gamma}, m'_{\delta})}(\mu_{1}',\mu_{2}',\mathfrak{g}')\,,
\end{align}
for appropriate choices of holonomies $\mu_i$, $\mu_i^\prime$ and singularities $\fg$, $\fg^\prime$ as summarized in Table~\ref{tab:TCMSummary}. Interestingly, we find that all cases which we can realize geometrically have solutions with $e_2=1$ and $e_1=n$, i.e., the order of the twist. Furthermore, the existence of an untwisted $\mathfrak{e}_{8} \times \mathfrak{e}_{8}$ dual with singularity $\mathfrak{g}'$ fixes the shape of the base quiver to be that of the Dynkin diagram of $\fg'^{(1)}$ in the $\mathfrak{so}_{32}$ duals. In the geometric context, this type of T-duality is known as~\textit{fiber-base} duality, and we will use this term throughout the paper. 

Since most $\fg'$ singularities admit an affine outer automorphism, they come with another $\fso_{32}$ T-dual \cite{DelZotto:2022xrh}. These are genuine untwisted 6D theories given in the second column of Table~\ref{tab:so32top}. Interestingly, the same type of $\fso_{32}$ theories appear when switching off vector structure, which on the F-theory side corresponds to the presence of an $O7^+$ plane~\cite{Oehlmann:2024cyn}. To summarize, we often find a chain of dualities that starts from a twisted $\mathfrak{e}_{8} \times \mathfrak{e}_{8}$ theory and goes via an untwisted $\mathfrak{e}_{8} \times \mathfrak{e}_{8} $ theory to $\fso_{32}$ theories,
\begin{align}
\label{twisTdual2}
\mathcal{K}^{(n_{\text{CM}}, n_{\alpha}, m_{\beta})}_{n_{\text{CM}}M}(\mathfrak{g}) \quad \xleftrightarrow{\text{T-dual}~} \quad \mathcal{K}_{M}^{(1,n'_{\gamma}, m'_{\delta})}(\mathfrak{g}') \quad \xleftrightarrow{\text{T-dual~}} \quad \mathcal{\phantom{}_+\Tilde{K}}_{M}^{(k_{i})}(\mathfrak{g'})\quad
\xleftrightarrow{\text{T-dual~}} \quad
\mathcal{\phantom{}_-\Tilde{K}}_{M}^{(k_{i}')}(\mathfrak{g}')\,,
\end{align}
which can be extended even further for special types of singularities. For example, when starting with $\mathfrak{g}=\mathfrak{so}_{2N+6}^{(2)}$, one can deduce from Table~\ref{tab:twistuntwist} that this chain should also include an $\mathfrak{e}_{8} \times \mathfrak{e}_{8}$ theory with $\mathfrak{g}=\mathfrak{su}_{4N+4}^{(2)}$, giving rise to the duality chain
\begin{align}
\label{twisTdual21}
\mathcal{K}^{(2, n_{\alpha}, m_{\beta})}_{2M}(\mathfrak{so}_{2N+6}) \quad & 
\xleftrightarrow{\text{T-dual~}} \quad \mathcal{K}_{M}^{(1,n'_{\gamma}, m'_{\delta})}(\mathfrak{so}_{4N+8}) \quad 
\xleftrightarrow{\text{T-dual~}} \quad \mathcal{K}^{(2, n_{\alpha}'', m_{\beta}'')}_{2M}(\mathfrak{su}_{4N+4}) \nonumber \\ \quad &
\xleftrightarrow{\text{T-dual~}} \quad \mathcal{\phantom{}_+\Tilde{K}}_{M}^{(k_{i})}(\mathfrak{so}_{4N+8})\quad\hspace{5mm}
\xleftrightarrow{\text{T-dual~}} \quad \mathcal{\phantom{}_-\Tilde{K}}_{M}^{(k_{i}')}(\mathfrak{so}_{4N+8})\,.
\end{align}
There are even larger duality chain that start from special twisted $\mathfrak{so}_{32}$ theories, such as the $\fso_{4N+10}$ theories with vector structure given in the first column of Table~\ref{tab:so32top}. Following the discussion below \eqref{ancase}, we know that the $ \mathbbm{Z}_{2}$ outer automorphism of the $\mathfrak{su}_{2M}$ gauge algebra at the end of the chain is a symmetry of the theory, and hence it can be used to twist upon circle compactification to obtain a twisted $\mathfrak{so}_{32}$ LST. We can easily check that this theory has the same Coulomb branch dimension and $\kappa_{R}$ as a twisted $\mathfrak{e}_{8} \times \mathfrak{e}_{8}$ LST with a $\mathfrak{so}_{4N+10}^{(2)}$ singularity and $M$ NS5 branes.The full duality chain then follows from Table~\ref{tab:twistuntwist} (for even $M$),
\begin{align}
\label{twisTdual22}
\mathcal{\phantom{}_+\Tilde{K}}_{M}^{(2)}(\mathfrak{so}_{4N+10})\quad&
\xleftrightarrow{\text{T-dual~}} \quad \mathcal{K}^{(2, n_{\alpha}, m_{\beta})}_{M}(\mathfrak{so}_{4N+10}) \quad 
\xleftrightarrow{\text{T-dual~}} \quad \mathcal{K}_{M/2}^{(1,n'_{\gamma}, m'_{\delta})}(\mathfrak{so}_{8N+16})\nonumber \\ \quad &
\xleftrightarrow{\text{T-dual~}} \quad \mathcal{K}^{(2, n_{\alpha}'', m_{\beta}'')}_{M}(\mathfrak{su}_{8N+12}) \quad 
\xleftrightarrow{\text{T-dual~}} \quad \mathcal{\phantom{}_+\Tilde{K}}_{M/2}^{(k_{i})}(\mathfrak{so}_{8N+16})\\\quad &
\xleftrightarrow{\text{T-dual~}} \quad \mathcal{\phantom{}_-\Tilde{K}}_{M/2}^{(k_{i}')}(\mathfrak{so}_{8N+16})\,, \nonumber
\end{align}
where we now have at least 6 dual theories. One can construct a similar chain by starting from a twisted $\mathfrak{so}_{32}$ theory of $\mathfrak{e}_{6}$ type, where the $\mathfrak{su}$ gauge algebras are twisted:
\begin{align}
\label{twisTdual23}
\mathcal{\phantom{}_+\Tilde{K}}_{M}^{(2)}(\mathfrak{e}_{6})\quad &
\xleftrightarrow{\text{T-dual~}} \quad \mathcal{K}^{(2, n_{\alpha}, m_{\beta})}_{M}(\mathfrak{e}_{6}) \quad 
\xleftrightarrow{\text{T-dual~}} \quad \mathcal{K}_{M/2}^{(1,n'_{\gamma}, m'_{\delta})}(\mathfrak{e}_{7})\nonumber \\ \quad &
\xleftrightarrow{\text{T-dual~}} \quad \mathcal{\phantom{}_+\Tilde{K}}_{M/2}^{(k_{i})}(\mathfrak{e}_{7})\quad\hspace{3.5mm}
\xleftrightarrow{\text{T-dual~}} \quad \mathcal{\phantom{}_+\Tilde{K}}_{M/2}^{(k_{i}')}(\mathfrak{e}_{7})
\end{align}
We illustrate these observations in a few explicit examples.

\subsubsection{\texorpdfstring{$\boldsymbol{\mathbbm{Z}_2}$}{Z2}-twisted example with \texorpdfstring{${\fg=\fe^{(2)}_6}$}{g=e62}}
We choose $\fg=\fe_6$ on $\tilde{M}$ NS5 branes and a non-trivial flavor holonomy on the $\fe_8$ side, which results in an LST with quiver
\begin{align}
\mathcal{K}^{(1,1,1)}_{\tilde{M}} (\fe_6): \quad 
\lbrack \mathfrak{e}_{7}\rbrack  \, \, 
1 \, \,
{\overset{\mathfrak{su}_{2}}{2}} \, \,
{\overset{\mathfrak{so}_{7}}{3}} \, \,
{\overset{\mathfrak{su}_{2}}{2}} \, \,
\overset{{\left[\mathfrak{u}_{1}\right]}}{1} \, \,
\underbrace{{\overset{\mathfrak{e}_{6}}{6}}\,\, 
1\,\,
\overset{\mathfrak{su}_{3}}{3}\,\,
1\,\,
{\overset{\mathfrak{e}_{6}}{6}}
\ldots
{\overset{\mathfrak{e}_{6}}{6}}\,\,
1\,\,
\overset{\mathfrak{su}_{3}}{3}\,\,
1\,\,
{\overset{\mathfrak{e}_{6}}{6}}}_{\times \tilde{M}}\,\,
\overset{{\left[\mathfrak{u}_{1}\right]}}{1} \,\,
{\overset{\mathfrak{su}_{2}}{2}}
\, \,
{\overset{\mathfrak{so}_{7}}{3}} \,\,
{\overset{\mathfrak{su}_{2}}{2}} \,\,
 1 \,\,\lbrack \mathfrak{e}_{7}\rbrack\,.
\end{align}
The theory admits a unique twist that comes from the $\fe_6$ conformal matter spine  connecting the two orbi-instanton ends. The twisted theory reads
\begin{align}
\label{ex1p}
\mathcal{K}^{(2,1,1)}_{\tilde{M}} ( \fe_6): \quad 
\lbrack \mathfrak{e}^{(1)}_{7}\rbrack  \, \, 
1 \, \,
{\overset{\mathfrak{su}^{(1)}_{2}}{2}} \, \,
{\overset{\mathfrak{so}^{(1)}_{7}}{3}} \, \,
{\overset{\mathfrak{su}^{(1)}_{2}}{2}} \, \,
1 \, \,
\underbrace{{\overset{\mathfrak{e}^{(2)}_{6}}{6}}\,\, 
1\,\,
\overset{\mathfrak{su}^{(2)}_{3}}{3}\,\,
1\,\,
{\overset{\mathfrak{e}^{(2)}_{6}}{6}}
\ldots
{\overset{\mathfrak{e}^{(2)}_{6}}{6}}
1\,\,
\overset{\mathfrak{su}^{(2)}_{3}}{3}\,\,
1\,\,
{\overset{\mathfrak{e}^{(2)}_{6}}{6}}}_{\times \tilde{M}}\,\,
1\,\,
{\overset{\mathfrak{su}^{(1)}_{2}}{2}}
\, \,
{\overset{\mathfrak{so}^{(1)}_{7}}{3}} \,\,
{\overset{\mathfrak{su}^{(1)}_{2}}{2}} \,\,
 1 \,\,\lbrack \mathfrak{e}^{(1)}_{7}\rbrack \,,
\end{align}
and its matching data is
\begin{align}
\label{match1}
\text{rk}(\ff)=14\, , \qquad  \text{dim(CB)}=9\tilde{M} +15 \, , \qquad \kappa_{R}=24\tilde{M}+18\, .
\end{align}
Note that we had to twist the two U(1) flavor factors, which we will discuss in more detail in the next section. 

To find a dual theory, we first note that there is no singularity that fractionalizes an NS5 brane with $c_2=9$ and $r_2=24$. However, the singularity $\fg=\fe_7$ fractionalizes an NS5 brane with $c_2=18$ and $r_2=48$. Thus, we may interpret the fractionalization of one NS5 brane on a $\fg=\fe_7$ singularity as two NS5 branes on a twisted $\fe_6$ singularity. A matching $\fe_8$ quiver on an $\fe_7$ singularity quiver is readily constructed and given by 
\begin{align}
\label{ex2un2}
\mathcal{K}_{M}^{(1,1,1)}(\mathfrak{e}_{7}): \quad 
\lbrack \mathfrak{e}_{6}\rbrack  \, \, 
1 \, \,
{\overset{\mathfrak{su}_{3}}{3}} \, \,
1 \, \,
{\overset{\mathfrak{e}_{6}}{6}} 
\underbrace{{\overset{{\left[\mathfrak{u}_{1}\right]}}{1}} \, \
{\overset{\mathfrak{su}_{2}}{2}} \, \,
{\overset{\mathfrak{so}_{7}}{3}} \, \,
{\overset{\mathfrak{su}_{2}}{2}} \, \,
1 \, \,
{\overset{\mathfrak{e}_{7}}{8}} }_{\times (M)} \, \,
1 \, \,
{\overset{\mathfrak{su}_{2}}{2}} \, \,
{\overset{\mathfrak{su}_{4}}{2}} \, \,
{\overset{\mathfrak{su}_{6}}{2}} \, \,
\lbrack \mathfrak{su}_{8}\rbrack  \, , 
\end{align}
The two theories become identical under the identification $\tilde{M}=2M$. There is (at least) one untwisted $\fso_{32}$ dual for each $\fe_8\times\fe_8$ LST, which in this case is given by
\begin{align}
\label{un2so2}
\mathcal{\phantom{}_+\Tilde{K}}_{M}^{(1)}(\mathfrak{e}_{7}): \quad 
\lbrack \mathfrak{so}_{8}\rbrack  \, \,
{\overset{\mathfrak{sp}_{M}}{1}} \, \,
{\overset{\mathfrak{so}_{4M+8}}{4}} \, \, 
{\overset{\mathfrak{sp}_{3M}}{1}} \, \,
\myoverset{\overset{\overset{{\left[ \mathfrak{so}_{4} \right]}}{\mathfrak{sp}_{2M-1}}}{1}}
{\overset{\mathfrak{so}_{8M+8}}{4}} \, \, 
{\overset{\mathfrak{sp}_{3M+1}}{1}} \, \,
{\overset{\mathfrak{so}_{4M+12}}{4}} \, \,
{\overset{\mathfrak{sp}_{M+3}}{1}} \, \,
\lbrack \mathfrak{so}_{16}\rbrack  \,. 
\end{align}
In the above identification of twisted T-duals, it was crucial that the number $\tilde{M}$ of twisted $\fe_6^{(2)}$ instantons was even, such that we could identify them pairwise with one $\fe_7$ five-brane. Cases with $\tilde{M}$ odd also have an $\fe_7$ T-dual as NS5 brane fractionalization may be absorbed into M9 brane fractionalization encoded in the two orbi-instanton ends. For $\tilde{M}=2M-1$, we have the LST 
\begin{align}
\label{ex2un}
 \mathcal{K}_{M-1}^{(1,1,1)}(\mathfrak{e}_{7}): \quad 
\lbrack \mathfrak{e}_{6}\rbrack  \, \, 
1 \, \,
{\overset{\mathfrak{su}_{3}}{3}} \, \,
1 \, \,
{\overset{\mathfrak{e}_{6}}{6}} 
\underbrace{{\overset{{\left[\mathfrak{u}_{1}\right]}}{1}} \, \
{\overset{\mathfrak{su}_{2}}{2}} \, \,
{\overset{\mathfrak{so}_{7}}{3}} \, \,
{\overset{\mathfrak{su}_{2}}{2}} \, \,
1 \, \,
{\overset{\mathfrak{e}_{7}}{8}} }_{\times (M-1)} \, \,
1 \, \,
{\overset{\mathfrak{su}_{2}}{2}} \, \,
{\overset{\mathfrak{so}_{7}}{3}} \, \,
{\overset{\mathfrak{su}_{2}}{2}} \, \,
{\overset{{\left[\mathfrak{u}_{1}\right]}}{1}} \, \
{\overset{\mathfrak{e}_{6}}{6}} \, \,
1 \, \,
{\overset{\mathfrak{su}_{3}}{3}} \, \,
1 \, \,
\lbrack \mathfrak{e}_{6}\rbrack  \, , \qquad 
\end{align}
which, consistent with expectations from the arguments in Section~\ref{section:twisthet}, is an untwisted $\mathfrak{e}_{8} \times \mathfrak{e}_{8}$ dual. Interestingly, the two classes of duals~\eqref{ex2un2} and~\eqref{ex2un} differ by a Higgs branch deformation~\cite{Baume:2021qho} of the right-hand-side orbi-instanton piece, plus a shift in the number of instantons. We will see this pattern repeatedly in all such families of duals. Higgs branches of orbi-instanton theories and LSTs were studied in detail in~\cite{Baume:2023onr, Fazzi:2022hal ,Lawrie:2023uiu, Lawrie:2024zon}.

The $\fso_{32}$ dual for $M>1$ is readily constructed and reads
\begin{align}
\label{un2so}
\phantom{=}_+\mathcal{\Tilde{K}}_{M}^{(1)}(\mathfrak{e}_{7}):\quad
\lbrack \mathfrak{so}_{4}\rbrack  \, \,
{\overset{\mathfrak{sp}_{M-2}}{1}} \, \,
{\overset{\mathfrak{so}_{4M+4}}{4}} \, \, 
{\overset{\mathfrak{sp}_{3M-2}}{1}} \, \,
\myoverset{\overset{\overset{{\left[ \mathfrak{so}_{4} \right]}}{\mathfrak{sp}_{2M-2}}}{1}}
{\overset{\mathfrak{so}_{8M+4}}{4}} \, \, 
{\overset{\mathfrak{sp}_{3M}}{1}} \, \,
{\overset{\mathfrak{so}_{4M+12}}{4}} \, \,
{\overset{\mathfrak{sp}_{M+4}}{1}} \, \,
\lbrack \mathfrak{so}_{20}\rbrack  \, .
\end{align}
Note that for $\mathfrak{g}'=\mathfrak{e}_{7}$, we expect another $\mathfrak{so}_{32}$ theory (without vector structure) according to Table~\ref{tab:so32top}. It is easy to find this as the dual of \eqref{ex2un},
\begin{align}
\label{chaincomp}
\phantom{=}_-\mathcal{\Tilde{K}}_{M}^{(1)}(\mathfrak{e}_{7}):
\lbrack \mathfrak{sp}_{2} \rbrack \,
{\overset{\mathfrak{so}_{4M+8}}{4}} \, \, 
{\overset{\mathfrak{sp}_{4M-2}}{1}} \, \,  
{\overset{\mathfrak{su}_{6M}}{2}} \, \, 
{\overset{\mathfrak{su}_{4M+2}}{2}} \, \,
{\overset{\mathfrak{su}_{2M+8}}{2}} \, \,
\lbrack \mathfrak{su}_{12} \rbrack \, , \qquad 
\end{align}
which completes the T-duality chain outlined in~\eqref{twisTdual2}. Interestingly, this $\mathfrak{so}_{32}$ dual has the base topology of an $\mathfrak{e}_{6}^{(2)}$ algebra as one can verify explicitly via the LS charges of \eqref{chaincomp},
\begin{align}
\vec{\ell}_{\text{LST}}=(1, 4, 3, 2,1) \, .
\end{align}
Hence, this $\mathfrak{so}_{32}$ dual is a consequence of fiber-base duality applied to~\eqref{ex1p}. The matching data in terms of the $M$ instanton theory is 
 \begin{align}
\label{match12}
\text{rk}(\ff)=14\, , \qquad  \text{dim(CB)}=18M +24 \, , \qquad \kappa_{R}=48M+42\, .
\end{align}
Summarizing this example, we have identified a $ \mathbbm{Z}_2$-twisted theory with untwisted T-duals: The match exploits fractionalization of NS5 branes, which requires an identification of two five-branes in the twisted theory with a single five-brane in the untwisted theory. This completes the conjectured duality chain \eqref{twisTdual2}, matching all the relevant data and expectations from fiber-base duality. 

Note that it was important to identify the number of five-branes to find the two inequivalent T-duals on the untwisted side. This shift in the number of five-branes $M$ on the untwisted side is such that the number of five-branes on the twisted side remains even for even $M$ and odd for odd $M$. Moreover, either twisted theory with even/odd number of five-branes has distinct untwisted T-duals. This suggests a possible type of discrete symmetry that interlinks the discrete 0-form symmetry we twist by with the number of five-branes and appears to distinguish untwisted T-dual theories. We find that this behavior is universal for all twisted T-dual LSTs we construct geometrically, and it would be interesting to study this observation in more detail in the future. 

The shift in the number of NS5 branes generalizes to twists of order $k$, where a single 5-brane on the untwisted side is identified with $k$ five-branes on the twisted side. As a consequence, the number of distinct T-dual families is then labeled by $M$ mod $k$ five-branes on the twisted side. Let us illustrate this in a $ \mathbbm{Z}_3$ twisted example.

\subsubsection{\texorpdfstring{$\boldsymbol{\mathbbm{Z}_3}$}{Z3}-twisted example with \texorpdfstring{$\fg=\fso_8^{(3)}$}{g=so83}}
We start from an untwisted $\mathfrak{e}_{8} \times \mathfrak{e}_{8}$ LST with $N$ copies of $\mathfrak{so}_{8}-\mathfrak{so}_{8}$ conformal matter and two copies of the orbi-instanton theory
\begin{align}
\label{Z3ex1}
\lbrack \mathfrak{e}_{7}\rbrack  \, \, 
1 \, \,
{\overset{\mathfrak{su}_{2}}{2}} \, \,
{\overset{\mathfrak{g}_{2}}{3}} \, \,
1 \, \,
{\overset{\mathfrak{so}_{8}}{4}} \, \,
\cdots
\lbrack \mathfrak{so}_{8}\rbrack  \, .
\end{align}
By fusing these components, we twist the circle compactification of this theory by the $ \mathbbm{Z}_{3}$ outer automorphism of the $\mathfrak{so}_8$ algebras, resulting in the 5D theory
\begin{align}
\label{ex5t}
\mathcal{K}^{(3,1,1)}_{N}(\mathfrak{so}_{8}):\quad
\lbrack \mathfrak{e}_{7}\rbrack  \, \, 
1 \, \,
{\overset{\mathfrak{su}_{2}}{2}} \, \,
{\overset{\mathfrak{g}_{2}}{3}} \, \,
1 \, \,
\underbrace{
\overset{\mathfrak{so}_{8}^{(3)}}{4} \, \,
1 \, \,}_{\times (N)} \, \,
{\overset{\mathfrak{g}_{2}}{3}} \, \,
{\overset{\mathfrak{su}_{2}}{2}} \, \,
1 \, \,
\lbrack \mathfrak{e}_{7}\rbrack  \, .
\end{align}
We find three distinct classes of T-dual families for this twisted theory. The first theory in each class of families is the $\mathfrak{e}_{8} \times \mathfrak{e}_{8}$ dual theory with singularity type $\mathfrak{e}_{6}$, in agreement with Table~\ref{tab:twistuntwist}. Starting with the class of families that has $N=3M-1$ we find the theory
\begin{align}
\label{ex5un}
\mathcal{K}^{(1)}_{M-1}(\mathfrak{e}_{6}):\quad
{\left[\mathfrak{u}_{1}\right]}\times\bigg(\lbrack \mathfrak{su}_{6}\rbrack  \, \, 
{\overset{\mathfrak{su}_{3}}{2}} \, \,
1 \,\,
\myoverset{{\overset{{\left[\mathfrak{su}_{3}\right]}}{1}}}
{\overset{\mathfrak{e}_{6}}{6}} \, \,
\underbrace{1 \, \, 
{\overset{\mathfrak{su}_{3}}{3}} \, \,
1 \, \,
{\overset{\mathfrak{e}_{6}}{6}} \, \,
}_{\times (M-1)} \, \,
1 \, \,
{\overset{\mathfrak{su}_{3}}{2}} \, \,
{\overset{\mathfrak{su}_{5}}{2}} \, \,
\lbrack \mathfrak{su}_{7}\rbrack \bigg)  \, , 
\end{align}
where the overall $\mathfrak{u}_{1}$ factor corresponds to the non-anomalous linear combination of the $\mathfrak{u}_{1}$'s in the quiver. As was shown in~\cite{DelZotto:2022xrh}, an $\mathfrak{e}_{6}$ type heterotic $\mathfrak{e}_{8} \times \mathfrak{e}_{8}$ LST has two $\mathfrak{so}_{32}$ LSTs with distinct choices of flat connection and an $F_{4}$ toric polytope in the naming convention of~\cite{Klevers:2014bqa}. The two $\fso_{32}$ theories are given by
\begin{align}
\label{ex25soun}
\mathcal{\phantom{=}_+\Tilde{K}}^{(1)}_{M-1}(\lambda'';\mathfrak{e}_{6}): \qquad
{\left[\mathfrak{u}_{1}^{2}\right]}\times\bigg(\lbrack \mathfrak{su}_{6}\rbrack  \, \, 
{\overset{\mathfrak{su}_{2M+3}}{2}} \, \,
{\overset{\mathfrak{su}_{4M}}{2}} \, \,
{\overset{\mathfrak{sp}_{3M-2}}{1}} \, \,
\underset{\left[\mathfrak{sp}_{1}\right]}
{\overset{\mathfrak{so}_{4M+8}}{4}} \, \,
{\overset{\mathfrak{sp}_{M+1}}{1}} \, \,
\lbrack \mathfrak{so}_{12}\rbrack \bigg)  \, , 
\end{align}
\begin{align}
\label{ex5soun}
\mathcal{\phantom{=}_-\Tilde{K}}^{(1)}_{M-1}(\lambda';\mathfrak{e}_{6}): \qquad
{\left[\mathfrak{u}_{1}^{2}\right]}\times\bigg(\lbrack \mathfrak{so}_{6}\rbrack  \, \, 
{\overset{\mathfrak{sp}_{M-1}}{1}} \, \,
\underset{\left[\mathfrak{sp}_{1}\right]}
{\overset{\mathfrak{so}_{4M+6}}{4}} \, \,
{\overset{\mathfrak{sp}_{3M-2}}{1}} \, \,
{\overset{\mathfrak{su}_{4M+1}}{2}} \, \,
{\overset{\mathfrak{su}_{2M+5}}{2}} \, \,
\lbrack \mathfrak{su}_{9}\rbrack \bigg)  \, , 
\end{align}
which completes the T-duality chain~\eqref{twisTdual2}. The matching data is
\begin{align}
\text{rk}(\ff)=14\, , \qquad  \text{dim(CB)}=12M+8 \, , \qquad \kappa_{R}=24M+8\, .
\end{align}

It is instructive to look at the other two classes of families with $N=3M-2$ and $N=3M$. The untwisted $\mathfrak{e}_{8} \times \mathfrak{e}_{8}$ duals are
\begin{align}
\label{ex52un}
\mathcal{K}^{(1,1,1)}_{M-2}(\mathfrak{e}_{6}): \qquad
\lbrack \mathfrak{su}_{6}\rbrack  \, \, 
{\overset{\mathfrak{su}_{3}}{2}} \, \,
1 \,\,
\myoverset{{\overset{{\left[\mathfrak{su}_{3}\right]}}{1}}}
{\overset{\mathfrak{e}_{6}}{6}} \, \,
\underbrace{1 \, \, 
{\overset{\mathfrak{su}_{3}}{3}} \, \,
1 \, \,
{\overset{\mathfrak{e}_{6}}{6}} \, \,
}_{\times (M-2)} \, \,
1 \, \, 
{\overset{\mathfrak{su}_{3}}{3}} \, \,
1 \, \,
\myoverset{{\overset{{\left[\mathfrak{su}_{3}\right]}}{1}}}
{\overset{\mathfrak{e}_{6}}{6}} \, \,
1 \, \,
{\overset{\mathfrak{su}_{3}}{2}} \, \,
\lbrack \mathfrak{su}_{6}\rbrack  \, , \qquad 
\end{align}
\begin{align}
\label{ex53un}
\mathcal{K}^{(1,1,1)}_{M-1}(\mathfrak{e}_{6}): \qquad
\lbrack \mathfrak{su}_{6}\rbrack  \, \, 
{\overset{\mathfrak{su}_{3}}{2}} \, \,
1 \,\,
\myoverset{{\overset{{\left[\mathfrak{su}_{3}\right]}}{1}}}
{\overset{\mathfrak{e}_{6}}{6}} \, \,
\underbrace{1 \, \, 
{\overset{\mathfrak{su}_{3}}{3}} \, \,
1 \, \,
{\overset{\mathfrak{e}_{6}}{6}} \, \,
}_{\times (M-1)} \, \,
1 \, \, 
{\overset{\mathfrak{su}_{3}}{3}} \, \,
{\overset{{\left[\mathfrak{u}_{1}\right]}}{1}} \, \,
\underset{{\left[\mathfrak{sp}_{1}\right]}}
{\overset{\mathfrak{so}_{10}}{4}} \, \,
{\overset{\mathfrak{sp}_{1}}{1}} \, \,
\lbrack \mathfrak{so}_{10}\rbrack  \, , \qquad 
\end{align}
As we saw in the $ \mathbbm{Z}_{2}$-twisted case with two families of duals, the untwisted duals differ by a Higgs branch deformation of one of the orbi-instanton components, plus shifts in the instanton number. The three orbi-instanton theories are (from \eqref{ex5un}, \eqref{ex52un}, \eqref{ex53un}, respectively):
\begin{align}
&\lbrack \mathfrak{su}_{7}\rbrack  \, \, 
{\overset{\mathfrak{su}_{5}}{2}} \, \,
\underset{{\left[\mathfrak{u}_{1}\right]}}
{\overset{\mathfrak{su}_{3}}{2}} \, \,
1 \, \,
{\overset{\mathfrak{e}_{6}}{6}} \, \,
\cdots
\lbrack \mathfrak{e}_{6}\rbrack  \, , 
&\lbrack \mathfrak{su}_{6}\rbrack  \, \, 
{\overset{\mathfrak{su}_{3}}{2}} \, \,
\underset{{\left[\mathfrak{u}_{1}\right]}}
{\overset{\mathfrak{su}_{3}}{2}} \, \,
1 \,\,
\myoverset{{1}}
{\overset{\mathfrak{e}_{6}}{6}} \, \,
\cdots
\lbrack \mathfrak{e}_{6}\rbrack  \, , 
\quad&\lbrack \mathfrak{so}_{10}\rbrack  \, \, 
{\overset{\mathfrak{sp}_{1}}{1}} \, \,
\underset{{\left[\mathfrak{sp}_{1}\right]}}
{\overset{\mathfrak{so}_{10}}{4}} \, \,
{\overset{{\left[\mathfrak{u}_{1}\right]}}{1}} \, \,
{\overset{\mathfrak{su}_{3}}{3}} \, \,
1 \, \,
{\overset{\mathfrak{e}_{6}}{6}} \, \,
\cdots
\lbrack \mathfrak{e}_{6}\rbrack  \, . 
\end{align}
Deforming one theory into another changes the number of twisted conformal matter theories in the $\mathfrak{e}_{8} \times \mathfrak{e}_{8}$ twisted LSTs. This happens due to the relation between the Coulomb branch dimension and $\kappa_{R}$ of the twisted conformal matter with the orbi-instanton theories. It would be interesting to investigate this further in future work. Similar to \eqref{ex5soun} and \eqref{ex25soun}, we also find untwisted $\mathfrak{so}_{32}$ duals for \eqref{ex52un} and \eqref{ex53un}.

For completeness, the matching data for $N=3M-2$ is
\begin{align}
\text{rk}(\ff)=14\, , \qquad  \text{dim(CB)}=12M+4 \, , \qquad \kappa_{R}=24M\,,
\end{align}
and for $N=3M$
\begin{align}
\text{rk}(\ff)=14\, , \qquad  \text{dim(CB)}=12M+12 \, , \qquad \kappa_{R}=24M+16\, .
\end{align}
From \eqref{ex5t}, one also expects via fiber-base duality, an $\mathfrak{so}_{32}$ theory with a $\mathfrak{g}_{2}$ base quiver topology. However, such a theory does not exist according to Table~\ref{tab:so32top}. As it turns out, this dual can be realized by twisting an $\mathfrak{so}_{32}$ theory with a $ \mathbbm{Z}_{3}$ \textit{base twist}. We will study this dual in Section~\ref{sec:basetwists2}.

\subsubsection{Longer duality chains}
Let us discuss the longer duality chains that appear for specific singularity types, as illustrated by the chains \eqref{twisTdual21}, \eqref{twisTdual22}, and \eqref{twisTdual23}. We start from a rank $K$ $A_{4N+3}$ type LST in 6D, with unbroken $\mathfrak{e}_{8} \times \mathfrak{e}_{8}$ flavor symmetry (plus a delocalized $\mathfrak{u}_{1}$):
\begin{align}
\label{exl1}
{\left[\mathfrak{u}_{1}\right]}\times\bigg(\lbrack \mathfrak{e}_{8}\rbrack  \, \, 
1 \, \,
2 \, \,
{\overset{\mathfrak{su}_{2}}{2}} \, \,
{\overset{\mathfrak{su}_{3}}{2}} \, \,
{\overset{\mathfrak{su}_{4}}{2}} \, \,
\cdots
\underbrace{{\overset{\mathfrak{su}_{4N+4}}{2}} \, \,}_{\times (K)} \, \,
\cdots
{\overset{\mathfrak{su}_{4}}{2}} \, \,
{\overset{\mathfrak{su}_{3}}{2}} \, \,
{\overset{\mathfrak{su}_{2}}{2}} \, \,
2 \, \;
1 \, \,
\lbrack \mathfrak{e}_{8}\rbrack \bigg) \,. 
\end{align}
This theory has a single $ \mathbbm{Z}_{2}$ outer automorphism symmetry, combining the individual outer automorphisms of the $\mathfrak{su}$ algebras. One can then compactify this theory on a circle and twist by this symmetry to get the 5D theory $\mathcal{K}^{(2,1,1)}_{M}(\mathfrak{su}_{4N+4})$ (for odd $K$)
\begin{align}
\label{exl1t}
\lbrack \mathfrak{e}_{8}\rbrack  \, \, 
1 \, \,
2 \, \,
{\overset{\mathfrak{su}_{2}}{2}} \, \,
{\overset{\mathfrak{su}_{3}^{(2)}}{2}} \, \,
{\overset{\mathfrak{su}_{4}^{(2')}}{2}} \, \,
\cdots
\underbrace{{\overset{\mathfrak{su}_{4N+4}^{(2/2')}}{2}} \, \,}_{\times (K)} \, \,
\cdots
{\overset{\mathfrak{su}_{4}^{(2')}}{2}} \, \,
{\overset{\mathfrak{su}_{3}^{(2)}}{2}} \, \,
{\overset{\mathfrak{su}_{2}}{2}} \, \,
2 \, \;
1 \, \,
\lbrack \mathfrak{e}_{8}\rbrack \, , 
\end{align}
where we pick the twists (2) and (2$'$) alternatingly to have consistent massless bi-fundamentals in 5D. Using Table~\ref{tab:twistuntwist}, one can easily deduce the untwisted $\mathfrak{e}_{8} \times \mathfrak{e}_{8}$ singularity and complete it into the LST $\mathcal{K}^{(1,1,1)}_{K}(\mathfrak{so}_{4N+8})$ (for $K=2M-1$):
\begin{align}
\label{exl1un}
\lbrack \mathfrak{so}_{16}\rbrack  \, \, 
 \overset{\mathfrak{sp}_{2}}{1} \, \,
{\overset{\mathfrak{so}_{7}}{3}} \, \,
1 \, \,
\overset{\mathfrak{so}_{9}}{4} \, \,
\cdots 
 \overset{\mathfrak{sp}_{2N-1}}{1} \, \,
 \overset{\mathfrak{so}_{4N+7}}{4} \, \,
\underbrace{
{\overset{\mathfrak{sp}_{2N}}{1}} \, \,
\overset{\mathfrak{so}_{4N+8}}{4} \, \,}_{\times (M)} \, \,
 {\overset{\mathfrak{sp}_{2N}}{1}} \, \,
 \overset{\mathfrak{so}_{4N+7}}{4} \, \,
 {\overset{\mathfrak{sp}_{2N-1}}{1}} \, \,
\cdots 
\overset{\mathfrak{so}_{9}}{4} \, \,
1 \, \,
{\overset{\mathfrak{so}_{7}}{3}} \, \,
\overset{\mathfrak{sp}_{2}}{1} \, \,
\lbrack \mathfrak{so}_{16}\rbrack  \,,
\end{align}
We can write down the following $\mathfrak{so}_{32}$ duals
\begin{align}
\label{exunsoc}
\mathcal{\phantom{=}_+\Tilde{K}}^{(1)}_{M}(\lambda;\mathfrak{so}_{4N+8}):\quad
\lbrack \mathfrak{so}_{16}\rbrack  \, \, 
{\overset{\mathfrak{sp}_{M+2N+2}}{1}} \, \, 
\myoverset{\overset{\mathfrak{sp}_{M+2N-2}}{1}}
{\overset{\mathfrak{so}_{4M+8N+8}}{4}} \, \, 
{\overset{\mathfrak{sp}_{2M+4N}}{1}} \, \,
\underbrace{{\overset{\mathfrak{so}_{4M+8N+8}}{4}} \, \, 
{\overset{\mathfrak{sp}_{2M+4N}}{1}}}_{\times N-1} \, \, 
\myoverset{\overset{\mathfrak{sp}_{M+2N-2}}{1}}
{\overset{\mathfrak{so}_{4M+8N+8}}{4}} \, \, 
{\overset{\mathfrak{sp}_{M+2N+2}}{1}} \, \,
\lbrack \mathfrak{so}_{16} \rbrack
\end{align}
\begin{align}
\label{exunsoc1}
\mathcal{\phantom{=}_-\Tilde{K}}^{(1)}_{M}(\lambda';\mathfrak{so}_{4N+8}):\quad
{\left[\mathfrak{u}_{1}\right]} \times \bigg(\myoverset{\overset{\myoverset{\left[\mathfrak{su}_{16} \right]}{\mathfrak{su}_{2M+6N+8}}}{2}}
{\myunderset{\overset{\mathfrak{su}_{2M+6N}}{2}}
{\overset{\mathfrak{su}_{4M+12N}}{2}}} \, \, 
\underbrace{{\overset{\mathfrak{su}_{4M+12N-8}}{2}} \, \,  
{\overset{\mathfrak{su}_{4M+12N-16}}{2}} \, \,
\dots \dots}_{\times (N-1)}
{\overset{\mathfrak{sp}_{2M+2N}}{1}} \, \, \bigg)\,.
\end{align}
We expect another twisted $\mathfrak{e}_{8} \times \mathfrak{e}_{8}$ dual of type $\mathfrak{so}_{2N+6}^{(2)}$, as shown in the chain~\eqref{twisTdual21}. This theory is $\mathcal{K}^{(2,1,1)}_{2M+3N+1}(\mathfrak{so}_{2N+6})$:
\begin{align}
\label{exl1un2}
\lbrack \mathfrak{so}_{16}\rbrack  \, \, 
 \overset{\mathfrak{sp}_{2}}{1} \, \,
{\overset{\mathfrak{so}_{7}}{3}} \, \,
1 \, \,
\overset{\mathfrak{so}_{9}}{4} \, \,
\cdots 
 \overset{\mathfrak{sp}_{N-2}}{1} \, \,
 \overset{\mathfrak{so}_{2N+5}}{4} \, \,
\underbrace{
{\overset{\mathfrak{sp}_{N-1}}{1}} \, \,
\overset{\mathfrak{so}_{2N+6}^{(2)}}{4} \, \,}_{\times (2M+3N+1)} \, \,
 {\overset{\mathfrak{sp}_{N-1}}{1}} \, \,
 \overset{\mathfrak{so}_{2N+5}}{4} \, \,
 {\overset{\mathfrak{sp}_{N-2}}{1}} \, \,
\cdots 
\overset{\mathfrak{so}_{9}}{4} \, \,
1 \, \,
{\overset{\mathfrak{so}_{7}}{3}} \, \,
\overset{\mathfrak{sp}_{2}}{1} \, \,
\lbrack \mathfrak{so}_{16}\rbrack  \,,
\end{align}
where the instanton shift depends on the type of twisted singularity. This completes the duality chain~\eqref{twisTdual21}, which has matching data
\begin{align}
\text{rk}(\ff)&=16\, , \quad  \text{dim(CB)}=8N^{2}+4MN+18N+6M+8 \, ,\nonumber\\
\kappa_{R}&=16N^{2}+8MN+24N+8M+10\, .
\end{align}

Let us next realize the T-duality chain~\eqref{twisTdual22}. We start from the twisted $\mathfrak{so}_{4N+10}$ type $\mathfrak{so}_{32}$ LSTs
\begin{align} 
\label{exl1tso}
\mathcal{\phantom{=}_+\Tilde{K}}^{(2)}_{M-1}(\mathfrak{so}_{4N+10}):\quad
\lbrack \mathfrak{so}_{16}\rbrack  \, \, 
{\overset{\mathfrak{sp}_{M+4N+4}}{1}} \, \,
\myoverset{\overset{\overset{{\left[ \mathfrak{so}_{16} \right]}}{\mathfrak{sp}_{M+4N+4}}}{1}}
{\overset{\mathfrak{so}_{4M+16N+16}}{4}} \, \,
\cdots
{\overset{\mathfrak{sp}_{2M+16}}{1}} \, \,
{\overset{\mathfrak{so}_{4M+32}}{4}} \, \,
{\overset{\mathfrak{sp}_{2M+8}}{1}} \, \,
{\overset{\mathfrak{so}_{4M+16}}{4}} \, \,
{\overset{\mathfrak{sp}_{2M}}{1}} \, \,
{\overset{\mathfrak{su}_{2M}^{(2')}}{2}} \,,
\end{align}
where we have twisted by the $ \mathbbm{Z}_{2}$ outer automorphism of the $\mathfrak{su}$ gauge algebra, which is the only discrete outer automorphism symmetry of this theory that is consistent with the massive (spinor) spectrum of the theory, cf.~\eqref{ancase}. One can now deduce from the Coulomb branch and $\kappa_{R}$ data that this theory has a twisted $\mathfrak{e}_{8} \times \mathfrak{e}_{8}$ dual with $\mathfrak{so}_{4N+10}^{(2)}$ type singularity, probed by $(M-1)$ NS5 branes. In particular, this is the class of T-dual theories $\mathcal{K}^{(2,1,1)}_{M-1}(\mathfrak{so}_{4N+10})$, given by the quiver:
\begin{align}
\label{exl1t1}
\lbrack \mathfrak{so}_{16}\rbrack  \, \, 
 \overset{\mathfrak{sp}_{2}}{1} \, \,
{\overset{\mathfrak{so}_{7}}{3}} \, \,
1 \, \,
\overset{\mathfrak{so}_{9}}{4} \, \,
\cdots 
 \overset{\mathfrak{sp}_{2N}}{1} \, \,
 \overset{\mathfrak{so}_{4N+9}}{4} \, \,
\underbrace{
{\overset{\mathfrak{sp}_{2N+1}}{1}} \, \,
\overset{\mathfrak{so}_{4N+10}^{(2)}}{4} \, \,}_{\times (M-1)} \, \,
 {\overset{\mathfrak{sp}_{2N+1}}{1}} \, \,
 \overset{\mathfrak{so}_{4N+9}}{4} \, \,
 {\overset{\mathfrak{sp}_{2N}}{1}} \, \,
\cdots 
\overset{\mathfrak{so}_{9}}{4} \, \,
1 \, \,
{\overset{\mathfrak{so}_{7}}{3}} \, \,
\overset{\mathfrak{sp}_{2}}{1} \, \,
\lbrack \mathfrak{so}_{16}\rbrack  \,.
\end{align}
This is the same theory as~\eqref{exl1un2}, with $N \rightarrow 2N+2$ and $M \rightarrow M/2 -4-3N$. Hence, the duality chain~\eqref{twisTdual22} can now be completed with \eqref{exl1t}, \eqref{exl1un}, \eqref{exunsoc}, and \eqref{exunsoc1}, with matching data 
\begin{align}
\text{rk}(\ff)&=16\, , \quad  \text{dim(CB)}=8N^{2}+4MN+26N+7M+20 \, , \nonumber\\ \kappa_{R}&=16N^{2}+8MN+40N+12M+26\, .
\end{align}
Interestingly, the untwisted 6D theories corresponding to \eqref{exl1tso} and \eqref{exl1t1} are also T-dual to each other. Hence, this is an example of untwisted T-dual 6D LSTs that are still T-dual after a discrete twist along the circle. 

Finally, we can realize the chain \eqref{twisTdual23} with the twisted $\mathfrak{so}_{32}$ dual of \eqref{ex1p} ($\tilde{M}=M-1$),
\begin{align}
\label{e6sot}
\mathcal{\phantom{=}_+\Tilde{K}}^{(2)}_{M-1}(\mathfrak{e}_{6}):\quad
\lbrack \mathfrak{su}_{4}^{(2)}\rbrack  \, \, 
{\overset{\mathfrak{su}_{2M+2}^{(2)}}{2}} \, \,
{\overset{\mathfrak{su}_{4M}^{(2)}}{2}} \, \,
{\overset{\mathfrak{sp}_{3M-1}}{1}} \, \,
{\overset{\mathfrak{so}_{4M+12}}{4}} \, \,
{\overset{\mathfrak{sp}_{M+5}}{1}} \, \,
\lbrack \mathfrak{so}_{24}\rbrack  \, .
\end{align}
The twist of the corresponding 6D theory is fixed uniquely by the massive matter of the theory. Note that \eqref{ex1p} and \eqref{e6sot} are again T-dual before and after twisting. Furthermore, the above theory requires a twisted flavor factor in 5D, which, according to Table~\ref{tab:twistalginv}, is $\mathfrak{sp}_{2}$ instead of $\mathfrak{su}_{4}$. Indeed, the rank reduction is required for the flavor rank to match across T-duals, which we discuss in more detail in Section~\ref{sec:flavmatch}.

The above analysis can be extended to further include the fractionalization coefficients of (twisted) $\fso_{32}$ theories, which results in an even richer type of duality. The summary of these dualities, based on the match of the fractionalization coefficients, is given in Table~\ref{tab:SO32-E8-SingMatch}, where we give the two respective heterotic LST types, including the vector structure in the $\fso_{32}$ case and the resulting fractionalization coefficients for $M$ five-branes. 

\begin{table}[t]
 \renewcommand{\arraystretch}{1.3}
 \centering
 \begin{tabular}{|c|c||cc|}\hline 
 $\mathcal{\phantom{}_{+/-}\Tilde{K}}_{M}^{(k)}(\mathfrak{g}) $ & $\mathcal{K}^{(k)}_{M}(\mathfrak{g}) $ & $c_2 M$ & $r_2 M$\\ \hline \hline

 $\mathcal{\phantom{}_{-}\Tilde{K}}_{M}^{(2)}(\mathfrak{su}_{24N})$ & $\mathcal{K}^{(1)}_{NM}(\mathfrak{e}_{6}) $ &$ 12N M$& $24N M$ \\ \hline

$\mathcal{\phantom{}_{+}\Tilde{K}}_{M}^{(2)}(\mathfrak{so}_{10+4N})$ & $\mathcal{K}^{(2)}_{M}(\mathfrak{so}_{10+4N}) $ &$(7+4N)M$& $(12+8N)M$ \\ \hline

 $\mathcal{\phantom{}_{-}\Tilde{K}}_{3M}^{(2)}(\mathfrak{so}_{12}) $ & $\mathcal{K}^{(1)}_{M}(\mathfrak{e}_{7})$ &$18M$& $48 M$ \\ \hline

 $\mathcal{\phantom{}_{-}\Tilde{K}}_{2M}^{(2)}(\mathfrak{e}_{6}) $ & $\mathcal{K}^{(1)}_{M}(\mathfrak{e}_{7}) $ &$18M$& $48 M$ \\ \hline
 
 $\mathcal{\phantom{}_{-}\Tilde{K}}_{5M}^{(2)}(\mathfrak{e}_{7}) $ & $\mathcal{K}^{(1)}_{2M}(\mathfrak{e}_{8}) $ &$60M$& $240 M$ \\ \hline
 \end{tabular}
\renewcommand{\arraystretch}{1.0}
 \caption{Combinations of coinciding fractionalization coefficients of $M$ five-branes in either heterotic LST with and without vector structure on a $\fg$ singularity and a twist.}
 \label{tab:SO32-E8-SingMatch}
\end{table}

\subsection{Matching twisted flavor}
\label{sec:flavmatch}
In untwisted LSTs, the rank of the flavor symmetry algebras matches across T-dual LSTs, as these correspond to the number of Wilson line parameters that can be turned on along the compactification circle~\cite{Ahmed:2023lhj}.In this section, we show how this match extends to twisted $\mathfrak{e}_{8} \times \mathfrak{e}_{8}$ and $\mathfrak{so}_{32}$ theories with twisted flavor algebras. So far, we considered $\mathfrak{e}_{8} \times \mathfrak{e}_{8}$ LSTs where the orbi-instanton pieces did not have a discrete outer automorphism that would lead to twisted flavor algebras in 5D. To remedy this, we still consider a theory of $N$ NS5 branes probing an ALE singularity $\mathbbm{C}^{2}/\Gamma_{\mathfrak{e}_{6}}$, but pick the orbi-instanton theory
\begin{align}
\lbrack \mathfrak{e}_{6}\rbrack  \, \, 
1 \, \,
{\overset{\mathfrak{su}_{3}}{3}} \, \,
1 \, \,
\myoverset{{\overset{{\left[\mathfrak{su}_{3}\right]}}{1}}}
{{\overset{\mathfrak{e}_{6}}{6}}} \, \,
\cdots \lbrack \mathfrak{e}_{6}\rbrack\,.
\end{align}
This theory has discrete symmetries from the outer automorphisms of the $\mathfrak{e}_{6}$ and $\mathfrak{su}_{3}$ gauge and flavor algebras. Using~\eqref{eq:orbiconforbi}, we can construct the 6D LST. In the resulting theory, we have one $ \mathbbm{Z}_{2}$ symmetry, corresponding to the combined outer automorphism of all the $\mathfrak{e}_{6}$ and $\mathfrak{su}_{3}$ algebras. Compactifying the 6D LST on a circle and twisting by this symmetry, we get the theory
\begin{align}
\label{ex3t}
\mathcal{K}^{(2, 2, 2)}_{N}(\mathfrak{e}_{6}):\quad
\lbrack \mathfrak{e}_{6}^{(2)}\rbrack  \, \, 
1 \, \,
{\overset{\mathfrak{su}_{3}^{(2)}}{3}} \, \,
1 \, \,
\myoverset{{\overset{{\left[\mathfrak{su}_{3}^{(2)}\right]}}{1}}}
{{\overset{\mathfrak{e}_{6}^{(2)}}{6}}} \, \,
\underbrace{1 \, \,
{\overset{\mathfrak{su}_{3}^{(2)}}{3}} \, \,
1 \, \,
{\overset{\mathfrak{e}_{6}^{(2)}}{6}} \, \,}_{\times (N)} \, \,
1 \, \,
{\overset{\mathfrak{su}_{3}^{(2)}}{3}} \, \,
1 \, \,
\myoverset{{\overset{{\left[\mathfrak{su}_{3}^{(2)}\right]}}{1}}}
{{\overset{\mathfrak{e}_{6}^{(2)}}{6}}} \, \,
1 \, \,
{\overset{\mathfrak{su}_{3}^{(2)}}{3}} \, \,
1 \, \,
\lbrack \mathfrak{e}_{6}^{(2)}\rbrack  \, .
\end{align}
For $N=2M-1$, we find, as expected, the untwisted $\mathfrak{e}_{8} \times \mathfrak{e}_{8}$ dual theory
\begin{align}
\label{ex3un}
\mathcal{K}_{M}^{(1,1,1)}(\mathfrak{e}_{7}):\quad
\lbrack \mathfrak{f}_{4}\rbrack  \, \, 
1\, \,
{\overset{\mathfrak{g}_{2}}{3}} \, \,
{\overset{\mathfrak{su}_{2}}{2}} \, \,
2 \, \,
{\overset{{\left[\mathfrak{su}_{2}\right]}}{1}} \, \, 
\underbrace{{\overset{\mathfrak{e}_{7}}{8}} \, \,
1 \, \,
{\overset{\mathfrak{su}_{2}}{2}} \, \,
{\overset{\mathfrak{so}_{7}}{3}} \, \,
{\overset{\mathfrak{su}_{2}}{2}} \, \,
{\overset{{\left[\mathfrak{su}_{2}\right]}}{1}} \, \, }_{\times (M)} \, \,
{\overset{\mathfrak{so}_{12}}{4}} \, \,
\lbrack \mathfrak{sp}_{4}\rbrack  \, .
\end{align}
This dual now has non-simply laced flavor factors, which will appear generically when the original $\mathfrak{e}_{8} \times \mathfrak{e}_{8}$ twisted theory has twisted flavor algebras, as could be expected from the arguments presented in~\cite{Ahmed:2023lhj}: Non-simply laced flavor algebras can only arise in circle compactifications of 6D $\mathfrak{e}_{8} \times \mathfrak{e}_{8}$ LSTs if there is some non-trivial flux in the M-theory realization that ``folds'' the ADE algebras. It was then argued in~\cite{Ahmed:2023lhj} that this implies a non-trivial flux in the T-dual theory, and hence non-simply laced flavor algebras in all duals (with vector structure). The appearance of non-simply laced flavors in the dual~\eqref{ex3un} is consistent with the fact that the twisted flavor symmetries in \eqref{ex3t} lead to non-simply laced algebras in 5D according to Table~\ref{tab:twistalginv}.

We expect a T-dual $\mathfrak{so}_{32}$ theory of~\eqref{ex3un}, with the base topology of $\mathfrak{e}_{7}$. However, we find the twisted $\mathfrak{so}_{32}$ LST
\begin{align} 
\label{ex3tso}
\mathcal{\phantom{=}_+\Tilde{K}}_{M}^{(2)}(\mathfrak{e}_{7}):\quad
\lbrack\mathfrak{so}_{6}^{(2)}\rbrack
{\overset{\mathfrak{sp}_{M-1}}{1}} \, \,
{\overset{\mathfrak{so}_{4M+6}^{(2)}}{4}} \, \,
{\overset{\mathfrak{sp}_{3M-1}}{1}} \, \,
\myoverset{\overset{\overset{{\left[ \mathfrak{so}_{6}^{(2)} \right]}}{\mathfrak{sp}_{2M-1}}}{1}}
{\overset{\mathfrak{so}_{8M+6}^{(2)}}{4}} \, \,
{\overset{\mathfrak{sp}_{3M}}{1}} \, \,
{\overset{\mathfrak{so}_{4M+10}^{(2)}}{4}} \, \,
{\overset{\mathfrak{sp}_{M+2}}{1}} \, \,
\lbrack \mathfrak{so}_{14}^{(2)}\rbrack  \, , 
\end{align}
where all $\mathfrak{so}$-algebras are twisted. The uplift of this theory obtained by ``untwisting'' the twisted algebras also leads to a consistent 6D theory. The twisted flavor algebras are fixed since the dual theories both have non-simply laced algebras in 5D. In fact, this by itself requires that at least one $\mathfrak{so}$ gauge algebra in the quiver must be twisted! This follows from the arguments in Section~\ref{section:frev}, around equation~\eqref{eq:spinex1}, where we note that the massive spectrum on a (-1)-curve with an $\mathfrak{sp}$ algebra transforms in spinor representations of the neighboring $\mathfrak{so}$ algebras, and if there are two such neighbors, both of them must be twisted for a consistent theory. This implies that \eqref{ex3tso} is the only consistent 5D outer automorphism twisted theory with the 6D uplift discussed above. The matching data is
\begin{align}
\label{match2}
\text{rk}(\ff)=10\, , \qquad  \text{dim(CB)}=18M -4 \, , \qquad \kappa_{R}=48M-22\, .
\end{align}
and consistency of the massive spectrum uniquely fix, for a fixed flavor holonomy $\lambda$, the $\mathfrak{so}_{32}$ dual with $e_{7}$ base topology! Note, however, that we can have other $\mathfrak{so}_{32}$ duals with this base topology as well, but with different $\lambda$, as we will see in Section~\ref{sec:Geometry}.
In conclusion, we see that twisted flavor algebras on the $\mathfrak{e}_{8} \times \mathfrak{e}_{8}$ side imply twisted $\mathfrak{so}_{32}$ dual theories.

There is a subtle exception to this when we twist by a discrete symmetry of an $\mathfrak{e}_{8} \times \mathfrak{e}_{8}$ LST whose only effect on the flavor is twisting $\mathfrak{u}_{1}$ flavor factors to ``nothing'', as in equation~\eqref{3so}. In that case, there is no non-simply laced flavor factor in 5D, and hence we only expect untwisted $\mathfrak{so}_{32}$ duals. We illustrate this using the untwisted $\mathfrak{e}_{8} \times \mathfrak{e}_{8}$ theory already introduced in \eqref{exagain}, with a $D_{4}$ singularity. We consider a particular twisted version of this theory, which we repeat here again for convenience:
\begin{align}
\label{ex4t1}
\mathcal{K}^{(2,2_{1},2_{1})}_{N}(\mathfrak{so}_{8}): \qquad 
\lbrack \mathfrak{e}_{6}\rbrack  \, \, 
1 \, \,
{\overset{\mathfrak{su}_{3}}{3}} \, \,
{\overset{{\left[\mathfrak{u}_{1}\right]}}{1}} \, \,
\underbrace{
\overset{\mathfrak{so}_{8}^{(2)}}{4} \, \,
{\overset{{\left[\mathfrak{u}_{1}\right]}}{1}} \, \ \, \,}_{\times (N)} \, \,
{\overset{\mathfrak{su}_{3}}{3}} \, \,
1 \, \,
\lbrack \mathfrak{e}_{6}\rbrack  \, .
\end{align}
The $\mathfrak{u}_{1} \times \mathfrak{u}_{1}$ factors on the E-strings in \eqref{exagain} are each twisted to a single $\mathfrak{u}_{1}$ factor. This is required for a consistent matter spectrum and due to the fact that this theory only has a combined $ \mathbbm{Z}_{2}$ outer automorphism symmetry of the $\mathfrak{so}_{8}$ and $\mathfrak{u}_{1}$ factors. We can then write down the expected untwisted $\mathfrak{e}_{8} \times \mathfrak{e}_{8}$ dual for $N=2M$,
\begin{align}
\label{ex4un}
\mathcal{K}^{(1,1,1)}_{M-2}(\mathfrak{so}_{12}):\quad
{\left[\mathfrak{u}_{1}^{2}\right]}\times\bigg(
\lbrack \mathfrak{su}_{6}\rbrack  \, \, 
{\overset{\mathfrak{su}_{4}}{2}} \, \,
{\overset{\mathfrak{sp}_{1}}{1}} \, \,
\underset{\left[\mathfrak{sp}_{1}\right]}
{\overset{\mathfrak{so}_{12}}{4}} \, \,
\underbrace{{\overset{\mathfrak{sp}_{2}}{1}} \, \,
\overset{\mathfrak{so}_{12}}{4} \, \,
}_{\times (M-2)} 
{\overset{\mathfrak{sp}_{2}}{1}} \, \,
\underset{\left[\mathfrak{sp}_{1}\right]}
{\overset{\mathfrak{so}_{12}}{4}} \, \,
{\overset{\mathfrak{sp}_{1}}{1}} \, \,
{\overset{\mathfrak{su}_{4}}{2}} \, \,
\lbrack \mathfrak{su}_{6}\rbrack \bigg) \,.
\end{align}
This fixes the topology of the untwisted $\mathfrak{so}_{32}$ dual
\begin{align}
\label{ex4t1so}
\mathcal{\phantom{=}_+\Tilde{K}}^{(1)}_{M-2}(\mathfrak{so}_{12}):\quad
\lbrack \mathfrak{so}_{8}\rbrack  \, \, 
{\overset{\mathfrak{sp}_{M-1}}{1}} \, \,
\myoverset{\overset{\mathfrak{sp}_{M-3}}{1^{*}}}
{\overset{\mathfrak{so}_{4M+4}}{4^{*}}} \, \,
\underset{{\left[\mathfrak{so}_{4}\right]}}
{\overset{\mathfrak{sp}_{2M}}{1}} \, \,
\myoverset{\overset{\overset{{\left[ \mathfrak{so}_{4} \right]}}{\mathfrak{sp}_{M-1}}}{1}}
{\overset{\mathfrak{so}_{4M+8}}{4}} \, \,
{\overset{\mathfrak{sp}_{M+1}}{1}} \, \,
\lbrack \mathfrak{so}_{12}\rbrack  \, ,
\end{align}
where the nodes with a $*$ only occur for $M>2$. The matching data is 
\begin{align}
\label{matcht1}
\text{rk}(\ff)=14\, , \qquad  \text{dim(CB)}=10M +8 \, , \qquad \kappa_{R}=16M+10\, .
\end{align}
Note that one of the $\mathfrak{u}_{1}$ factors twisting to nothing in the 5D twisted theory~\eqref{ex4t1} is crucial for matching flavor ranks across T-duality. Hence, the field theory arguments regarding $\mathfrak{u}_{1}$'s on E-strings fits nicely with flavor rank being an invariant of T-dual LSTs.

\subsection{Multiple twists and T-duals}
\label{subsec:mtduals}
As explained in Section~\ref{sec:Twisted}, 6D LSTs can have multiple discrete symmetries, and hence multiple twists upon circle reduction. Consider again the 6D LST~\eqref{exagain}, whose eight inequivalent 5D twisted theories are given in \eqref{ex4t1am}-\eqref{ex4t8am}, and whose T-duals of a particular instance in the family we studied in the last section. In this section, we consider the duality chain of a differently twisted 5D theory, and its relation to the previous one, given by
\begin{align}
\label{ex4t2}
\mathcal{K}^{(2,2_{2},2_{2})}_{N}(\mathfrak{so}_{8}): \qquad \qquad 
\lbrack \mathfrak{e}_{6}^{(2)}\rbrack  \, \, 
1 \, \,
{\overset{\mathfrak{su}_{3}^{(2)}}{3}} \, \,
{\overset{{\left[\mathfrak{u}_{1}\right]}}{1}}\, \,
\underbrace{
\overset{\mathfrak{so}_{8}^{(2)}}{4} \, \,
{\overset{{\left[\mathfrak{u}_{1}\right]}}{1}} \, \ \, \,}_{\times (N)} \, \,
{\overset{\mathfrak{su}_{3}^{(2)}}{3}} \, \,
1 \, \,
\lbrack \mathfrak{e}_{6}^{(2)}\rbrack  \,.
\end{align}
The untwisted $\mathfrak{e}_{8}\times \mathfrak{e}_{8}$ dual theory is for $N=2M$
\begin{align}
\label{ex4un2}
\mathcal{K}^{(1,1,1)}_{M-1}(\mathfrak{so}_{12}):\quad
\lbrack \mathfrak{sp}_{4}\rbrack  \, \, 
\underset{\left[\mathfrak{u}_{1}\right]}
{\overset{\mathfrak{so}_{12}}{2}} \, \,
{\overset{\mathfrak{sp}_{2}}{1}} \, \,
\underbrace{
\overset{\mathfrak{so}_{12}}{4} \, \,
{\overset{\mathfrak{sp}_{2}}{1}} \, \,}_{\times (M-1)} \underset{\left[\mathfrak{u}_{1}\right]}
{\overset{\mathfrak{so}_{12}}{2}} \, \,
\lbrack \mathfrak{sp}_{4}\rbrack  \, ,
\end{align}
Since both \ref{ex4t2} and \eqref{ex4un2} have non-simply laced flavor symmetries in 5D, we expect the same for any $\mathfrak{so}_{32}$ dual. This enables us to write down the twisted $\mathfrak{so}_{32}$ dual given by $\mathcal{\phantom{}_+\Tilde{K}}^{(2_{1})}_{M-1}(\mathfrak{so}_{12})$:
\begin{align}
\label{ex4t2so}
\lbrack \mathfrak{so}_{10}^{(2)}\rbrack  \, \, 
{\overset{\mathfrak{sp}_{M}}{1}} \, \,
\myoverset{\overset{\mathfrak{sp}_{M-2}}{1^{*}}}
{\overset{\mathfrak{so}_{4M+6}^{(2)}}{4^{*}}} \, \,
\underset{{\left[\mathfrak{so}_{4}\right]}}
{\overset{\mathfrak{sp}_{2M}}{1}} \, \,
\myoverset{\overset{\mathfrak{sp}_{M-2}}{1^{*}}}
{\overset{\mathfrak{so}_{4M+6}^{(2)}}{4^{*}}} \, \,
{\overset{\mathfrak{sp}_{M}}{1}} \, \,
\lbrack \mathfrak{so}_{10}^{(2)}\rbrack  \,,
\end{align}
where nodes with a $*$ only occur for $M>1$. The twisted flavor again necessitates twisting the gauge algebras to obtain a consistent massive matter spectrum. Note that the $\mathfrak{so}_{4} \simeq \mathfrak{su}_{2} \times \mathfrak{su}_{2}$ algebra also admits a $ \mathbbm{Z}_{2}$ twist that corresponds to the exchanging the two $\mathfrak{su}_{2}$ factors. In fact, due to the highly symmetric form of this quiver, it is consistent to twist by this outer automorphism as well, as outlined in Section~\ref{section:frev}. In this case, however, the $\mathfrak{so}_{4}$ flavor cannot be twisted since this would lead to a mismatch of flavor ranks across duals. Hence, the duality predicts the 5D twisted theory given in~\eqref{ex4t2so}. The complete T-duality matching data for \eqref{ex4t2so} is given by
\begin{align}
\label{matcht2}
\text{rk}(\ff)=10\, , \qquad  \text{dim(CB)}=10M +6 \, , \qquad \kappa_{R}=16M+10\, .
\end{align}
Finally, we note that \eqref{ex4t1so} and \eqref{ex4t2so} are related by a Higgs branch deformation in addition to twisting by the $ \mathbbm{Z}_{2}$ outer automorphisms of the $\mathfrak{so}$ algebras. It would be interesting to further investigate the relation between T-dual chains of twisted $\mathfrak{e}_{8} \times \mathfrak{e}_{8}$ theories that uplift to the same 6D theory, which is, however, beyond the scope of the paper.

Before concluding the section, it is instructive to look at the other 5D twisted theories whose 6D uplift is~\eqref{ex4t2so}. According to~\eqref{n5d}, the 6D uplift should have two more distinct discrete symmetries one can twist by. These are given by
\begin{align}
\label{ex4t1d1}
\mathcal{\phantom{=}_+\Tilde{K}}^{(2_{2})}_{M-1}(\mathfrak{so}_{12}): \qquad \qquad
\lbrack \mathfrak{so}_{10}^{(2)}\rbrack  \, \, 
{\overset{\mathfrak{sp}_{M}}{1}} \, \,
\myoverset{\overset{\mathfrak{sp}_{M-2}}{1^{*}}}
{\overset{\mathfrak{so}_{4M+6}^{(2)}}{4^{*}}} \, \,
\underset{{\left[\mathfrak{so}_{4}^{(2)}\right]}}
{\overset{\mathfrak{sp}_{2M}}{1}} \, \,
\myoverset{\overset{\mathfrak{sp}_{M-2}}{1^{*}}}
{\overset{\mathfrak{so}_{4M+6}^{(2)}}{4^{*}}} \, \,
{\overset{\mathfrak{sp}_{M}}{1}} \, \,
\lbrack \mathfrak{so}_{10}^{(2)}\rbrack  \, ,
\end{align}
\begin{align}
\label{ex4t1d2}
\mathcal{\phantom{=}_+\Tilde{K}}^{(2_{3})}_{M-1}(\mathfrak{so}_{12}): \qquad \qquad
\lbrack \mathfrak{so}_{10}\rbrack  \, \, 
{\overset{\mathfrak{sp}_{M}}{1}} \, \,
\myoverset{\overset{\overset{{\left[ \mathfrak{u}_{1} \right]}}{\mathfrak{sp}_{M-2}}}{1^{*}}}
{\overset{\mathfrak{so}_{4M+6}}{4^{*}}} \, \,
\underset{{\left[\mathfrak{so}_{4}^{(2)}\right]}}
{\overset{\mathfrak{sp}_{2M}}{1}} \, \,
\myoverset{\overset{\overset{{\left[ \mathfrak{u}_{1} \right]}}{\mathfrak{sp}_{M-2}}}{1^{*}}}
{\overset{\mathfrak{so}_{4M+6}}{4^{*}}} \, \,
{\overset{\mathfrak{sp}_{M}}{1}} \, \,
\lbrack \mathfrak{so}_{10}\rbrack  \, .
\end{align}
Even though $\kappa_{R}$ for these theories is still the same, the rest of the matching data, dim(CB) and $\text{rk}(\ff)$, are different from \eqref{match2}, and hence \eqref{ex4t2so} is indeed the unique twisted dual we can have in this case. Note further that both the dual theories \eqref{ex4t2} and \eqref{ex4t2so} have 6D uplifts with multiple possible 5D twisted theories. It would be interesting to investigate this further in the future.

\section{Twists acting on tensor multiplets}
\label{sec:basetwists}
In the sections before we have focused on gauge algebra outer automorphism twists $\mathcal{O}$ along the compactification circle. In this section, we want to extend the discussion to cases with twists $\mathcal{B}$ acting on the tensor multiplets as reviewed Section~\ref{ssec:reviewDiscrete}. Even though such twisted LSTs cannot be constructed using our toric methods, some low-rank examples were constructed and their T-duals were identified in~\cite{Bhardwaj:2022ekc}. We study these examples and show that the matching data~\eqref{eq:invdata} still provides invariants among twisted T-duals for $\mathcal{N}=(1,0)$ theories.\footnote{For LSTs with 16 supercharges, such a match has been originally been proposed in \cite{DelZotto:2020sop}.} In Section~\ref{sec:basetwists2} and~\ref{ssec:Kappa1examples}, we use this to propose novel families of twisted T-dual LSTs.

\subsection{Known base-twisted LSTs}
\label{sec:basetwists1}
In order to establish that the matching data~\eqref{eq:invdata} are also base-twisted T-duality invariants, we first discuss the action of the base twist $\mathcal{B}$ on the Dirac pairing $\eta$ and obtain the corresponding LST charges $\ell_{\text{LST}}$. We denote the twisted Dirac pairing by $\eta^{\,\mathcal{B}}$ and call the LST charges after the twist $\vec{\ell}^{\,\mathcal{B}}_{\text{LST}}$. The computation of $\eta^{\,\mathcal{B}}$ proceeds in the same way as for Cartan matrices of non-simply laced Lie algebras: Take $\eta$ to be a symmetric $(n_T+1)\times (n_T+1)$ matrix, and $\mathcal{B}$ an order $n$ symmetry with $k$ orbits $O_\alpha$. For each orbit $\alpha$, the diagonal entry $\eta^{\,\mathcal{B}}_{\alpha\alpha}$ is the value that $\eta$ takes within this orbit,
\begin{align}
 \eta^{\,\mathcal{B}}_{\alpha,\alpha}= \eta_{I_\alpha,I_\alpha}\quad ,~\alpha=1\ldots k \, .
\end{align}
Off-diagonal entries of $\eta^{\mathcal{B}}$ are the same as before with the only exception that 
\begin{align}
\eta^{\,\mathcal{B}}_{\alpha,\beta} = n_\beta \, \eta_{I_\alpha,I_\beta} \, ,
\end{align}
if $I_\alpha$ is a one-dimensional orbit having an off-diagonal entry with nodes that are part of a non-trivial orbit $I_\beta$ of order $n_\beta$. On the other hand, we also have 
\begin{align}
 \eta^{\,\mathcal{B}}_{\beta,\alpha} = n_\alpha \eta_{I_\beta,I_\alpha} = \eta_{I_\beta,I_\alpha} \, ,
\end{align}
which results in $\eta^{\,\mathcal{B}}$ not being a symmetric matrix.

Secondly, we need to compute the LST charges of the twisted Dirac pairing, which is a $k$-dimensional charge vector $\vec{\ell}^{\,\mathcal{B}}_{\text{LST}}$. Recall that $\vec{\ell}$ 
is defined as the multiplicities of the null curve $\Sigma^{0}$. Since $\eta$ is a pairing on divisors, but $\Sigma^{0}$ is a curve \textit{dual} to a divisor, $\vec{\ell}$ is actually the null co-vector $(\vec{\ell})^T$ of $\eta$, or equivalently the null-vector of $(\eta)^T$. If $\eta$ is symmetric, this distinction is irrelevant, but it becomes important for non-symmetric $\eta^{\,\mathcal{B}}$. Hence, we have
\begin{align}
 (\eta^{\,\mathcal{B}})^T \cdot \vec{\ell}^{\,\mathcal{B}}= 0 \, .
\end{align}
Similarly, one may obtain $\ell^\mathcal{B}$ from
\begin{align}
 \ell^{\mathcal{B}}_\alpha = \frac{1}{N} \sum_{I_{\alpha}} \ell_{i}\quad \text{ for } i \in I_\alpha,\quad N=\text{gcd}(\sum_{I_{\alpha}} \ell_{i})\,.
\end{align}
The twisted 2-group structure constants are then
\begin{align}
 \kappa_P= - \sum_\alpha^k \ell^\mathcal{B}_\alpha (\eta^{\,\mathcal{B}}_{\alpha,\alpha}-2) \, , \qquad \kappa_R= \sum_\alpha^k \ell^\mathcal{B}_\alpha h^\vee_{\fg_\alpha}\, .
\end{align}
In the following, we study the base-twisted LSTs constructed in \cite{Bhardwaj:2022ekc}, together with their T-duals, and explicitly confirm the matching of their 5D twisted data. 

We first consider the worldvolume theory of two NS5 branes probing a $\mathbbm{C}^{2}/ \mathbbm{Z}_{2}$ singularity in the $\mathfrak{e}_{8} \times \mathfrak{e}_{8}$ heterotic string. We pick the flat connections $\mu_{1}=\mu_{2}=\mu$ that act as a $ \mathbbm{Z}_2$ reflection in $\fe_8$ such that the heterotic gauge bundle is broken to a the maximal subgroup $\mathfrak{so}_{16} \times \mathfrak{so}_{16}$. This leads to the quiver
\begin{align}
\label{bt1}
\lbrack \mathfrak{so}_{16}\rbrack  \, \, 
{\overset{\mathfrak{sp}_{1}}{1}} \, \,
\overset{{\left[\mathfrak{sp}_{1}\right]}}
{\overset{\mathfrak{su}_{2}}{2}} \, \,
{\overset{\mathfrak{sp}_{1}}{1}} \, \,
\lbrack \mathfrak{so}_{16}\rbrack  \, , \qquad \text{ with }\quad 
\vec{\ell}_{\text{LST}}=(1,1,1) \, .
\end{align} 
This theory has the following tensor permutation symmetry:
\begin{align}
\label{bt11}
\lbrack \mathfrak{so}_{16}\rbrack  \, \, 
{\overset{\mathfrak{sp}_{1}}{\tikzmarknode[black!70!black]{c}{1}}} \, \,
\overset{\left[\mathfrak{sp}_{1}\right]}
{\overset{\mathfrak{su}_{2}}{2}} \, \,
{\overset{\mathfrak{sp}_{1}}{\tikzmarknode[black!50!black]{d}{1}}} \, \,
\lbrack \mathfrak{so}_{16}\rbrack  \, ,
\begin{tikzpicture}[remember picture,overlay]
\draw[red,<->]([yshift=0.000009ex]c.south) to[bend right]node[below]{\scriptsize} ([yshift=0.000009ex]d.south);
\end{tikzpicture}
\end{align} \\
The quotient has two distinct orbits; one that maps the central node to itself, and one that swaps the two $(-1)$-curves. Reducing the LST on a circle and twisting by the symmetry identifies the two $\mathfrak{sp}_{1}$ gauge algebras along with their $\mathfrak{so}_{16}$ flavor factors. The twisted Dirac pairing matrix and LST charges are then
\begin{align}
\eta^{\,\mathcal{B}}_{\alpha \beta}=
\left(\begin{array}{cc}
2 & -2 \\
-1 & 1
\end{array}\right)\, \quad \text{ with } \quad \vec{\ell}_\text{LST}^{\,\mathcal{B}}=(1,2) \, . 
\end{align}
Note that the twisted LST charges of the tensor multiplets that are part of the degree two orbit is now 2 instead of 1. The twisted matching data is summarized as 
\begin{align}
\label{eq:matchbase}
\text{rk}(\ff)=9\, , \qquad  \text{dim(CB)}=4 \, , \qquad \kappa_{P}=2\, ,\qquad \, \kappa_{R}=6\,,
\end{align}
which shows that the 2-group structure constant in theory~\eqref{bt11} is unchanged upon base-twisting, just as in the fiber-twisted cases.

Finally we turn to the twisted T-dual LST which was determined geometrically in \cite{Bhardwaj:2022ekc} as 
\begin{align}
\label{Lakprop}
\lbrack \mathfrak{su}_{16}^{(2)} \times \mathfrak{u}_{1}\rbrack  \, \, 
{\overset{\mathfrak{su}_{6}^{(2)}}{0}} \, \, \qquad \text{ with }\, \, 
\vec{\ell}_{\text{LST}}=(1) \, .
\end{align}
The finite flavor subgroup is given by $\mathfrak{so}_{16} \times \mathfrak{u}_{1}$, as can be inferred from the 6D theory: 6D~anomaly cancellation for an $\mathfrak{su}_{6}$ gauge theory with 0 Dirac pairing requires sixteen fundamental and two anti-symmetric hypermultiplets. Thus the 6D flavor symmetry is given by $\mathfrak{u}_{16} \times \mathfrak{u}_{2}$. Both $\mathfrak{u}_1$ flavor factors are ABJ-anomalous, but one linear combination is anomaly-free. Under the outer automorphism twist of the $\mathfrak{su}_{6}$ gauge algebra, the sixteen fundamentals and two antisymmetrics descend to eight fundamentals and one antisymmetric of $\mathfrak{sp}_{3} \subset \fsu_6^{(2)}$ in the 5D theory. Finally, when this matter is viewed as half-hypermultiplets of $\mathfrak{sp}_{3}$, we obtain the flavor group of $\mathfrak{so}_{16} \times \mathfrak{so}_{2} \simeq \mathfrak{so}_{16} \times \mathfrak{u}_{1}$, where $\mathfrak{so}_{16}\subset \mathfrak{su}_{16}^{(2')}$, as seen from Table~\ref{tab:twistalginv}.
Therefore, we find a match of flavor rank, Coulomb branch and 2-group structure constants among the two twisted T-dual LSTs. 

We consider a second example with tensor twist of order three to further establish match of the invariants across T-duals. Such an LST can be constructed by considering the worldvolume theory of a single NS5 brane in the $\mathfrak{e}_{8} \times \mathfrak{e}_{8}$ heterotic string, probing a $\mathbbm{C}^{2}/ \mathbbm{Z}_{3}$ singularity, with a flavor symmetry $\mathfrak{e}_{6}^{3}$~\cite{DelZotto:2022xrh,DelZotto:2023ahf},
\begin{align} 
\lbrack \mathfrak{e}_{6}\rbrack  \, \,
1 \, \,
\myoverset{{\myoverset{{\left[\mathfrak{e}_{6}\right]}}{1}}} 
{\overset{\mathfrak{su}_{3}}{3}} \, \,
1 \, \,
\lbrack \mathfrak{e}_{6}\rbrack  \, \, \quad 
\text{ with } \quad 
\vec{\ell}_{\text{LST}}=(1,\myoverset{1}{1},1) \,.
\end{align}
This theory admits both outer automorphism twists $\mathcal{O}^{(2)}$ for the $\mathfrak{su}_{3}$ gauge and $\mathfrak{e}_{6}$ flavor algebras as well as an $S_{3}$ permutation symmetry $\mathcal{B}^{(3)}$ of the three E-strings. We can compactify this theory on a circle and twist by the combined $ \mathbbm{Z}_6$ symmetry $\mathcal{S}^{(2,3)}=\mathcal{O}^{(2)}\circ \mathcal{B}^{(3)}$. The resulting 5D theory is
\begin{align} 
\label{eq:Z6Twistquiver}
\lbrack \mathfrak{e}_{6}^{(2)}\rbrack  \, \,
{\overset{}{\tikzmarknode[black!50!black]{c}{1}}} \, \,
\myoverset{{\myoverset{{\left[\mathfrak{e}_{6}^{(2)}\right]}}{\tikzmarknode[black!50!black]{f}{1}}}} 
{\overset{\mathfrak{su}_{3}^{(2)}}{3}} \, \,
{\overset{}{\tikzmarknode[black!50!black]{d}{1}}} \, \,
\lbrack \mathfrak{e}_{6}^{(2)}\rbrack  \, 
\begin{tikzpicture}[remember picture,overlay]
\draw[red,->]([yshift=0.000009ex]c.south) to[bend right]node[below]{\scriptsize} ([yshift=0.000009ex]d.south);
\draw[red,<-]([yshift=0.0009ex]f.south) to[bend left]node[below]{\scriptsize} ([yshift=0.000009ex]d.north);
\draw[red,->]([yshift=0.0009ex]f.south) to[bend right]node[below]{\scriptsize} ([yshift=0.0009ex]c.north);
\end{tikzpicture}\,.
\end{align}
There are two orbits of the modded $ \mathbbm{Z}_{3}$ symmetry: one that leaves the $(-3)$-curve invariant, and one which cyclically permutes the three $(-1)$-curves. The 5D twisted Dirac pairing matrix is given by
\begin{align}
\eta^{\,\mathcal{B}^{(3)}}=
\left(\begin{array}{cc}
3 & -3 \\
-1 & 1
\end{array}\right)\, , \quad \text{ with } \quad \vec{\ell}^{\,\,\mathcal{B}^{(3)}}_{\text{LST}}=(1,3) \,,
\end{align}
and the twisted 5D data is
\begin{align}
\label{matchbasegen2}
\text{rk}(\ff)=4\, , \qquad  \text{dim(CB)}=3 \, , \qquad \kappa_{P}=2\, , \qquad \kappa_{R}=6\, .
\end{align}
The T-dual theory of \eqref{eq:Z6Twistquiver} was constructed geometrically in \cite{Bhardwaj:2022ekc}, and is given by the 5D KK theory
\begin{align} 
\lbrack \mathfrak{sp}_{4}\rbrack  \, \, 
{\overset{\mathfrak{so}_{8}^{(3)}}{0}} \, \, \qquad \text{ with }\qquad
\vec{\ell}_{\text{LST}}=(1) \, .
\end{align}
The above theory descends from a $ \mathbbm{Z}_3$-twisted $\mathfrak{so}_{8}$ gauge theory, with matter transforming in four fundamentals, spinors and co-spinors, which leads to an $\fsp_4^3$ flavor symmetry. The twist in the 5D theory identifies the three $\fsp_4$ factors, leading to a single $\mathfrak{sp}_{4}^{(1)}$ flavor symmetry acting on the four $\fg_2\subset \mathfrak{so}_{8}^{(3)}$ fundamental $\mathbf{7}$-plet representation. Again, the flavor symmetry and 5D data~\eqref{matchbasegen2} matches between T-duals.

\subsection{New families of base-twisted LSTs}
\label{sec:basetwists2} 
In this section, construct new families of base-twisted 5D LSTs and propose new twisted T-dual theories by imposing a match of the data~\eqref{eq:invdata}. We start by generalizing the quiver~\eqref{bt1} to include $(2M-1)$ NS5 branes,
\begin{align}
\label{eq:SNquiver}
{\left[\mathfrak{su}_{2}\right]}\times\bigg(
\lbrack \mathfrak{so}_{16}\rbrack  \, \, 
{\overset{\mathfrak{sp}_{1}}{\tikzmarknode[black!70!black]{c}{1}}} \, \,
\overbrace{
\overset{\mathfrak{su}_{2}}{\tikzmarknode[black!70!black]{l}{2}} \, \,
\cdots
\overset{\mathfrak{su}_{2}}{\tikzmarknode[black!70!black]{a}{2}} \, \,
{\overset{\mathfrak{su}_{2}}{2}} \, \,
\overset{\mathfrak{su}_{2}}{\tikzmarknode[black!50!black]{b}{2}} \, \,
\cdots
\overset{\mathfrak{su}_{2}}{\tikzmarknode[black!70!black]{m}{2}} \, \,}^{\times (2M-3)} \, \,
{\overset{\mathfrak{sp}_{1}}{\tikzmarknode[black!50!black]{d}{1}}} \, \,
\lbrack \mathfrak{so}_{16}\rbrack\bigg)  \, ,
\begin{tikzpicture}[remember picture,overlay]
\draw[red,<->]([yshift=0.000009ex]a.south) to[bend right]node[below]{\scriptsize} ([yshift=0.000009ex]b.south);
\draw[red,<->]([yshift=0.000009ex]l.south) to[bend right]node[below]{\scriptsize} ([yshift=0.000009ex]m.south);
\draw[red,<->]([yshift=0.000009ex]c.south) to[bend right]node[below]{\scriptsize} ([yshift=0.000009ex]d.south);
\end{tikzpicture}
\end{align} \\
where the common $\mathfrak{su}_{2}$ flavor factor acts on all $\mathfrak{su}_{2}-\mathfrak{su}_{2}$ bi-fundamentals~\cite{Apruzzi:2020eqi}. There are $M$ distinct orbits, and we can compute the matrix $\eta_{\alpha \beta}^{\,\mathcal{B}^{(2)}}$ just as before. The reduced LST charges in the twisted theory are 
\begin{align}
\vec{\ell}^{\,\,\mathcal{B}^{(2)}}_{\text{LST}}=(1,\underbrace{2}_{\times (M-1)}) \, .
\end{align} 
The matching data is then summarized as
\begin{align}
\label{matchbasegen}
\text{rk}(\ff)=9\, , \qquad  \text{dim(CB)}=2M \, , \qquad \kappa_{R}=4M-2\, .
\end{align}
The analog of the T-dual theory~\eqref{eq:SNquiver} for the case with $M$ instantons is
\begin{align}
\lbrack \mathfrak{su}_{16}^{(2)} \times \mathfrak{u}_{1}\rbrack  \, \, 
{\overset{\mathfrak{su}_{4M-2}^{(2)}}{0}}\,.
\end{align}
The flavor symmetry is $\fso_{16} \times \mathfrak{u}_1$ for the $\mathfrak{sp}_{2M-1}$ 5D gauge symmetry, which means that the data \eqref{matchbasegen} matches. 

From Table~\ref{tab:so32top}, one can see that certain $\mathfrak{so}_{32}$ theories have base topologies that may allow for a discrete permutation symmetry of the tensor multiplets. Of course, we have already seen this for the $A_{n}$ type base. The $D_{n}$ and $E_{7}$ base topologies also have (multiple) symmetries we may twist by, as long as we fix a flavor holonomy $\lambda$ such that the flavor algebras match. As we discuss in the next section, some of those twists can lead to $\kappa_P=1$. We first discuss a $D_{4}$ singularity with a $ \mathbbm{Z}_{3}$ base twist. In fact, one could have already anticipated such an LST to exist from~\eqref{ex5t}, upon flipping fiber and base. This would predict an $\mathfrak{so}_{32}$ LST with a $\mathfrak{g}_{2}$ base topology, which is impossible for standard $\mathfrak{so}_{32}$ LSTs, cf.\ Table~\ref{tab:so32top}. However, it may be possible for a base-twisted $\mathfrak{so}_{32}$ LST. In particular, consider the quiver
\begin{align} 
{\overset{\mathfrak{sp}_{M+1}}{1}} \, \, 
\myoverset{\overset{\mathfrak{sp}_{M+1}}{1}}
{\myunderset{\underset{\mathfrak{sp}_{M+1}}{1}}
{\underset{\left[\mathfrak{sp}_{2}\right]}{\overset{\mathfrak{so}_{4M+20}}{4}}}} \, \, 
{\overset{\mathfrak{sp}_{M+7}}{1}} \, \,
\lbrack \mathfrak{so}_{24}\rbrack\,.
\end{align}
The above quiver admits an $S_3$ symmetry.We focus on a $ \mathbbm{Z}_3$ subgroup that acts as
\begin{align}
\label{Z3ex2sym} 
{\overset{\mathfrak{sp}_{M+1}}{\tikzmarknode[black!50!black]{c}{1}}} \, \, 
\myoverset{\overset{\mathfrak{sp}_{M+1}}{\tikzmarknode[black!50!black]{f}{1}}}
{\myunderset{\underset{\mathfrak{sp}_{M+1}}{\tikzmarknode[black!50!black]{d}{1}}}
{\underset{\left[\mathfrak{sp}_{2}\right]}{\overset{\mathfrak{so}_{4M+20}}{4}}}} \, \,
{\overset{\mathfrak{sp}_{M+7}}{1}} \, \,
\lbrack \mathfrak{so}_{24}\rbrack  \, \,
\begin{tikzpicture}[remember picture,overlay]
\draw[red,->]([yshift=-0.4ex]c.south) to[bend right]node[below]{\scriptsize} ([xshift=-0.7ex, yshift=-0.4ex]d.south);
\draw[red,<-]([xshift=0.8ex, yshift=2ex]f.south) to[bend left=80]node[below]{\scriptsize} ([xshift=0.9ex, yshift=-0.4ex]d.south);
\draw[red,->]([xshift=-0.9ex, yshift=2ex]f.south) to[bend right=60]node[below]{\scriptsize} ([yshift=1.5ex]c.north);
\end{tikzpicture}\,,
\end{align}
i.e., one orbit permuting the three nodes indicated, and then two orbits for the fixed nodes. We obtain the twisted 5D twisted Dirac pairing matrix:
\begin{align}
\eta^{\,\mathcal{B}^{(3)}}=
\left(\begin{array}{ccc}
1 & -1 & 0 \\
-3 & 4 & -1 \\
0 & -1 & 1
\end{array}\right)\, , \quad \text{ with } \quad \vec{\ell}^{\,\,\mathcal{B}^{(3)}}_{\text{LST}}=(3,1,1) \, ,
\end{align}
which is precisely what one would expect from a ``$\mathfrak{g}_{2}$ base''. In fact, the data of this theory matches that in~\eqref{ex5t} for $N=M+1$, and is given by
\begin{align}
\text{rk}(\ff)=14\, , \qquad  \text{dim(CB)}=4M+20 \, , \qquad \kappa_{R}=8M+32 , \qquad \kappa_{P}=2 \, .
\end{align}
Although we do not have a geometric construction of the T-dual in this case, the matching provides strong evidence.

\subsection{CHL-like twisted LSTs}
\label{ssec:Kappa1examples}
A characteristic feature of all (twisted) LSTs constructed to date is that their twisted 2-group structure is unchanged under twisting. The value of $\kappa_P$ is central for characterizing LSTs and their duals as heterotic ($\kappa_P=2$) or Type II ($\kappa_P=0$) LSTs. In this section, we propose a third type of LST, realized by specific twists of $\mathfrak{e}_{8} \times \mathfrak{e}_{8}$ or $\mathfrak{so}_{32}$ LSTs, that have $\kappa_P=1$. Since $\kappa_P$ is related to the number of M9 branes in M-theory, we will call these LSTs \textit{CHL-like} twisted LSTs, where the twist identifies two M9 brane stacks similar to the 9D CHL string~\cite{Chaudhuri:1995fk}. Since $\kappa_P$ is still a quantity we expect to match across (twisted) T-duals, CHL-like twisted LSTs may only be dual to other CHL-like LSTs, but not to heterotic or Type II LSTs. 

The starting point of our construction will be a simple $ \mathbbm{Z}_{2}$ reflection symmetry $\mathcal{B}^{(2)}$ of the base quiver, similar to~\eqref{eq:SNquiver}. Generally, we may consider a quiver of type
\begin{align}
{\overset{\mathfrak{g}_{1}}{\tikzmarknode[black!70!black]{c}{n_{1}}}} \, \,
\overset{\mathfrak{g}_{2}}{\tikzmarknode[black!70!black]{a}{n_{2}}} \, \,
\cdots
\overset{\mathfrak{g}_{N-1}}{\tikzmarknode[black!70!black]{l}{n_{N-1}}} \, \,
\overset{ \overset{ \textstyle\overset{\fg_{N+F}}{\textstyle n_{N+F}\mathstrut}\mathstrut }{ \vdots} }
{{\overset{\mathfrak{g}_{N}}{n_{N}}}} \, \,
\overset{\mathfrak{g}_{N-1}}{\tikzmarknode[black!50!black]{m}{n_{N-1}}} \, \,
\cdots
\overset{\mathfrak{g}_{2}}{\tikzmarknode[black!70!black]{b}{n_{2}}} \, \,
{\overset{\mathfrak{g}_{1}}{\tikzmarknode[black!50!black]{d}{n_{1}}}} \, \,, 
\begin{tikzpicture}[remember picture,overlay]
\draw[red,<->]([yshift=0.000009ex]a.south) to[bend right]node[below]{\scriptsize} ([yshift=0.000009ex]b.south);
\draw[red,<->]([yshift=0.000009ex]l.south) to[bend right]node[below]{\scriptsize} ([yshift=0.000009ex]m.south);
\draw[red,<->]([yshift=0.000009ex]c.south) to[bend right]node[below]{\scriptsize} ([yshift=0.000009ex]d.south);
\end{tikzpicture}
\end{align}\\
where $F+1$ quiver nodes, denoted by $n_N \ldots n_{N+F}$, are fixed under the quotient action. The $ \mathbbm{Z}_2$ quotient action $\mathcal{B}^{(2)}$ changes the 5D LS charges $\ell_i$ to
\begin{align}
\vec{\ell}^{\,\,\mathcal{B}^{(2)}}_{\text{LST}}=
\frac{1}{N(\ell)}
(2l_{1},2l_{2},\cdots ,2l_{N-1}, l_{N}, \ldots l_{N+F})\,,
\end{align}
with 
\begin{align}
 N(\ell)=\text{gcd}(2l_{1},2l_{2},\cdots ,2l_{N-1}, l_{N})\,.
\end{align}
We can now compute the twisted 2-group structure constant $\kappa_{P}^{5D}$,
\begin{align}
\kappa_{P}^{5D}&=-\frac{1}{N(\ell)} \bigg(2l_{1}(n_{1}-2)+2l_{2}(n_{2}-2) + \cdots + l_{N}(n_{N}-2)+\ldots l_{N}(n_{N+F}-2) \bigg) \nonumber \\ 
&=\frac{1}{N(\ell)}(\kappa_{P}^{6D}) 
 =\frac{2}{N(\ell)} \, ,
\end{align}
where we assumed a 6D heterotic LST to start with. The change in $\kappa_P$ only depends on the LST charges of the fixed curves; if they are even, $N(\ell)=\text{gcd}(l_{N}, \ldots l_{N+F})=2$, and thus $\kappa_{P} = 1$.

In the following, we discuss the CHL-like quotients in $\mathfrak{e}_8$ and $\fso_{32}$ LSTs, starting with $\mathfrak{e}_8$. These LSTs generically have a $ \mathbbm{Z}_2$ symmetry for any number of instantons and any singularity $\fg$, as long as both orbi-instanton theories are identical and hence $\mu_1=\mu_2=\mu$. The only distinction is whether the number of five-branes is odd or even:
\begin{align}
 \ldots \overset{\mathfrak{g}_{N-1}}{\tikzmarknode[black!70!black]{l}{n_{N-1}}} \, \,
{\overset{\mathfrak{g}_{N}}{n_{N}}} \, \,
\overset{\mathfrak{g}_{N-1}}{\tikzmarknode[black!50!black]{m}{n_{N-1}}}
 \ldots \text{ for } M=2N+1 \quad \text{ and } \quad 
\ldots {\overset{\mathfrak{g}_{N-1}}{\tikzmarknode[black!70!black]{a}{n_{N-1}}}} \, \,
{\overset{\mathfrak{g}_{N}}{\tikzmarknode[black!70!black]{c}{n_{N}}}} \, \,
{\overset{\mathfrak{g}_{N}}{\tikzmarknode[black!70!black]{d}{n_{N}}}} \, \,
{\overset{\mathfrak{g}_{N-1}}{\tikzmarknode[black!50!black]{b}{n_{N-1}}}} \ldots \text{ for } M=2N
\begin{tikzpicture}[remember picture,overlay]
\draw[red,<->]([yshift=0.000009ex]l.south) to[bend right]node[below]{\scriptsize} ([yshift=0.000009ex]m.south);
\draw[red,<->]([yshift=0.000009ex]a.south) to[bend right]node[below]{\scriptsize} ([yshift=0.000009ex]b.south);
\draw[red,<->]([yshift=0.000009ex]c.south) to[bend right]node[below]{\scriptsize} ([yshift=0.000009ex]d.south);
\end{tikzpicture}
\end{align} 
Let us further assume that we are on a partial tensor branch, where we have not introduced $\mathcal{T}(\fg,\fg)$ conformal matter. Note that the LST charges are $l_i=1$ for all curves, which remains unchanged when introducing conformal matter~\cite{DelZotto:2020sop,Baume:2024oqn}. For odd M, the curve $n_N$ is always invariant under the quotient and since its LST charge is one, we have $\kappa_p=2$ in 5D. When $M$ is even, we need to take the contribution of conformal matter into account, and in particular its LST charges. Recall that the Dirac and LST charges of tensors under smooth blow-ups/blow-downs change as
\begin{align}
\label{eq:blowupLScharge}
 \overset{\fg_1}{n_1}\, \, \overset{\fg_2}{n_2} \quad\text{ with }\quad \ell=(\ell_1,\ell_2) \qquad \leftrightarrow \qquad 
 \overset{\fg_1}{(n_1+1)} \, \, \overset{\fg_E}{1} \, \, \overset{\fg_2}{(n_2+1)} \quad\text{ with }\quad \ell= (\ell_1 , \ell_1+\ell_2,~\ell_2) \, .
\end{align}
Hence, adding conformal matter will always introduce an odd number of curves, of which the middle one has LST charge two. This can be also confirmed for any $\mathcal{T}(\fg,\fg)$ conformal matter using Table~\ref{tab:TCMSummary}. We conclude that for an even number of instantons, the $ \mathbbm{Z}_2$ quotient reduces $\kappa_p$ to 1, except for the case with $\fg=\fsu_n$, which has no conformal matter. Note that these arguments continue to hold even if we include outer automorphism twists in addition to base twists, since the former do not affect the LST charges. 

Next we consider $ \mathbbm{Z}_{2}$ base twists for $\mathfrak{so}_{32}$ LSTs. From Table~\ref{tab:so32top}, we can infer that the base topologies that admit a reflection symmetry are those with a chosen singularity $\fsu_{2n}, \fso_{2n+4}$ and $\fe_7$. Indeed, $\kappa_P=1$ is possible when choosing $\fg=\fso_{2n+4}$ bases without vector structure. The same is true for an $\fe_{7}$ base, since both fixed nodes have LST charge 2. These and all other base twists for $\mathfrak{so}_{32}$ theories are summarized in Table~\ref{tab:conjduals}, along with the putative $\mathfrak{e}_{8} \times \mathfrak{e}_{8}$ T-duals based on the matching data. Furthermore, we have proposed twisted T-duals with $\kappa_{P}=1$ for which so far no geometric realization exists to the best of our knowledge. It would be interesting to realize these duals geometrically in future work.

We illustrate the construction in an example where we pick an $\mathfrak{so}_{32}$ theory with $\fg=\fe_{7}$ and choose $\lambda$ such that the quiver admits a $ \mathbbm{Z}_{2}$ symmetry,
\begin{align}
\lbrack \mathfrak{so}_{16}\rbrack  \, \,
{\overset{\mathfrak{sp}_{M+4}}{{\tikzmarknode[black!70!black]{c}{1}}}} \, \, {\overset{\mathfrak{so}_{4M+16}}{{{\tikzmarknode[black!70!black]{a}{4}}}}} \, \,  {\overset{\mathfrak{sp}_{3M+4}}{{{\tikzmarknode[black!70!black]{l}{1}}}}} \, \, 
\myoverset{{\overset{\mathfrak{sp}_{2M}}{1}}}
{\overset{\mathfrak{so}_{8M+16}}{4}} \, \,
{\overset{\mathfrak{sp}_{3M+4}}{{{\tikzmarknode[black!70!black]{m}{1}}}}} \, \,
{\overset{\mathfrak{so}_{4M+16}}{{{\tikzmarknode[black!70!black]{b}{4}}}}} \, \,
{\overset{\mathfrak{sp}_{M+4}}{{{\tikzmarknode[black!70!black]{d}{1}}}}} \, \,
\lbrack \mathfrak{so}_{16} \rbrack \, , \qquad 
\vec{\ell}_{\text{LST}}=(1, 1, 3,\myoverset{2}{2},3,1, 1) \,.
\begin{tikzpicture}[remember picture,overlay]
\draw[red,<->]([yshift=0.000009ex]a.south) to[bend right]node[below]{\scriptsize} ([yshift=0.000009ex]b.south);
\draw[red,<->]([yshift=0.000009ex]l.south) to[bend right]node[below]{\scriptsize} ([yshift=0.000009ex]m.south);
\draw[red,<->]([yshift=0.000009ex]c.south) to[bend right]node[below]{\scriptsize} ([yshift=0.000009ex]d.south);
\end{tikzpicture}
\end{align}\\[12pt]
Note that the LST charges of the two fixed curves are even. Compactifying this theory on a circle and twisting by the $ \mathbbm{Z}_2$ symmetry, we obtain the 5D twisted Dirac pairing matrix
\begin{align}
\eta^{\mathcal{B}^{(2)}}=
\left(\begin{array}{ccccc}
1 & -1 & 0 & 0 & 0 \\
-1 & 4 & -1 & 0 & 0 \\
0 & -1 & 1 & -1 & 0 \\
0 & 0 & -2 & 4 & -1 \\
0 & 0 & 0 & -1 & 1 
\end{array}\right)\, , \quad \text{ with } \quad \vec{\ell}^{\,\,\mathcal{B}^{(2)}}_{\text{LST}}=(1,1,3,1,1) \, .
\end{align}
The resulting LST charges suggest that there exists an $\mathfrak{e}_{8} \times \mathfrak{e}_{8}$ type of LST probing an $\mathfrak{e}_{6}$ singularity. Since $\kappa_{P}=1$, the theory must be base-twisted as well. The full matching data is
\begin{align}
\label{eq:basekp1}
\text{rk}(\ff)=8\, , \qquad  \text{dim(CB)}=12M+28 \, , \qquad \kappa_{R}=24M+49 \, ,
\end{align}
which matches the data of the theory resulting from a base twist of the $\mathfrak{e}_{8} \times \mathfrak{e}_{8}$ LST
\begin{align}
\lbrack \mathfrak{e}_{8}\rbrack  \, \, 
1 \, \,
2 \, \,
{\overset{\mathfrak{su}_{2}}{2}} \, \,
{\overset{\mathfrak{g}_{2}}{3}} \, \,
1 \, \,
{\overset{\mathfrak{f}_{4}}{5}} \, \,
1 \, \,
{\overset{\mathfrak{su}_{3}}{3}} \, \,
1 \, \,
\underbrace{{\overset{\mathfrak{e}_{6}}{6}} \, \,\ldots 1 \, \,
{\overset{\mathfrak{su}_{3}}{3}} \, \,
1 \, \,
{\overset{\mathfrak{e}_{6}}{6}} }_{\times M} 
\,\,1 \, \,
{\overset{\mathfrak{su}_{3}}{3}} \, \,
1 \, \, 
\underbrace{\overset{\fe_6}{6}\, \, 1 \, \,
{\overset{\mathfrak{su}_{3}}{3}} \, \,
1 \, \,
\ldots 
{\overset{\mathfrak{e}_{6}}{6}} \, \,}_{\times M} 
1 \, \,
{\overset{\mathfrak{su}_{3}}{3}} \, \,
1 \, \,
{\overset{\mathfrak{f}_{4}}{5}} \, \,
1 \, \,
{\overset{\mathfrak{g}_{2}}{3}} \, \,
{\overset{\mathfrak{su}_{2}}{2}} \, \,
2 \, \,
1 \, \,
\lbrack \mathfrak{e}_{8}\rbrack  \,,
\end{align} 
where the fixed curve is the middle $\fsu_3$ node. Using~\eqref{eq:blowupLScharge}, we can check that its LS charge is 2. Hence, the twisted 5D LS charge is 
\begin{align}
\vec{\ell}^{\mathcal{B}^{(2)}}_{\text{LST}}=(1,1,1,1,2,1,\underbrace{3,2,3,1}_{\times (M+1)},3,1) \, .
\end{align} 
From the twisted LST charges and 2-group structure constants, it follows that $\kappa_{P}=1$, and the other data in~\eqref{eq:basekp1} matches as well. Hence, fiber-base duality coupled with T-duality data is enough to motivate the dual LST.

\subsubsection{M9 branes and anomaly inflow} 
In \cite{DelZotto:2020sop}, $\kappa_P$ was given the interpretation as the number of M9 branes in the HW construction of the heterotic string leading to the two $\fe_8$ factors.
Type II LSTs have $\kappa_P=0$, which is the only other possible value for 6D LSTs in general \cite{Bhardwaj:2015oru,Baume:2024oqn}. The structure constant $\kappa_P$ characterizes parts of the $\mathcal{N}=(0,4)$ 2D worldsheet CFT of the LST. It contributes $n_{3-7}=8 \kappa_P$ string defects that couple to the bulk flavor symmetry, which are readily interpreted as the two M9 brane factors of the heterotic string. In total, one obtains the total left-handed central charge \cite{Kim:2019vuc,Tarazi:2021duw,Baume:2024oqn} 
\begin{align}
 c_l = 9 \kappa_P + 2 \, . 
\end{align}
Unitarity of the worldsheet CFT then constrains the (bulk) flavor symmetry $\ff= \prod_I \ff_{I}$ to
\begin{align}
\label{eq:unitarity}
 c_l \geq \sum_I \ell^I_\text{LST} \frac{\text{dim}(\ff_{I})}{\ell^I_\text{LST} + h^\vee_{\ff_{I}}}\geq \text{rk}(\ff)\, , 
\end{align}
where $\ell^I_\text{LST}$ denotes the LST charge of the quiver node $\overset{\fg_I}{n_I}$ which the flavor factor $\ff_I$ attaches to. Assuming that the above considerations also apply for the twisted worldsheet theory, the flavor rank would be bounded by $\text{rk}(\ff)\leq 11$ for theories with $\kappa_P=1$. Generically, the quotient that identifies the M9-branes acts freely on the 6D flavor group, since the fixed curve is part of a conformal matter chain and hence has no flavor factor.\footnote{An interesting exception might be delocalized $\mathfrak{u}_1$ flavor factors under which conformal matter may be charged and that survive under twisting.} 

While the same is true for the $\fso_{32}$ theory, the argument is more intricate since the O9/D9-branes sit on top of each other and hence must be split by some non-trivial flavor holonomy such as 
\begin{align}
 \lambda: \fso_{32} \rightarrow \fso_{16} \times \fso_{16}\, ,
\end{align}
to allow for twisting. This is also clear from the quiver, where an $\fso_{32}$ flavor symmetry leads to gauge symmetry decorations that are incompatible with the tensor multiplet symmetry we need to twist by. There is one further interesting exception: $\fso_{32}$ LSTs allow for flavor symmetry factors on nodes that are \textit{invariant} under the $ \mathbbm{Z}_2$ twist. However, such flavor symmetry factors $G^{\text{inv}}_{F_I}$ attach to LST curves with $\ell^I_\text{LST}$ even and are hence strongly constrained by the unitarity bounds~\eqref{eq:unitarity} of the untwisted theory. For example, an $\ff=\fso_{32}$ flavor symmetry factor attached to an $\ell^{I}_\text{LST}=2$ LST curve contributes $c_l=31$, which overshoots the LST unitarity condition. An $\fso_{16}$ flavor factor, on the other hand, has $c_l=15$, consistent with the untwisted bound. The same configuration stays consistent upon twisting, as the LST charges are divided by two, and we expect in the twisted worldsheet theory a left-moving contribution $c_l=8$, consistent with the left-moving current for $\kappa_p=1$. These arguments give evidence for the consistency of $\kappa_p=1$ twisted LSTs. It would be very interesting to rigorously derive the twisted worldsheet theory from anomaly inflow in the future.

\section{Geometry and twisting}
\label{sec:Geometry}
In this section, we discuss the geometric setup we use to construct the T-dual models for the LSTs discussed in this paper. The geometry of twisted compactifications in the context of SCFTs and LSTs has been explored in detail in \cite{Bhardwaj:2019fzv} and \cite{Braun:2021lzt,Bhardwaj:2022ekc,Anderson:2023wkr}.

\subsection{Genus-one polarized K3s}
Circle compactification of an F-theory model is equivalent to M-theory on a torus-fibered CY threefold. $ \mathbbm{Z}_n$-twisted circle reductions, on the other hand, are related to torus-fibrations that do not have a section but only an $n$-section \cite{Braun:2014oya,Mayrhofer:2014opa,Anderson:2014yva}, and are called genus-one fibrations. Moreover, as we are considering heterotic LSTs, the threefold also has to be K3 fibered. Thus, the starting point of our construction of twisted LSTs is a genus-one fibered polarized K3 surfaces $S$. The K3 fibers $S$ must hence have a Picard lattice of rank at least two, to accommodate the fiber and base. If the fibration is elliptic, the respective NS lattice is just the hyperbolic lattice 
\begin{align}
 \text{NS}(S)= U=\left( \begin{array}{cc}
 0& 1\\ 1 & 0
 \end{array}\right) \, .
\end{align}
More generally, a 2D Picard lattice has the form \cite{Braun:2016sks}
\begin{align}
 \text{NS}(S)= H_n=\left( \begin{array}{cc}
 2a& b \\ b & 2c
 \end{array}\right) \, \quad \text{ with } a,b,c \in  \mathbbm{Z} \, . 
\end{align}
If the determinant of $H_n$ is a perfect square, $b^2-4ac = n^2$, the K3 surface $S$ admits an $n$-section. This becomes evident when choosing a basis in which one of the generators is a Kollar divisor $D_K$ corresponding to the $ \mathbbm{P}^1$ base of $S$. This divisor must square to zero and hence fixes $D^2_K=2c=0$. The second generator must therefore contain the torus class with an $n$-section $D_n$, which indeed intersects the base $D_n \cdot D_K=b=n$ times. 

The general geometric construction of an $n$-section geometry is not known in general but for small $n$ it is given by~\cite{Braun:2014oya} 
\begin{itemize}
	\item $n=1$: a degree six polynomial in $F= \mathbbm{P}^2_{1,2,3}$
 \item $n=2$: a quartic curve in $F= \mathbbm{P}^2_{1,1,2}$.
 \item $n=3$: a cubic curve in $F= \mathbbm{P}^2$.
 \item $n=4$: a complete intersection of two quadrics in $F= \mathbbm{P}^3$. 
\end{itemize}
Higher order twists require more sophisticated constructions.\footnote{A detailed discussion of genus one fibrations with $n=5$-sections can be found in~\cite{Knapp:2021vkm,Schimannek:2021pau}.} The K3 surface can be constructed torically as a hypersurface in an ambient space $\mathcal{A}$, that is a fibration $F$ over a $ \mathbbm{P}^1$, using the Batyrev construction. Toric data and Mori-cone charges for the $n=2$ case are
\begin{align}
\label{eq:K3n=2}
  \begin{array}{c} X\\Y\\Z\\z_0\\ z_1\\ \\ \end{array} \left(\begin{array}{ccc|cc}
  -1&1&0& 1 & -1 \\
  -1&-1&0&1& -1 \\
  1&0&0&2& 0\\
  -1&0&1&0& 1 \\
   -1&0&-1&0& 1 \\
   0 & 0 &0 &-4 &0
 \end{array} \right)\,.
\end{align}
The anticanonical hypersurface is given by the zero locus of the quartic polynomial
\begin{align}
\label{eq:K3Z2}
 p=a_{4,1} X^4+Y^4 + a_{4,2} X^2 Y^2 + a_{4,3} X^3 Y+ a_{4,4} Y^3 + a_{2,1} X Y Z + a_{2,2}X^2 Z+ a_{2,3} Y^2 Z + a_{0} Z^2\,.
\end{align}
Similarly, the case $n=3$ is given by the vanishing of the cubic
\begin{align}
\label{eq:K3n=3}
 \begin{array}{c} X_0\\X_1\\X_2\\z_0\\ z_1\\ \\ \end{array} \left(\begin{array}{ccc|cc}
  1&0&0& 1 & -1 \\
  0&1&0&1& -1 \\
  -1&-1&0&1& 0\\
  1&0&1&-1& 1 \\
   0&1&-1&-1& 1 \\
   0 & 0 &0 &-3 &0
 \end{array} \right)\,.
\end{align} 
with the hypersurface
\begin{align}
\label{eq:K3Z3}
\begin{split}
 p= &\phantom{+}a_{3,1} X_0^3 + a_{3,2}X_1^3 + a_{0} X_1^3 +a_{3,3} X_0^2 X_1+a_{3,4} X_0 X_1^2 \\ &+ a_{2,1} X_0 X_1 X_2 +a_{2,2} X_0^2 X_2 
 +a_{2,3} X_1 X_2^2
 +a_{2,4} X_1^2 X_2 + a_{1} X_0 X_2^2 \,.
 \end{split}
\end{align}
The $a_{k,j}$ are degree $k$ polynomials in the ambient space coordinates $z_i$. Choosing $z_0$ as the Kollar divisor and $[X]$ and $X_1$ in the quartic and cubic respectively, as fiber generators results in the NS lattice
\begin{align}
H_n=\left(\begin{array}{cc}
 0&n\\
n &0
 \end{array} \right) \, ,
\end{align} 
The two types of K3 geometries \eqref{eq:K3n=2} and \eqref{eq:K3n=3} serve as the starting points for $ \mathbbm{Z}_2$- and $ \mathbbm{Z}_3$-twisted compactifications.

The elliptic threefolds from which we engineer the LSTs are constructed by fibering the K3 geometry over a non-compact base $\mathbbm{C}$. The non-trivial dynamics of the theory are obtained from the compact divisors in which the central K3 becomes further reducible and/or admits non-trivial fibers as reviewed for example in~\cite{DelZotto:2022xrh,DelZotto:2023ahf}. The conceptual novelty of genus-one fibrations is that they can have new types of monodromies~\cite{Braun:2014oya}, as analyzed in detail in \cite{Anderson:2023wkr}. 

As discussed in \cite{Ahmed:2023lhj}, we expect to find a maximal polarization for the K3 surface $S$, which realizes the full field theory flavor symmetry group in the NS lattice. However, $S$ might not be realized as a smooth toric hypersurface with flavor factors encoded in its NS lattice. Similarly, in our construction we restrict ourselves to torus fibers of $S$ that descend from an ambient space fibration, which limits the amount of possible duals. This can be overcome with \textit{extended duality models}, as discussed in Section~\ref{ssec:overlap}. If the full flavor algebra is torically realizable, the rank of the flavor algebra is related to the rank of the toric Picard group, $\text{rk}(\ff)=\text{Pic}(S)_{\text{tor}}-2$, which can be computed from the toric rays. Twisted flavor algebras arising in genus-one-fibered geometries behave similarly to the non-simply laced flavor factors discussed in~\cite{Ahmed:2023lhj}: a toric ray corresponding to a divisor on the K3 ambient space $\mathcal{A}_{K3}$ that intersects the K3 hypersurface $S$ multiple times becomes reducible on $S$ and enforces the volume of all curves to be the same.

The full threefold $X$ in an ambient space $\mathcal{A}_{X}$ is then constructed by fibering the K3 ambient space $\mathcal{A}_{K3}$ over $\mathbbm{C}$. This results in a semi-convex polytope $\Delta_4$ analogous to the Batyrev construction as defined in \cite{DelZotto:2022xrh}. Just as the torus fibration is inherited from a nested fibration $F$ in the K3 polytope $\Delta_3$, we obtain a K3 fibration by nesting the K3 polytope in $\Delta_4$ as 
\begin{align}
 \Delta_4 = \left( \begin{array}{c|c}
  ( \Delta_{K3} & 0) \\ \hline
  (x_i,y_i,z_i & d_i) 
 \end{array} \right) \, ,\qquad \text{ for } i=0 \ldots \text{dim(CB})\, , 
\end{align}
with $d_i \in  \mathbbm{N}^+$. The fibral part of the vertices $v_i=(x_i,y_i,z_i,d_i)$ in the K3 direction 
encode its respective degeneration, and yields the compact divisors of the geometry.

From this, twisted algebras can be constructed from an ADE singularity that is folded according to the generalized monodromies arising in genus-one fibrations~\cite{Bouchard:2003bu,Braun:2014oya,Anderson:2023wkr}. The geometric construction deviates from the naive expectation in that engineering a $\fg^{(n)}$ twisted algebra does not start with a $\fg$ type of singularity. Instead, the twisted algebra originates from the monodromy action induced by an $n$-section on the affine node of the $\fg^{(1)}$ extended Dynkin diagram that resolves the singularity. The geometric dictionary of such monodromies for toric hypersurfaces \cite{Bouchard:2003bu} is
\begin{align}
 \fe_7^{(1)} \rightarrow \fe_6^{(2)} \,,\quad
 \fe_6^{(1)} \rightarrow \fso_8^{(3)} \,,\quad
 \fso_{2N+8}^{(1)} \rightarrow \fso_{2N+6}^{(2)}\,,\quad
 \fso_{8}^{(1)} \rightarrow \fsu_{3}^{(2)} \,.
\end{align}
Hence, there are only two $\fsu_N^{(2)}$ type of algebras, both obtained from twisting $\fso_8^{(1)}$, $\fsu_3$ and $\fsu_4\sim\fso_6 $. We find that the first two affine folded singularities correspond to the exchange symmetry of twisted T-dual singularities summarized in Table~\ref{tab:TCMSummary}.

\subsubsection*{Examples}
To illustrate the geometric construction introduced above, we discuss several examples. We consider the following 5D twisted LST:
\begin{align}
\text{LST}_1: \quad
\lbrack \mathfrak{e}_{6}^{(2)}\rbrack  \, \, 
1 \, \,
{\overset{\mathfrak{su}_{3}^{(2)}}{3}} \, \, 
1 \, \,
{\overset{\mathfrak{e}_{6}^{(2)}}{6}} \, \, 
1 \, \,
{\overset{\mathfrak{su}_{3}^{(2)}}{3}} \, \, 
1 \, \,
{\overset{\mathfrak{f}_4}{5}} \, \, 
1 \, \,
{\overset{\mathfrak{g}_2}{3}} \, \,
{\overset{\mathfrak{su}_2}{2}} \, \,
2 \, \,
1 \, \,
{\overset{\mathfrak{e}_8}{12}} \, \,
\underbrace{\ldots 
\overset{\mathfrak{e}_{8}}{12} \, \,}_{\times (M-3)} \, \, 
1 \, \, 
2 \, \, 
{\overset{\mathfrak{su}_2}{2}} \, \,
{\overset{\mathfrak{g}_2}{3}} \, \,
1 \, \,
{\overset{\mathfrak{f}_{4}}{5}} \, \,
1 \, \,
{\overset{\mathfrak{su}_{3}^{(2)}}{3}} \, \, 
1 \, \,
{\overset{\mathfrak{e}_{6}^{(2)}}{6}} \, \, 
1 \, \,
{\overset{\mathfrak{su}_{3}^{(2)}}{3}} \, \,
1 \, \,
\lbrack \mathfrak{e}_{6}^{(2)}\rbrack  \, \, 
\end{align}
This is a $ \mathbbm{Z}_2 \times  \mathbbm{Z}_2$-twisted compactification of the $\fe_{8} \times \fe_{8}$ heterotic string with $(M-3)$ NS5 branes probing an $\fg=\fe_8$ singularity. Each $ \mathbbm{Z}_2$ corresponds to the discrete symmetry of one of the orbi-instanton pieces, and twisting by both of them results in a $\ff=\fe_6^{(2)}\times \fe_6^{(2)}$ 5D flavor symmetry.

\begin{figure}
 \centering 
\includegraphics[width=0.15\textwidth,angle=90]{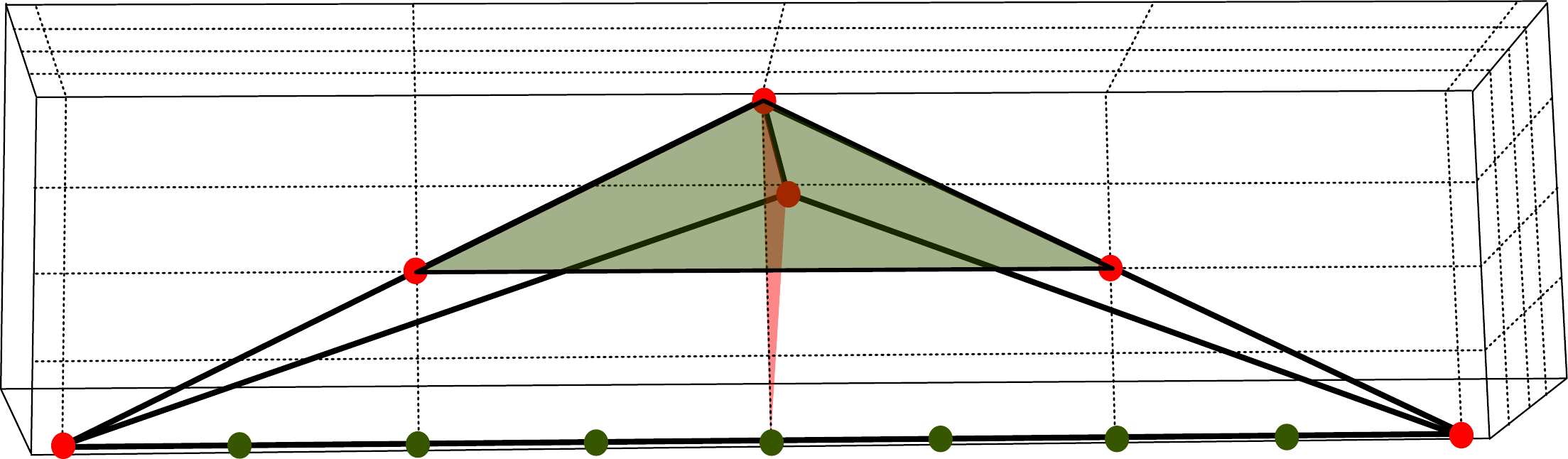}\\
$
\begin{array}{|ccccc|}
\multicolumn{5}{c}{\Delta_{K3}} \\ \hline
 (-1, -1, \phantom{-}1)   & (-1, -1, -4) & (\phantom{-}1, \phantom{-}0, \phantom{-}0) & (-1, -1, \phantom{-}4) & (-1, \phantom{-}1, \phantom{-}0) \\
(-1, -1, -3) & (-1, -1, -2) & (-1, -1, -1) & (-1, -1, \phantom{-}0) & \\
(-1, -1, \phantom{-}2) & (-1, -1, \phantom{-}3) & (-1, \phantom{-}0, -2) & (-1, \phantom{-}0, \phantom{-}2) & \\ \hline
\end{array}
$
 \caption{The K3 flavor polytope and rays that correspond to divisors that resolve the underling hypersurface. Rays in green intersect the K3 twice and are folded by a $\mathbbm{Z}_2$ monodromy. The red and green 2D sub-polytope give rise to the two torus fibrations.}
 \label{fig:TwistedK3}
\end{figure}

The model has an untwisted $\fso_{32}$ dual with quiver
\begin{align} 
\text{LST}_2: \quad
 \overset{\mathfrak{sp}_{M-5}}{1}\, \, 
{\overset{\mathfrak{so}_{4M-4}}{4}} \, \,
{\overset{\mathfrak{sp}_{3M-7}}{1}} \, \,
{\overset{\mathfrak{so}_{8M-8}}{4}} \, \,
{\overset{\mathfrak{sp}_{5M-9}}{1}} \, \,
\myoverset{\overset{\overset{{}}{\mathfrak{sp}_{3M-7}}}{1}}
{\overset{\mathfrak{so}_{12M-12}}{4}} \, \,
{\overset{\mathfrak{sp}_{4M-4}}{1}} \, \,
{\overset{\mathfrak{so}_{4M+12}}{4}} \, \,
\lbrack \mathfrak{sp}_{8}\rbrack  \,.
\end{align}
The flavor rays of the respective polytope are
\begin{align}
 \rho_1=(-1, -1, -4)\, , \quad
\rho_2=(-1, -1, 4)\, , \quad
 \rho_3=( 1, 0, 0)\, , \quad
\rho_4=(-1, 1, 0)\, .
\end{align}
We show all rays of the polytope in Figure~\ref{fig:TwistedK3}, and highlight the ones with a $ \mathbbm{Z}_2$ monodromy in green. 
The dimension of the Picard group of $S$ can be computed via the Batyrev formulas as 
\begin{align}
 \text{Pic}(S)=\text{Pic}(S)_{\text{tor}}+\text{Pic}(S)_{\text{n-tor}}=10+7 \, .
\end{align}
The seven non-toric contributions are due to the seven non-toric divisors, depicted in Figure~\ref{fig:TwistedK3} in green. The 2D sub-polytopes that defines an ambient torus fibration are given by a 2D hyperplane equation
\begin{align}
\label{eq:projection}
 \langle \vec{c}_\perp , \vec{v} \rangle = C=0 \quad \text{ for }\quad \vec{v} \in \Delta_{K3} \, ,
\end{align}
where $\vec{c}_\perp$ is a primitive lattice vector and $(0,0,0)$ is the unique interior point within 2D sub-polytope spanned by the rays $\vec{v}$ that satisfy \eqref{eq:projection} which is hence reflexive.
There may be additional singularities in the base $ \mathbbm{P}^1_{z_0,z_1}$ at $\mathcal{D}_i:\{z_i=0\}$, resolved by additional fibral divisors. The corresponding rays are above and below the 2D plane defined by \eqref{eq:projection}, respectively, and therefore satisfy
\begin{align}
\begin{split}
 z_0=0& \quad\text{ for }\quad \langle \vec{c}_\perp, \vec{v} \rangle = C>0 \;,~~ \vec{v} \in \Delta_{K3} \, , \\
  z_1=0& \quad\text{ for }\quad \langle \vec{c}_\perp, \vec{v} \rangle = C<0 \;,~~ \vec{v} \in \Delta_{K3} \, ,
  \end{split}
\end{align}
Reading off the flavor factors is then straight forward: For the genus-one fibration, we need to keep the red subpolytope given by $\vec{c}_\perp=(0,0,1)$ in~\eqref{eq:projection} at finite size, which leaves two $\fe_6^{(2)}$ fibers. 
For the second fibration, we consider the green 2D subpolytope defined by $\vec{c}_\perp=(0,1,0)$ in \eqref{eq:projection}, which yields an $F_{13}$ fiber type which hosts a $ \mathbbm{Z}_2$ MW group and an $\fsp_8^{(1)}$ fiber. Thus, the geometric construction proves that the two LSTs are dual. 

We can also use the geometry to read off the maximal flavor symmetry common to both theories (MCFS) in 5D, as introduced in \cite{Ahmed:2023lhj}: Note that both 5D KK theories have different flavor symmetries, which may become identical upon choosing certain Wilson line backgrounds. The generic choice of flavor Wilson lines breaks both symmetries to their Cartan subalgebras. The minimal choice of flavor Wilson lines defines the MCFS. To read off the MCFS, we shrink only those divisors that correspond to rays that do not lie in either of the two 2D sub-polytopes that encode the elliptic fiber. This results in conditions for which curves are not blown down, and hence a flavor symmetry breaking. For the two LSTs discussed above, the MCFS is
\begin{align}
 \text{LST}_1 \cup \text{LST}_2: \quad \fsp_4 \times \fsp_4 \leftarrow \left\{ \begin{array}{l} \fsp_8^{(1)} \\ \fe_6^{(2)} \times \fe_6^{(2)} \end{array} \right. \, .
\end{align}

\subsection{Extended duality models}
\label{ssec:overlap}
For all duality chains discussed in Section~\ref{sec:MatchingDuals}, except for the theories below~\eqref{exl1}, we have proven T-duality via geometry by finding inequivalent torus fibrations. Our constructions are somewhat limited, since they assume that the torus fibrations descend from fibrations in the toric ambient space. This can be partially overcome via \textit{extended duality} models, which realizes the same non-compact CY in different ambient spaces. Consider $m$ ambient spaces 
$\mathcal{A}_1 , \mathcal{A}_2, \ldots \mathcal{A}_m$ with $k_1, k_2, \ldots k_m$
inequivalent fibrations, each such that the non-compact threefolds $X_1,X_2, \ldots X_m$, have $k_1, k_2, \ldots k_m$ torus fibrations. If an LST is realized in (at least one) pair of threefolds $X_i$ and $X_j$, all LSTs encoded $X_i$ and $X_j$ are T-dual. 
 
Recall from Section~\ref{sec:MatchingDuals} that $M$ NS5 branes fractionalizing on a $\fg=\fso_{8n-4}$ singularity, fractionalize in the same way as $2M$ NS5 branes on a twisted $\fg=\fso_{4n}^{(2)}$ singularity and are hence T-dual. Moreover, $\fg=\fso_{8n-4}$ generally admits two $\fso_{32}$ duals, one with $\fso_{8n-4}$ base topology and another with a folded base topology (see Figure~\ref{tab:so32top}).
We therefore expect (at least) four T-duals. 
Our starting point is the $\fe_8 \times \fe_{8}$ type LST $\mathcal{K}^{(1)}_{M-(n-1)}(\mathfrak{so}_{8n-4})$ given by the quiver
\begin{align}
\label{ex7un}
\quad \lbrack \mathfrak{so}_{7}\rbrack  \, \, 
{\overset{\mathfrak{su}_{2}}{2}} \, \,
{\overset{{\left[\mathfrak{su}_{2}\right]}}{1}} \, \, 
{\overset{\mathfrak{so}_{12}}{4}} \, \,
{\overset{\mathfrak{sp}_{4}}{1}} \, \,
{\overset{\mathfrak{so}_{20}}{4}} \, \,
{\overset{\mathfrak{sp}_{8}}{1}} \, \,
\cdots
\underbrace{\overset{{\left[\mathfrak{sp}_{1}\right]}}
{\underset{\left[\mathfrak{sp}_{1}\right]}
{\overset{\mathfrak{so}_{8n-4}}{4}}} \, \,
{\overset{\mathfrak{sp}_{4n-6}}{1}} \, \,}_{\times (M-(n-1))} \, \,
\overset{{\left[\mathfrak{sp}_{1}\right]}}
{\underset{\left[\mathfrak{sp}_{1}\right]}
{\overset{\mathfrak{so}_{8n-4}}{4}}} \, \,
\cdots
{\overset{\mathfrak{sp}_{8}}{1}} \, \,
{\overset{\mathfrak{so}_{20}}{4}} \, \,
{\overset{\mathfrak{sp}_{4}}{1}} \, \,
{\overset{\mathfrak{so}_{12}}{4}} \, \,
{\overset{{\left[\mathfrak{su}_{2}\right]}}{1}} \, \,
{\overset{\mathfrak{su}_{2}}{2}} \, \,
\lbrack \mathfrak{so}_{7}\rbrack  \, \, .
\end{align}
The respective twisted $\fso_{32}$ T-dual quiver $\mathcal{\phantom{}_+\Tilde{K}}^{(2)}_{M-(n-1)}( \mathfrak{so}_{8n-4})$ is
\begin{align}
\label{ex7sot}
 \lbrack \mathfrak{so}_{10}^{(2)}\rbrack  \, \, 
{\overset{\mathfrak{sp}_{M-n+1}}{1}} \, \,
\myoverset{\overset{\mathfrak{sp}_{M-n-1}}{1^{*}}}
{\overset{\mathfrak{so}_{4M-4n+10}^{(2)}}{4^{*}}} \, \,
\underbrace{
{\overset{\mathfrak{sp}_{2M-2n+2}}{1}} \, \,
{\overset{\mathfrak{so}_{4M-4n+14}^{(2)}}{4}} \, \,
{\overset{\mathfrak{sp}_{2M-2n+4}}{1}} \, \,
\cdots}_{(4n-7) \text{ nodes}}
\myoverset{\overset{\overset{{\left[ \mathfrak{so}_{6}^{(2)} \right]}}{\mathfrak{sp}_{M+n-3}}}{1^{*}}}
{\overset{\mathfrak{so}_{4M+4n-2}^{(2)}}{4}} \, \,
{\overset{\mathfrak{sp}_{M+n-1}}{1}} \, \,
\lbrack \mathfrak{so}_{14}^{(2)}\rbrack  \, ,
\end{align}
and an $\fso_{32}$ type of untwisted quiver is 
\begin{align}
\mathcal{\phantom{=}_-\Tilde{K}}^{(1)}_{M-(n-1)}(\mathfrak{so}_{8n-4}): \quad
{\lbrack \mathfrak{u}_{1} \rbrack } \times \bigg(\myoverset{\overset{\myoverset{ }{\mathfrak{su}_{2M+2n-4}}}{2}}
{\myunderset{\overset{\myunderset{2}{\mathfrak{su}_{2M+2n}}}{\left[\mathfrak{su}_{8} \right]}}
{\overset{\mathfrak{su}_{4M+4n-8}}{2}}} \, \, 
\underbrace{{\overset{\mathfrak{su}_{4M+4n-12}}{2}} \, \,  
{\overset{\mathfrak{su}_{4M+4n-16}}{2}} \, \,
\dots \dots}_{(2n-4) \text{ nodes}}
{\overset{\mathfrak{sp}_{2M-2n+2}}{1}} \lbrack \mathfrak{so}_{8}\rbrack \, \,\bigg) \, .
\end{align}
Finally there is the twisted $\fe_8 \times \fe_8$ T-dual on a twisted $\fg=\fso_{4n}$ singularity, given by
\begin{align}
\label{ex7t}
 \mathcal{K}^{(2)}_{2M}(\fso_{4n}):\quad 
\lbrack \mathfrak{so}_{12}^{(2)}\rbrack  \, \, 
{\overset{\mathfrak{sp}_{n-1}}{1}} \, \,
\myoverset{\overset{\overset{{\left[ \mathfrak{so}_{4}^{(2)} \right]}}{\mathfrak{sp}_{n-3}}}{1}}
{\overset{\mathfrak{so}_{4n}^{(2)}}{4}} \, \,
{\overset{\mathfrak{sp}_{2n-4}}{1}} \, \,
\underbrace{{\overset{\mathfrak{so}_{4n}^{(2)}}{4}} \, \,
{\overset{\mathfrak{sp}_{2n-4}}{1}} \, \,}_{\times (2M-2)} \, \,
\myoverset{\overset{\overset{{\left[ \mathfrak{so}_{4}^{(2)} \right]}}{\mathfrak{sp}_{n-3}}}{1}}
{\overset{\mathfrak{so}_{4n}^{(2)}}{4}} \, \,
{\overset{\mathfrak{sp}_{n-1}}{1}} \, \,
\lbrack \mathfrak{so}_{12}^{(2)}\rbrack  \, . 
\end{align} 
All four models share the following matching data
\begin{align}
\text{rk}(\ff)=12\, , \qquad  \text{dim(CB)}=8Mn+2n-6M-2 \, , \qquad \kappa_{R}=16Mn-16M+2\, .
\end{align}
Three of the T-dual models admit a realization within one threefold $X_1$, while the last is realized in a second threefold $X_2$, which also realizes one of the other three models.

For concreteness, we choose $M=3$ and $n=2$ in the following. The toric rays of the flavor K3 and the compact divisors are summarized in Table~\ref{tab:A1ToricRays}.

\begin{table}[t!]
\centering
\begin{picture}(0,200)
\put(-230,20){\includegraphics[scale=0.6]{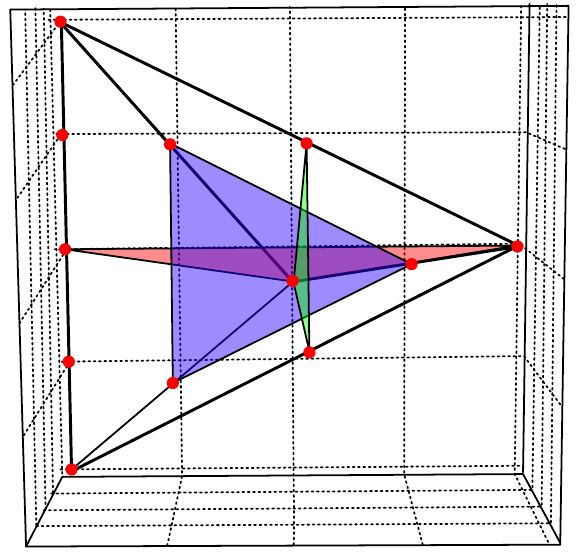}}
\put(-50,100){
 \begin{tabular}{ccc}
 \begin{tabular}{c}
  $(\Delta_{K3},0)$ \\ \hline
 $(-1, -2, -2,0)$\\
 $(\phantom{-}1, \phantom{-}0, \phantom{-}0,0)$\\
 $(-1, \phantom{-}2, \phantom{-}0,0)$ \\
 $(-1, -2, \phantom{-}2,0) $\\
 $(-1, -2, -1,0)$ \\
 $(-1, -2, \phantom{-}0,0)$ \\
 $(-1, -2, \phantom{-}1,0)$ \\ 
 $(-1, \phantom{-}0, -1,0)$ \\
 $(-1, \phantom{-}0, \phantom{-}1,0)$ \\ 
 $(\phantom{-}0, -1, -1,0)$ \\
 $(\phantom{-}0, -1, \phantom{-}1,0)$ \\
 \end{tabular} & 
\begin{tabular}{ c}
Compact rays \\ \hline 
 $(-1, -2, \phantom{-}0, 2)$ \\
 $(-1, -1, \phantom{-}0, 2)$ \\
 $(-1, \phantom{-}0, \phantom{-}0, 2)$ \\
 $(-1, -2, \phantom{-}0, 1)$ \\
 $(-1, \phantom{-}1, \phantom{-}0, 1)$ \\
 $(\phantom{-}0, -1, \phantom{-}0, 1)$ \\
 $(\phantom{-}0, \phantom{-}0, \phantom{-}0, 1)$
\end{tabular}
 &
\begin{tabular}{c}
Compact rays \\ \hline 
 $(-1, -2, \pm 3, 2)$ \\
 $(-1, \phantom{-}0, \pm 2, 2)$ \\
 $(-1, -2, \pm 2, 2)$ \\
 $(-1, -2,\pm 1, 2)$ \\
 $(-1, -1, \pm2, 2)$ \\
 $(-1, -1, \pm1, 2)$ \\
 $(-1, \phantom{-}0, \pm1, 2)$ \\
 $(-1, -2,\pm 2, 1)$ \\
 $(-1, -2, \pm1, 1)$ \\
 $(-1, -1,\pm 2, 1)$ \\
 $(-1, \phantom{-}1, \pm1, 1)$ \\
 $(\phantom{-}0, -1, \pm1, 1)$ \\
 $(\phantom{-}0, \phantom{-}0, \pm 1, 1)$
\end{tabular}
 \end{tabular}}
 \end{picture} 
 \caption{The toric vertices of the ambient space $\mathcal{A}_1$ that yield non-compact and compact divisors. The K3 polytope and its 3 2D subpolytopes of $\Delta_{K3}$ is illustrated on the left. }
 \label{tab:A1ToricRays}
\end{table}
\begin{table}[t!]
\centering
\begin{picture}(0,230)
\put(-230,30){\includegraphics[scale=0.25]{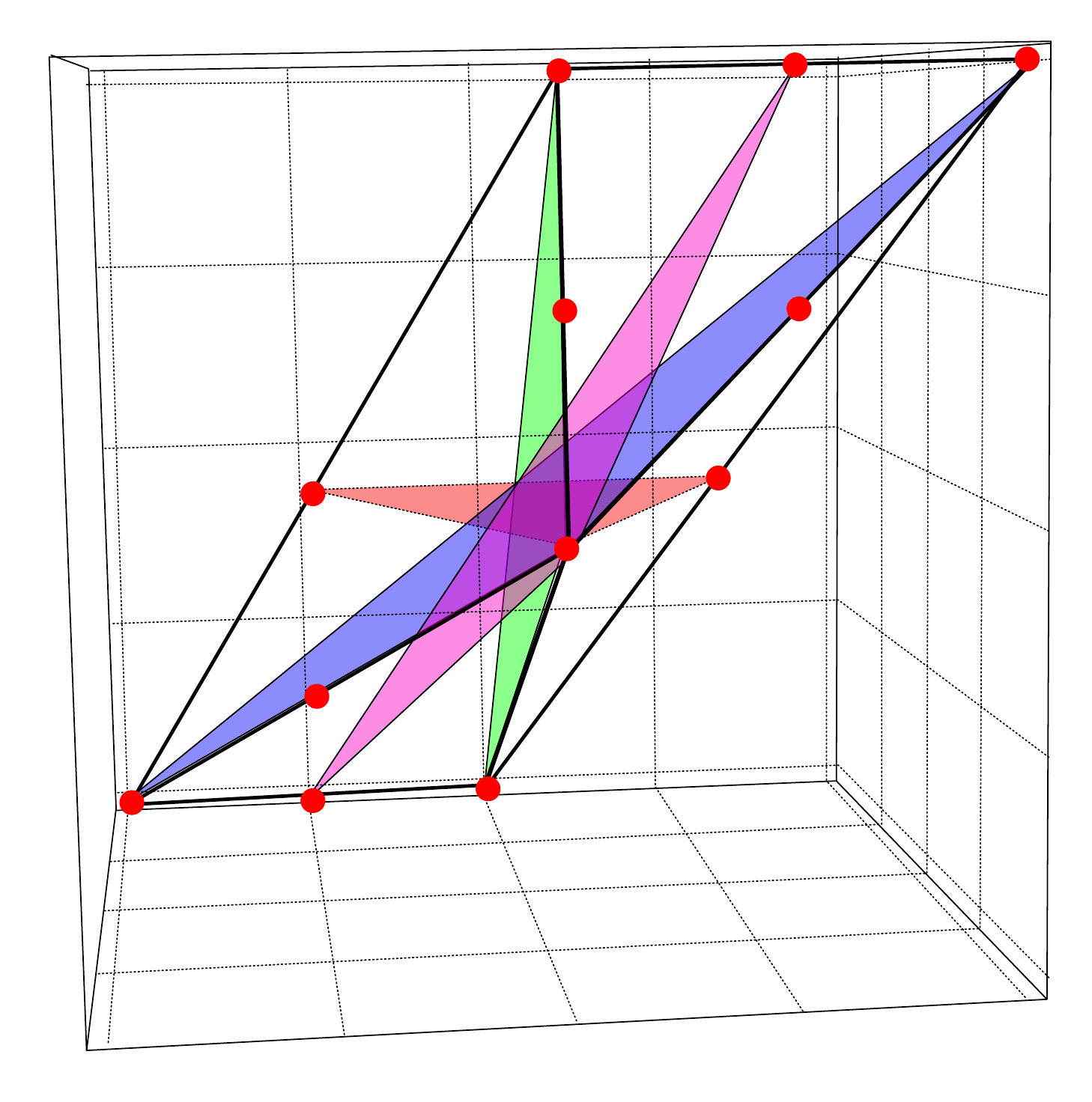}}
\put(-50,120){
 \begin{tabular}{ccc}
 \begin{tabular}{c}
  $(\Delta_{K3},0)$ \\ \hline
$(\phantom{-}1,\phantom{-}0,\phantom{-}0,0)$ \\
$(-3,-2,-2,0)$ \\
$(\phantom{-}1,\phantom{-}0,\phantom{-}2,0)$ \\
$(\phantom{-}1,\phantom{-}2,\phantom{-}2,0)$ \\
$(-3,\phantom{-}0,-2,0)$ \\
$(-1,\phantom{-}1,\phantom{-}0,0)$ \\
$(\phantom{-}1,\phantom{-}1,\phantom{-}2,0)$ \\
$(-3,-1,-2,0)$ \\
$(-1,-1,\phantom{-}0,0)$ \\
$(\phantom{-}1,\phantom{-}1,\phantom{-}1,0)$ \\
$(-1,\phantom{-}0,-1,0)$ \\
$(\phantom{-}1,\phantom{-}0,\phantom{-}1,0)$ \\
$(-1,-1,-1,0)$ 
 \end{tabular} & 
\begin{tabular}{ c}
Compact rays \\ \hline 
$(\phantom{-}4,\phantom{-}2,\phantom{-}5,1)$ \\
$(\phantom{-}2,\phantom{-}1,\phantom{-}3,1)$ \\
$(\phantom{-}0,\phantom{-}0,\phantom{-}1,1)$ \\
$(-2,-1,-1,1)$ \\
$(\phantom{-}5,\phantom{-}2,\phantom{-}6,1)$ \\
$(-2,-2,-1,1)$ \\
$(\phantom{-}5,\phantom{-}1,\phantom{-}6,1)$ \\
$(\phantom{-}3,\phantom{-}0,\phantom{-}4,1)$ \\
$(\phantom{-}1,-1,\phantom{-}2,1)$ \\
$(-1,-2,\phantom{-}0,1)$ \\
$(\phantom{-}4,\phantom{-}1,\phantom{-}4,1)$ \\
$(\phantom{-}2,\phantom{-}0,\phantom{-}2,1)$ \\
$(\phantom{-}0,-1,\phantom{-}0,1)$ \\
$(\phantom{-}5,\phantom{-}1,\phantom{-}5,1)$ \\
$(\phantom{-}3,\phantom{-}0,\phantom{-}3,1)$ \\
$(\phantom{-}1,-1,\phantom{-}1,1)$ \\
\end{tabular}
 &
\begin{tabular}{c}
Compact rays \\ \hline 
$(\phantom{-}9,\phantom{-}3,10,2)$ \\
$(\phantom{-}9,\phantom{-}2,10,2)$ \\
$(-1,-3,\phantom{-}0,2)$ \\
$(-1,-2,\phantom{-}0,2)$ \\
$(\phantom{-}7,\phantom{-}2,\phantom{-}8,2)$ \\
$(\phantom{-}5,\phantom{-}1,\phantom{-}6,2)$ \\
$(\phantom{-}3,\phantom{-}0,\phantom{-}4,2)$ \\
$(\phantom{-}1,-1,\phantom{-}2,2)$ \\
$(\phantom{-}8,\phantom{-}2,\phantom{-}9,2)$ \\
$(\phantom{-}6,\phantom{-}1,\phantom{-}7,2)$ \\
$(\phantom{-}4,\phantom{-}0,\phantom{-}5,2)$ \\
$(\phantom{-}2,-1,\phantom{-}3,2)$ \\
$(\phantom{-}0,-2,\phantom{-}1,2)$ \\
$(\phantom{-}7,\phantom{-}1,\phantom{-}8,2)$ \\
$(\phantom{-}5,\phantom{-}0,\phantom{-}6,2)$ \\
$(\phantom{-}3,-1,\phantom{-}4,2)$ \\
$(\phantom{-}1,-2,\phantom{-}2,2)$ \\
\end{tabular}
 \end{tabular}}
 \end{picture} 
 \caption{The toric vertices of the ambient space $\mathcal{A}_2$ that yield non-compact and compact divisors. The K3 polytope and its 4 2D subpolytopes of $\Delta_{K3}$ are illustrated on the left. }
 \label{tab:A2ToricRays}
\end{table}

The K3 has three different torus-fibrations, two of them being of genus-one type. The three fibrations are obtained via the projection vertices 
\begin{align}
 \vec{c}_{1,\perp}=(0,0,1,0)\, , \quad \vec{c}_{2,\perp}=(0,1,0,0) \, ,\quad \vec{c}_{3,\perp}=(1,0,0,0)\, , 
\end{align}
defining the three 2D sub-polytopes shown in 
Figure~\ref{tab:A1ToricRays}. The fiber types are given by ambient spaces $F_{13}$ and $F_4$ respectively in the nomenclature used in \cite{Klevers:2014bqa}. In Table~\ref{tab:A1ToricRays} we have summarized the additional compact divisors of the threefold ambient space that we will analyze in the following. 

The fan of the non-compact threefold base $B_2$ is obtained by projecting the toric ambient space rays. Recall that each $\vec{c}_\perp$ defines an inequivalent torus fibration with a projection $\pi$. Hence we may define $\langle \vec{v}_i , \vec{c} \rangle=C_i$ with $\vec{v}_i \in \Delta_4$ as before. The toric diagram of the base is then given as 
\begin{align}
\label{eq:baseFan}
 \pi(\vec{v}_i)= (\langle \vec{v}_i , \vec{c} \rangle, z_i)=(C_i, z_i) \, .
\end{align}
In this way, the vectors $\vec{v}_i$ with $z_i=0$ define the K3 ambient space $\Delta_{K3}$, and the vectors with $C_i=z_i=0$ yields the torus fiber.

The base fan and their respective fibers can then be read off easily, which results in the quiver
\begin{align}
\label{eq:dualM3-1}
\lbrack \mathfrak{so}_{7} \rbrack 
 \,\, \overset{\fsu_2}{2}\,\, \underset{[\fsu_2]}{1} 
 \,\, \overset{\fso_{12}}{4} \,\, \overset{\fsp_2}{1} 
 \,\, \overset{\fso_{12}}{4} \,\, \overset{\fsp_2}{1} \,\, \overset{\fso_{12}}{4} \,\, \underset{[\fsu_2]}{1} \,\, \overset{\fsu_2}{2} \,\, \lbrack \mathfrak{so}_{7} \rbrack 
\end{align}
Note that the two $\fsu_2$ flavor factors are realized non-torically in the K3 geometry. 

Similarly, we can find the second and third fibration by repeating the above analysis and using the new projection vertices. In the former case we find the quiver
\begin{align}
\label{ex6sot}
\lbrack \mathfrak{so}_{10}^{(2)}\rbrack  \, \, 
{\overset{\mathfrak{sp}_{2}}{1}} \, \,
\myoverset{{ 1 }}
{\overset{\mathfrak{so}_{14}^{(2)}}{4^{*}}} \, \,
{\overset{\mathfrak{sp}_{4}}{1}} \, \,
\myoverset{\overset{\overset{{\left[ \mathfrak{so}_{6}^{(2)} \right]}}{\mathfrak{sp}_{2}}}{1 }}
{\overset{\mathfrak{so}_{18}^{(2)}}{4}} \, \,
{\overset{\mathfrak{sp}_{4}}{1}} \,\,
\lbrack \mathfrak{so}_{14}^{(2)}\rbrack  \, .
\end{align}
The configuration admits so called non-flat fibers that are birational to resolving the $1$ curve and three non-flat fibers that resolve $\fsp_2$ in the second case (see \cite{Dierigl:2018nlv} for a general toric description).
Finally we have another genus-one fibration. Note that despite the absence of a section, no twisted algebra appears:
\begin{align}
\label{ex6oun}
{\left[\mathfrak{u}_{1}\right]}\times\bigg({\overset{\mathfrak{su}_{6}}{2}} \, \, 
\myoverset{\overset{\overset{{\left[ \mathfrak{so}_{8} \right]}}{\mathfrak{sp}_{4}}}{1}}
{\overset{\mathfrak{su}_{12}}{2}} \, \, 
{\overset{\mathfrak{su}_{10}}{2}} \, \, 
\lbrack \mathfrak{su}_{8} \rbrack \bigg) \, , 
\end{align}
Again the rays are non-flat fibers which resolve the $\fsp_4$ gauge algebra. 

To get access to more T-duals, we must find a second geometric description of one of the LSTs via a different ambient space $\mathcal{A}_2$ in such a way that it contains another inequivalent fibration. This is given by the toric vertices in Table~\ref{tab:A2ToricRays}. The key difference is that the different K3 fiber admits several torus fibrations. The first fibration is given via the projection vertex
\begin{align}
 \vec{c}_{1,\perp}=(0,1,0,0) \, .
\end{align}
It is an elliptic fibration given by the green polytope in Table~\ref{tab:A2ToricRays}. Taking into account the compact divisors, this configuration exactly reproduces the LST in \eqref{eq:dualM3-1}. However, the fiber type is different from that of $X_1$, and also both $\fsu_2$ flavor factors are torically realized. 

There exists another fibration, given by the projection vertex $\vec{c}_{2,\perp}=(0,0,1,0)$ corresponding to the red polytope in Table~\ref{tab:A2ToricRays}.
This fibration did not appear in $X_1$ and is a genus one fibration with $F_4$ fiber type and corresponding twisted quiver
\begin{align}
\label{eq:Overlap_4}
\lbrack \mathfrak{so}_{12}^{(2)}\rbrack  \, \, 
{\overset{\mathfrak{sp}_{1}}{1}} \, \,
\underset{[\fsp_1]}{
\overset{\fso_8^{(2)}}{3}}
\, \, 
1\, \, 
\overset{\fso_8^{(2)}}{4}
\, \,
1\, \, 
\overset{\fso_8^{(2)}}{4}
\, \,
1 \, \, 
\underset{[\fsp_1]}{
\overset{\fso_8^{(2)}}{3}} \, \,
\overset{\mathfrak{sp}_{1}}{1} \, \, 
\lbrack \mathfrak{so}_{12}^{(2)}\rbrack  \, .
\end{align} 
There are two more fibrations of the threefold $X_2$, induced from the violet and pink 2D sub-polytopesin Table~\ref{tab:A2ToricRays}. The first one corresponds to the projection vector $\vec{c}_{3,\perp}=(0,1,-1,0)$ and yields an untwisted LST given by 
 \begin{align}
\label{eq:Overlap_5}
\lbrack \mathfrak{so}_{9}\rbrack  \, \, 
1 \, \,
\overset{\fso_7}{3}
\, \, 
\underset{[\fso_{4}]}{
\overset{\fsp_2}{1}}\, \, 
\overset{\fso_{12}}{4}
\, \,
\overset{\fsp_2}{1}\, \, 
\overset{\fso_{12}}{4} \, \, 
\underset{[\fso_{4}]}{
\overset{\fsp_2}{1}}\, \, 
\overset{\fso_7}{3}
\, \, 
1 \, \, 
\lbrack \mathfrak{so}_{9}\rbrack  \, .
\end{align} 
The second fibration with projection vector $\vec{c}_{4,\perp}=(0,1,-2,0)$ yields another genus-one fibration with twisted quiver
\begin{align}
\label{eq:Overlap_6}
\lbrack \mathfrak{so}_{12}^{(2)} \rbrack  \, \, 
\overset{\fsp_3}{1}\, \, 
\overset{[\fso_{4}^{(2)}]}{\overset{
\overset{\fsp_1}{1}
}{\overset{\fso_{16}^{(2)}}{4}}} \, \, 
\overset{\fsp_4}{1} \, \, 
\overset{[\fso_{4}^{(2)}]}{\overset{
\overset{\fsp_1}{1}
}{\overset{\fso_{16}^{(2)}}{4}}} 
\overset{\fsp_3}{1}\, \, 
\lbrack \mathfrak{so}_{12}^{(2)} \rbrack 
\end{align} 
Note that the above quiver can be obtained 
via a (twisted) $\fso_{32}$ heterotic
theory on a $\fg=\fso_{12}$ singularity. Interestingly, we found a very similar model, but with a different flavor symmetry already in \eqref{eq:dualM3-1}.

\subsection{Twisting symmetries and the WC group}
\label{subsec:WC}
We want to close the geometry section by discussing a way to use the LST field theory to bootstrap potential bounds on the order of the so called Weil-Ch$\hat{\text{a}}$talet (WC) groups. The Weil-Ch$\hat{\text{a}}$talet group and its subgroup, the Tate-Shafarevich group $\Sha(X/B_2)$ (see \cite{1992alg.geom.10009D} and Appendix A of \cite{Cvetic:2015moa} for a mathematical discussion) are finite groups whose elements include threefold geometries that admit the same Jacobian $Y=\text{Jac}(X/B_2)$ as the genus-one fibration $X/B_2$. In particular, all geometries that are part of the WC group share the same base $B_2$ with the same discriminant $\Delta$ of the their respective torus fibers. 

The WC group is not widely discussed in the literature so far. It coincides with the more familiar Tate-Shafarevich group in the absence of \textit{multiple fibers}. However, in our study, multiple fibers appear as a generic feature of twisted fiber types (see \cite{Anderson:2018heq,Anderson:2019kmx,Anderson:2023wkr}).
As the WC group encodes the same base $B_2$ and discriminant $\Delta$, F-theory in either torus fibered threefold should be the same and thus $WC(X)$ be a symmetry of F-theory. However, M-theory differs on different WC geometries, since the geometries are topologically inequivalent. Therefore, the WC group describes the freedom to twist the 6D F-theory upon circle reduction, and it is natural to identify the WC group with the disconnected components of the F-theory symmetry group ,
\begin{align}
 \pi_0(G)=WC(X) \, .
\end{align} 
The TS/WC group has mostly been considered for compact threefolds, where the discrete symmetry must be gauged in order to use it in the twist. However, even in this well-studied setup, not much is known about the discrete symmetry sector, like bounds on the orders $ \mathbbm{Z}_n$ or how many independent factors it may have. 

Geometric engineering and the field theory perspective of SCFTs or LSTs may provide a bottom-up perspective of which type of bounds of the TS/WC group we may expect, at least for non-compact threefolds. Upon compactification, such symmetries must be gauged or broken according to the no global symmetry conjecture. Thus, the global symmetries in an LST may provide an upper bound on the discrete 0-form gauge sector $\pi_0(G)$ in a SUGRA theory. Such bounds can be obtained by taking an LST quiver and counting the number of consistent twists. Naively, there appear to be plenty of such twists, but interestingly the massive (BPS string) spectrum on the tensor branch constrains the discrete symmetries, as explained in Section~\ref{section:frev}. As an example, we consider the 6D quiver~\eqref{exagain} with the largest possible discrete symmetry we found:
 \begin{align}
\label{eq:again}
\mathcal{K}^{(1,1,1)}_{N}(\mu,\mu,\mathfrak{so}_{8}): \qquad \qquad 
\lbrack \mathfrak{e}_{6}\rbrack  \, \, 
1 \, \,
{\overset{\mathfrak{su}_{3}}{3}} \, \,
\underset{\left[\mathfrak{u}_{1} \times \mathfrak{u}_{1}\right]}
1 \, \,
\underbrace{
\overset{\mathfrak{so}_{8}}{4} \, \,
\underset{\left[\mathfrak{u}_{1} \times \mathfrak{u}_{1}\right]}
1 \, \,}_{\times (N)} \, \,
{\overset{\mathfrak{su}_{3}}{3}} \, \,
1 \, \,
\lbrack \mathfrak{e}_{6}\rbrack  \, , 
\end{align}
As a 5D KK theory, the above quiver represents the Jacobian of 5D twisted theories given in \eqref{ex4t1am}-\eqref{ex4t8am}: The left and right orbi-instanton blocks with $[\fe_6] \, \, 1 \overset{\fsu_3}{3} 1\ldots $ both support a $ \mathbbm{Z}_2$ discrete symmetry factor, which allows us to twist them individually. Similarly, the $\ldots \overset{\fso_8}{4}\, \, 1 \ldots $ conformal matter theory admits an $S_3$ symmetry, of which a $ \mathbbm{Z}_2$ or $ \mathbbm{Z}_3$ subgroup can be twisted. As both twists are possible, there is a $ \mathbbm{Z}_2 \times  \mathbbm{Z}_3$ symmetry possible for the block. We therefore expect a contribution of $ \mathbbm{Z}_2^2 \times  \mathbbm{Z}_6 $ to the WC group. Note that the above quiver also admits a $ \mathbbm{Z}_2$ base symmetry we could additionally twist by. However, base symmetries are not part of the WC group. It would be interesting to explore the possible group structure that captures such base twists geometrically in relation with F/M-theory duality. 

\section{Conclusion}
\label{sec:conclusion}
\begin{figure}[t!]
 \centering
 \begin{tikzpicture}[scale=1.7]
 \draw[very thick, ->] (-6,2.45) -- (-8,1.1); 
 	\node (3) at (-8.1,2) [ text width=3cm,align=center ] {$ \mathbbm{Z}_{2}$ outer automorphism twist }; 
\draw[very thick, ->] (-5.9,2.45) -- (-4,1.1); 
 	\node (3) at (-4.1,2) [ text width=3cm,align=center ] { $ \mathbbm{Z}_{2}$ base twist }; 
 	\node (1) at (-8,0.47) [draw,rounded corners,very thick,text width=5.6cm, align=center ] {{\bf \textcolor{black}{$
\lbrack \mathfrak{f}_{1}^{(1/2)}\rbrack  \, \, 
\cdots
\underbrace{{\overset{\mathfrak{e}_{6}^{(2)}}{6}} \, \,
1 \, \, 
\overset{\mathfrak{su}_{3}^{(2)}}{3} \,\,
1 \, \,}_{\times (2M)} 
\cdots 
\lbrack \mathfrak{f}_{1}^{(1/2)}\rbrack  \, \, 
\qquad $}}};
\node (4) at (-3.3,0.47) [draw,rounded corners,very thick,text width=8.7cm, align=center ] {{\bf \textcolor{black}{$
{{\tikzmarknode[black!70!black]{a}{\lbrack \mathfrak{f}_{1}^{(1)}\rbrack }}}\, \, 
{{\tikzmarknode[black!70!black]{f}{\cdots}}} \, \,
\underbrace{ 1 \, \,
{\overset{\mathfrak{su}_{3}^{(1)}}{3}} \, \,
1 \,\,
{\overset{\mathfrak{e}_{6}^{(1)}}{6}}}_{\times M} \,\,
{{\tikzmarknode[black!70!black]{a}{1}}} \, \,
{\overset{\mathfrak{su}_{3}^{(1)}}{3}} \, \,
{{\tikzmarknode[black!70!black]{c}{1}}} \, \,
\underbrace{ {\overset{\mathfrak{e}_{6}^{(1)}}{6}} \, \,
1 \, \,
{\overset{\mathfrak{su}_{3}^{(1)}}{3}} \, \,
1 \,\, }_{\times M} 
{{\tikzmarknode[black!70!black]{g}{\cdots}}} \, \,
{{\tikzmarknode[black!70!black]{b}{\lbrack \mathfrak{f}_{1}^{(1)}\rbrack }}} \, \, $
\begin{tikzpicture}[remember picture,overlay] 
\draw[red,<->]([yshift=-0.4ex]a.south) to[bend right]node[below]{\scriptsize} ([yshift=-0.4ex]c.south);
\draw[red,<->]([yshift=-0.4ex]f.south) to[bend right=15]node[below]{\scriptsize} ([yshift=-0.4ex]g.south);
\end{tikzpicture}
}}};

\draw[very thick, <->] (-3.5,-0.1) -- (-3.5,-1.7); 
 	\node (7) at (-4.05,-1) [ text width=3cm,align=center ] {T-duality};
\draw[very thick, <->] (-8,-0.05) -- (-8,-0.95); 
 	\node (5) at (-8.55,-0.5) [ text width=3cm,align=center ] {T-duality};
\node (6) at (-8,-1.5) [draw,rounded corners,very thick,text width=6cm, align=center ] {{\bf \textcolor{black}{$
\lbrack \mathfrak{f}_{2}^{(1)}\rbrack  \, \, 
\cdots
\underbrace{1 \, \overset{\fsu_2^{(1)}}{2}\, \overset{\fso_7^{(1)}}{3}\, \overset{\fsu_2^{(1)}}{2} \, 1 \overset{\fe_7^{(1)}}{8}}_{\times (M)} 
\cdots 
\lbrack \mathfrak{f}_{2}^{(1)}\rbrack  \, \, 
\qquad $}}};
\draw[very thick, ->] (-6.9,-3.5) -- (-3.5,-3.3); 
 	\node (6) at (-5.7,-3.7) [ text width=3cm,align=center ] {$ \mathbbm{Z}_{2}$ base twist};
\node (8) at (-3.5,-2.5) [draw,rounded corners,very thick, text width=8.2cm, align=center ] {{\bf \textcolor{black}{$
\lbrack \mathfrak{f}_{3}^{(1)}\rbrack  \, \,
{\overset{\mathfrak{sp}^{(1)}_{M..}}{{\tikzmarknode[black!70!black]{c}{1}}}} \, \, {\overset{\mathfrak{so}^{(1)}_{4M..}}{{{\tikzmarknode[black!70!black]{a}{4}}}}} \, \,  {\overset{\mathfrak{sp}^{(1)}_{3M..}}{{{\tikzmarknode[black!70!black]{l}{1}}}}} \, \, 
\myoverset{{\overset{\mathfrak{sp}^{(1)}_{2M..}}{1}}}
{\overset{\mathfrak{so}^{(1)}_{8M..}}{4}} \, \,
{\overset{\mathfrak{sp}^{(1)}_{3M..}}{{{\tikzmarknode[black!70!black]{m}{1}}}}} \, \,
{\overset{\mathfrak{so}^{(1)}_{4M..}}{{{\tikzmarknode[black!70!black]{b}{4}}}}} \, \,
{\overset{\mathfrak{sp}^{(1)}_{M..}}{{{\tikzmarknode[black!70!black]{d}{1}}}}} \, \, 
\lbrack \mathfrak{f}_{3}^{(1)}\rbrack  \, \,$
\begin{tikzpicture}[remember picture,overlay] 
\draw[red,<->]([yshift=-0.4ex]a.south) to[bend right=15]node[below]{\scriptsize} ([yshift=-0.4ex]b.south);
\draw[red,<->]([yshift=-0.4ex]l.south) to[bend right=15]node[below]{\scriptsize} ([yshift=-0.4ex]m.south);
\draw[red,<->]([yshift=-0.4ex]c.south) to[bend right=15]node[below]{\scriptsize} ([yshift=-0.4ex]d.south);
\end{tikzpicture}
}}};

\draw[very thick, <->] (-8,-2.0) -- (-8,-2.97); 
 	\node (5) at (-8.55,-2.5) [ text width=3cm,align=center ] {T-duality};
 \node (6) at (-8,-3.5) [draw,rounded corners,very thick,text width=3.4cm,align=center ] {{\bf \textcolor{black}{\underline{$\mathfrak{so}_{32}$ on a circle}\\ $\boldsymbol{M}$ NS5 branes on $\mathfrak{e}_{7}$\\}}};
\node (2) at (-6,3) [draw,rounded corners,very thick,text width=3.4cm,align=center ] {{\bf \textcolor{black}{\underline{$\mathfrak{e}_8 \times \mathfrak{e}_8$ on a circle}\\ 2$\boldsymbol{M}$ NS5 branes on $\mathfrak{e}_{6}$\\}}};
\node (9) at (-9,7) [draw,rounded corners,very thick,text width=3.4cm,align=center ] {{\bf \textcolor{black}{\underline{$\mathfrak{e}_8 \times \mathfrak{e}_8$ on a circle}\\ 6$\boldsymbol{M}$ NS5 branes on $\mathfrak{so}_{8}$ \\}}};
\node (10) at (-3,7) [draw,rounded corners,very thick,text width=3.4cm,align=center ] {{\bf \textcolor{black}{\underline{$\mathfrak{so}_{32}$ on a circle}\\ 6$\boldsymbol{M}$ NS5 branes on $\mathfrak{so}_{8}$\\}}};
 \draw[very thick, ->] (-9,6.45) -- (-9,5.2); 
 	\node (11) at (-8.0,5.8) [ text width=3cm,align=center ] {$ \mathbbm{Z}_{3}$ outer auto\-morphism twist }; 
\draw[very thick, ->] (-3,6.45) -- (-3,5.4); 
 	\node (12) at (-3.75,5.8) [ text width=3cm,align=center ] { $ \mathbbm{Z}_{3}$ base twist }; 
\node (13) at (-9,4.65) [draw,rounded corners,very thick,text width=4.8cm, align=center ] {{\bf \textcolor{black}{$
\lbrack \mathfrak{f}_{A}^{(1)}\rbrack  \, \, 
\cdots
\underbrace{{1 \, \,
\overset{\mathfrak{so}_{8}^{(3)}}{4}} \, \,
}_{\times (6M)} 
\cdots 
\lbrack \mathfrak{f}_{A}^{(1)}\rbrack  \, \, 
\qquad $}}}; 
\node (14) at (-3,4.47) [draw,rounded corners,very thick,text width=5.2cm, align=center ] {{\bf \textcolor{black}{$
{\overset{\mathfrak{sp}^{(1)}_{6M..}}{\tikzmarknode[black!50!black]{c}{1}}} \, \, 
\myoverset{\overset{\mathfrak{sp}^{(1)}_{6M..}}{\tikzmarknode[black!50!black]{f}{1}}}
{\myunderset{\underset{\mathfrak{sp}^{(1)}_{6M..}}{\tikzmarknode[black!50!black]{d}{1}}}
{\overset{\mathfrak{so}^{(1)}_{24M..}}{4}}} \, \, 
{\overset{\mathfrak{sp}^{(1)}_{6M..}}{1}} \, \,
\lbrack \mathfrak{f}_{C}^{(1)}\rbrack  \, \,$
\begin{tikzpicture}[remember picture,overlay]
\draw[red,->]([yshift=-0.4ex]c.south) to[bend right]node[below]{\scriptsize} ([xshift=-0.7ex, yshift=-0.4ex]d.south);
\draw[red,<-]([xshift=1.2ex, yshift=2ex]f.south) to[bend left=80]node[below]{\scriptsize} ([xshift=1.2ex, yshift=-0.4ex]d.south);
\draw[red,->]([xshift=-1.2ex, yshift=2ex]f.south) to[bend right=60]node[below]{\scriptsize} ([yshift=2.4ex]c.north);
\end{tikzpicture}}}}; 
\draw[very thick, <->] (-9,4) -- (-7.2,3); 
 	\node (15) at (-8.55,3.2) [ text width=3cm,align=center ] {T-duality};
\draw[very thick, <->] (-3,3.5) -- (-4.8,3); 
 \node (16) at (-3.4,3) [ text width=3cm,align=center ] {T-duality};
 \draw[very thick, <->] (-7.4,4.65) -- (-4.7,4.65); 
 \node (17) at (-6.05,4.8) [ text width=3cm,align=center ] {T-duality};
 \draw[very thick, <->] (-7.8,7) -- (-4.3,7); 
 \node (18) at (-6.05,7.15) [ text width=3cm,align=center ] {T-duality};
\end{tikzpicture}
\caption{Example of a duality network connecting multiple twisted theories.}
\label{fig:DualityNetwork}
\end{figure}
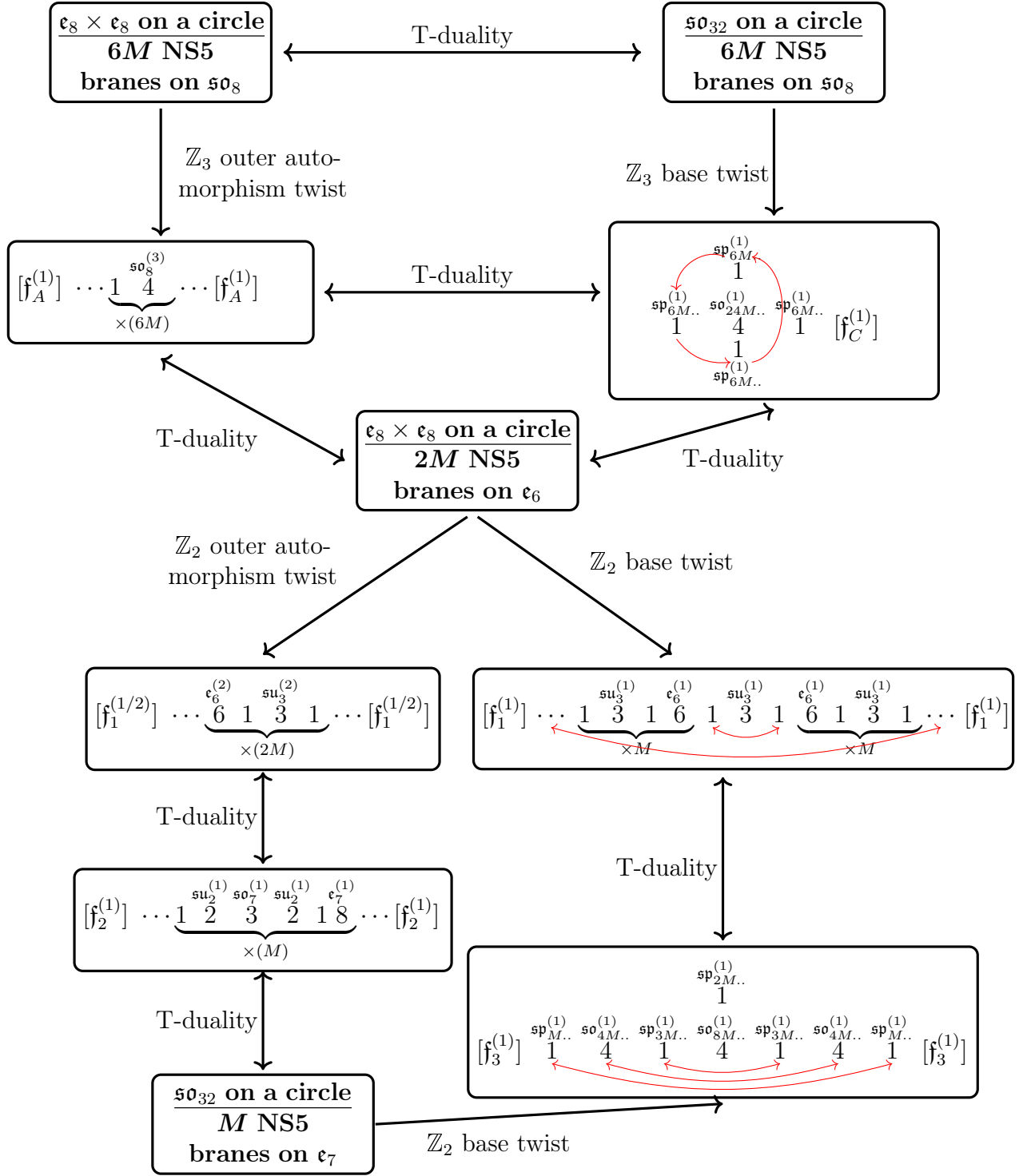

In this work we have systematically studied twisted dualities of heterotic little string theories. The rich possibilities to twist the 6D theory compactified on a circle by (combinations of) outer automorphisms of gauge/flavor algebras and permutation symmetries of tensor multiplets substantially enlarges the landscape of T-dual theories. In Figure~\ref{fig:DualityNetwork}, we illustrate an instance of this enlarged landscape, highlighting how the number of 5-branes, singularities $\fg$, and flavor symmetry algebras change across a vast network of extended twisted heterotic T-dual theories. 

We first consider outer automorphism twists both from a field theory and geometric point view. A key strategy to tame and study the large network of twisted theories, and to connect them via dualities, is to stratify theories by T-duality invariant data. This data, given in~\eqref{eq:invdata}, takes center stage in our study. 
We first characterize universal conditions for matching families of twisted heterotic LSTs via the fractionalization data of NS5 branes probing a twisted singularity $\fg^{(n)}$. The conditions for matching are captured by their 2-group and twisted Coulomb branch contributions, which need to be identical to the fractionalization data of stacks of NS5 branes probing a dual untwisted singularity $\tilde{\fg}$. To establish a full match we provide a careful analysis of the flavor symmetry sector, which requires taking Abelian flavor factors as well as contributions of massive fields and their ABJ anomalies into account~\cite{Apruzzi:2020eqi,Lee:2018ihr,Ahmed:2023lhj}. Inclusion of massive states in the form of non-tensionless strings that originate from E-strings and their generalizations is crucial, since the discrete symmetry we twist by acts on the full theory and not only the massless sector. 
 
Next we also consider tensor permutation twists. By imposing invariance of the matching data, we conjecture several new types of twisted-T-dual theories, including a novel kind we call \textit{CHL-twisted} LSTs. These 5D theories are obtained from special types of tensor multiplet twists
which identify the two $\fe_8$ flavor symmetry factors, similar to the 9D vacua due to Chaudhuri, Hockney and Lykken \cite{Chaudhuri:1995bf}. Importantly, these theories have a twisted Poincare 2-group structure constant with values $\kappa_P=1$, consistent with the anomaly inflow argument that $\kappa_P$ counts the number of M9 branes.

We prove the dualities of outer automorphism-twisted LSTs from geometry by exploiting F/M-theory duality. We consider M-theory on non-compact torus-fibered Calabi-Yau threefolds with an $m$-section, which engineers $m$-twisted F-theory circle compactificationa. If $m>1$, the torus fibration is not elliptic but only genus one, which allows for generalized monodromies acting on the fibers beyond those of the Tate classification. By identifying (multiple) other inequivalent torus fibrations
that give rise to other 6D F-theory lifts, we geometrically engineer twisted T-dual theories. Our geometric construction makes use of an algebraic description of the threefolds in terms of toric hypersurface. While simple and explicit, this setup has two restrictions. First, we can only describe gauge algebra outer automorphism twists, but not tensor twists. Second, the torus fibrations is by construction inherited from the ambient space, which is not a generic case~\cite{Grimm:2019bey}. To overcome the latter obstruction, we employ \textit{extended duality models}, which refers to torus-fibered threefolds that are as hypersurfaces in inequivalent ambient spaces that share a pair of the same LST description. In this way, the network of T-duals is extended to all theories that can be obtained in the union of all ambient space descriptions.
 
Our work opens up several interesting directions for future works, both from the field theory and the geometry perspective: while this work has shown how the network of (twisted-) T-dual LSTs systematically enhances the full structure of the T-duality map, incorporating different twists is not fully completed yet. Progress in that direction may be to systematically study the Higgs branch deformations of LSTs and their relation to T-duality, or their Hasse diagrams in the magnetic quiver description proposed in \cite{DelZotto:2023myd,DelZotto:2023nrb,Mansi:2023faa,Lawrie:2023uiu,Lawrie:2024zon}, and see how the twisted theories fit into this.

Similarly, it would be interesting to further study the precise relation between the 6D and 5D twisted theories when coupled to gravity. Indeed, in~\cite{Mayrhofer:2014opa,Braun:2014oya,Morrison:2014era,Anderson:2014yva,Klevers:2014bqa,Cvetic:2015moa} it was shown that in the 6D uplift of the genus-one fibrations, a discrete gauge symmetry exists that requires an intricate anomaly cancellation mechanism \cite{Monnier:2018nfs,Dierigl:2022zll}, even if no other non-Abelian gauge algebra is present. It would be very interesting to study these anomalies in the presence of additional (twisted) gauge algebras. Twisted LSTs could be useful in this regard, since LSTs are subsectors of SUGRA theories with at least one tensor multiplet. 

To study twisted compactifications directly in M-theory in both compact and non-compact cases, a deeper geometric understanding of the genus-one fibrations is required. 
The toric methods used in this work provide the benefit of an algebraic description of the threefold, but are not able to to describe general $\fsu_n^{(2)}$ twisted algebras as well as base twisted theories. In particular for the latter, it would be interesting to have geometric constructions of the \textit{CHL-like} twisted LSTs discussed in this work, which might also give new insights 
into theories coupled to gravity.

\section*{Acknowledgements}
We thank Florent Baume, Michele del Zotto and Muyang Liu for interesting conversations and useful discussions. The work of FR is supported by the NSF grants PHY-2210333 and PHY-2019786 (The NSF AI Institute for Artificial Intelligence and Fundamental Interactions). The work of HA, PKO, and FR is also supported by startup funding from Northeastern University. HA would like to thank the Perimeter Institute for Theoretical Physics and McGill University for their hospitality during the completion of this work.
PKO would like to thank the KITP and the program ``What is String Theory? Weaving Perspectives
Together'' during the completion of this work. This research was supported in part by grant
NSF PHY-2309135 to the Kavli Institute for Theoretical Physics (KITP).
\appendix

\section{Duality chain of \texorpdfstring{$A_{2N-1}^{(2)}$}{A(2N-1)(2)} LSTs from geometry}
\label{app:AnTwistDualities}

We discuss in Section~\ref{section:frev} that an $A_{N}$ type orbi-instanton theory has a single $ \mathbbm{Z}_{2}$ outer automorphism symmetry. One can then twist by this symmetry upon circle compactification and obtain a twisted 5D theory given by \eqref{eq:1ex3}. However, as shown in Table~\ref{tab:twistalginv}, one has two choices of 5D gauge algebras for $A_{2N-1}^{(2)}$, and we need to alternate between them to get massless bi-fundamentals of $\mathfrak{so}-\mathfrak{sp}$ in 5D. In this appendix, we give an example of the duality chain of an $A_{2N-1}^{(2)}$ type LST, and explicitly prove it via an F-theory realization. In particular, we consider the $A_{3}$ type LST, as this is the only such case with a toric realization. We start from the following orbi-instanton theory
\begin{align}
\mathfrak{u}_{1}^{2} \times \bigg(
\lbrack \mathfrak{e}_{6}\rbrack  \, \, 
1 \, \,
\underset{\left[\mathfrak{su}_{2}\right]}
{\overset{\mathfrak{su}_{3}}{2}} \, \,
{\overset{\mathfrak{su}_{4}}{2}} \,
\cdots
\lbrack \mathfrak{su}_{4}\rbrack  \, \,\bigg)\,,
\end{align}
with two delocalized non-ABJ anomalous $\mathfrak{u}_{1}$'s. We can complete this into an LST via~\eqref{eq:orbiconforbi}. This theory has a single $\mathbbm{Z}_{2}$ outer automorphism symmetry coming from the $\mathbbm{Z}_{2}$ outer automorphisms of all gauge and flavor algebras. Putting the theory on the circle, and twisting by this symmetry, we get the 5D theory
\begin{align}
\label{ex3at}
\lbrack \mathfrak{e}_{6}^{(2)}\rbrack  \, \, 
1 \, \,
\underset{\left[\mathfrak{u}_{1}\right]}
{\overset{\mathfrak{su}_{3}^{(2)}}{2}} \, \,
\underbrace{{\overset{\mathfrak{su}_{4}^{(2/2')}}{2}} \, \,}_{\times (2M-1)} \, \,
\underset{\left[\mathfrak{u}_{1}\right]}
{\overset{\mathfrak{su}_{3}^{(2)}}{2}} \, \,
1 \, \,
\lbrack \mathfrak{e}_{6}^{(2)}\rbrack  \, , 
\end{align}
where we choose the twist 2 or 2' alternatingly for the $\mathfrak{su}_{4}$ chain. The $\mathfrak{u}_{1}$ flavor factors arise on each $\mathfrak{su}_{3}^{(2)}$ since the two fundamentals of $\mathfrak{su}_{3}$ in the untwisted theory map to as single fundamental of $\mathfrak{sp}_{1}$ in the twisted theory (see Table~\ref{tab:twistalginv}). The delocalized $\mathfrak{u}_{1}$'s are twisted to ``nothing'' in 5D, since we project onto the invariant states. 

Using Table~\ref{tab:twistuntwist}, we see that the dual singularity is $\mathfrak{so}_{8}$ and hence we can write down an untwisted $\mathfrak{e}_{8} \times \mathfrak{e}_{8}$ dual,
\begin{align}
\label{ex3aun}
\lbrack \mathfrak{sp}_{4}\rbrack  \, \, 
\underset{\left[\mathfrak{sp}_{1}\right]}
{\overset{\mathfrak{so}_{7}}{2}} \, \,
1 \, \,
\underbrace{
{\overset{\mathfrak{so}_{8}}{4}} \, \,
1 \, \,}_{\times (M-1)} \, \,
\underset{\left[\mathfrak{sp}_{1}\right]}
{\overset{\mathfrak{g}_{2}}{3}} \, \,
1 \, \,
\lbrack \mathfrak{f}_{4}\rbrack  \, ,
\end{align}
which is exactly what we find in geometry. This fixes the base topology of the $\mathfrak{so}_{32}$ dual, and since both \eqref{ex3at} and \eqref{ex3aun} have non-simply laced flavor algebras in 5D, we can write down the twisted $\mathfrak{so}_{32}$ dual
\begin{align}
\label{ex3asot}
\lbrack \mathfrak{so}_{6}^{(2)}\rbrack  \, \, 
{\overset{\mathfrak{sp}_{M-1}}{1}} \, \, 
\myoverset{\overset{\mathfrak{sp}_{M-2}}{1}}
{\myunderset{\underset{\mathfrak{sp}_{M-2}}{\overset{\left[\mathfrak{sp}_{2}\right]}{1}}}
{\overset{\mathfrak{so}_{4M+6}^{(2)}}{4}}} \, \, 
{\overset{\mathfrak{sp}_{M+1}}{1}} \, \,
\lbrack \mathfrak{so}_{14}^{(2)}\rbrack \, , \, \nonumber
\end{align}
which is again geometrically realized in our construction.
One can easily check that the twisted T-duality data indeed matches for all three duals and is given by
\begin{align}
\text{rk}(\ff)=10\, , \qquad  \text{dim(CB)}=6M+2 \, , \qquad \kappa_{R}=8M+4\, .
\end{align}

\section{LST with \texorpdfstring{$ \mathbbm{Z}_{3}$}{Z3}-twisted flavor symmetry}
\label{app:Z3Twist}
We present an $\mathfrak{e}_{8} \times \mathfrak{e}_{8}$ theory where the flavor symmetry admits a $ \mathbbm{Z}_{3}$ twist. Instead of the orbi-instanton theory \eqref{Z3ex1}, we pick
\begin{align}
\label{Z3ex2}
\lbrack \mathfrak{so}_{8}\rbrack  \, \, 
1 \, \,
\myoverset{{1}}
{\overset{\mathfrak{so}_{8}}{4}} \, \,
1 \, \,
{\overset{\mathfrak{so}_{8}}{4}} \, \,
\cdots
\lbrack \mathfrak{so}_{8}\rbrack  \, \, \, .
\end{align}
We can twist by the $ \mathbbm{Z}_{3}$ symmetry corresponding to the combined outer automorphism of all $\mathfrak{so}_{8}$ algebras, and get the 5D theory $\mathcal{K}^{(3,3,3)}_{N}(\mathfrak{so}_{8})$,
\begin{align}
\label{ex52t}
\lbrack \mathfrak{so}_{8}^{(3)}\rbrack  \, \, 
1 \, \,
\myoverset{{\overset{{\left[\mathfrak{so}_{8}^{(3)}\right]}}{1}}}
{\overset{\mathfrak{so}_{8}^{(3)}}{4}} \, \,
1 \, \,
\underbrace{
\overset{\mathfrak{so}_{8}^{(3)}}{4} \, \,
1 \, \,}_{\times (N)} \, \,
\myoverset{{\overset{{\left[\mathfrak{so}_{8}^{(3)}\right]}}{1}}}
{\overset{\mathfrak{so}_{8}^{(3)}}{4}} \, \,
1 \, \,
\lbrack \mathfrak{so}_{8}^{(3)}\rbrack \, . 
\end{align}
Let us propose an untwisted $\mathfrak{e}_{8} \times \mathfrak{e}_{8}$ dual, inspired from the matching data. From our discussion in Section~\ref{sec:flavmatch}, we expect that this untwisted $\mathfrak{e}_{8} \times \mathfrak{e}_{8}$ dual has non-simply laced flavor factors. Furthermore, the rank of the flavor algebra in \eqref{ex52t} is 8. With these two facts in mind, we can consider an untwisted $\mathfrak{e}_{8}\times \mathfrak{e}_{8}$ LST with two copies of the orbi-instanton theory
\begin{align} 
\label{orbiZ31}
\lbrack \mathfrak{sp}_{4}\rbrack  \, \,
{\overset{\mathfrak{g}_{2}}{2}} \, \,
1 \, \,
{\overset{\mathfrak{f}_{4}}{5}} \, \,
1 \, \,
{\overset{\mathfrak{su}_{3}}{3}} \, \,
1 \, \,
{\overset{\mathfrak{e}_{6}}{6}} \, \,
\cdots
\lbrack \mathfrak{e}_{6}\rbrack  \, \, .
\end{align}
Fusing these with $(M-1)$ copies of $\mathfrak{e}_{6}-\mathfrak{e}_{6}$ conformal matter, we can write down the T-dual LST given by $\mathcal{K}^{(1)}_{M-1}(\mathfrak{e}_{6})$,
\begin{align} 
\label{ex52un2}
\lbrack \mathfrak{sp}_{4}\rbrack  \, \,
{\overset{\mathfrak{g}_{2}}{2}} \, \,
1 \, \,
{\overset{\mathfrak{f}_{4}}{5}} \, \,
1 \, \,
\underbrace{ 
{\overset{\mathfrak{su}_{3}}{3}} \, \,
1 \, \,
{\overset{\mathfrak{e}_{6}}{6}} \, \,
1 \, \,}_{\times (M-1)} \, \,
{\overset{\mathfrak{su}_{3}}{3}} \, \,
1 \, \,
{\overset{\mathfrak{f}_{4}}{5}} \, \,
1 \, \,
{\overset{\mathfrak{g}_{2}}{2}} \, \,
\lbrack \mathfrak{sp}_{4}\rbrack  \,.
\end{align}
One can also write down a twisted $\mathfrak{so}_{32}$ dual with an $F_{4}$ base,
\begin{align}
{\left[\mathfrak{u}_{1}^{2}\right]}\times\bigg(
\lbrack \mathfrak{so}_{8}^{(2)}\rbrack  \, \, 
{\overset{\mathfrak{sp}_{M+1}}{1}} \, \, 
{\underset{\left[\mathfrak{sp}_{3}\right]}{\overset{\mathfrak{so}_{4M+12}^{(2)}}{4}}} \, \, 
{\overset{\mathfrak{sp}_{3M}}{1}} \, \,
{\overset{\mathfrak{su}_{4M+1}}{2}} \, \, 
{\overset{\mathfrak{su}_{2M+1}}{2}} \, \, \bigg)\,.
\end{align}
Hence, this is a duality chain which includes each of $ \mathbbm{Z}_{1} (\text{trivial}),  \mathbbm{Z}_{2},  \mathbbm{Z}_{3} $ twisted 5D theories. For $N=3M$, the matching data is given by:
\begin{align}
\text{rk}(\ff)=8\, , \qquad  \text{dim(CB)}=12M+10 \, , \qquad \kappa_{R}=24M+18\, .
\end{align}

\section{LSTs with multiple twists}
\label{app:Mutltitwist}
We present the generalization of the twisted LST \eqref{ex4t2} with an $\mathfrak{so}_{2n}^{(2)}$ singularity. We have an $\mathfrak{e}_{8} \times \mathfrak{e}_{8}$ theory corresponding to $N$ NS5 branes probing a $D_{n}$ $(n>4)$ singularity, with the orbi-instanton theory being
\begin{align} 
\lbrack \mathfrak{e}_{6}\rbrack  \, \, 
1 \, \,
{\overset{\mathfrak{su}_{3}}{3}} \, \,
1 \, \,
\overset{\mathfrak{so}_{9}}{4} \, \,
\cdots \overset{\mathfrak{so}_{2n-1}}{4} \, \,
\overset{\mathfrak{sp}_{n-4}}{1} \, \,
\overset{\mathfrak{so}_{2n}}{4} \, \,
\cdots \lbrack \mathfrak{so}_{2n}\rbrack
\end{align}

As usual, we form an LST from two copies of this orbi-instanton theory, and $N$ copies of $\mathfrak{so}_{2n}-\mathfrak{so}_{2n}$ conformal matter theories. The resulting theory, just like \eqref{exagain}, has multiple possible 5D twisted theories. These correspond to five distinct $ \mathbbm{Z}_{2}$ outer automorphism symmetries which then lead to the 5D twisted theories:
\begin{align}
\label{appgen1}
\mathcal{K}^{(1,2_{1},1)}_{N}(\mathfrak{so}_{2n}): \qquad \lbrack \mathfrak{e}_{6}^{(2)}\rbrack  \, \, 
1 \, \,
{\overset{\mathfrak{su}_{3}^{(2)}}{3}} \, \,
1 \, \,
\overset{\mathfrak{so}_{9}}{4} \, \,
\cdots \overset{\mathfrak{so}_{2n-1}}{4} \, \,
\underbrace{
{\overset{\mathfrak{sp}_{n-4}}{1}} \, \,
\overset{\mathfrak{so}_{2n}}{4} \, \,}_{\times (N)} \, \,
 {\overset{\mathfrak{sp}_{n-4}}{1}} \, \,
 \overset{\mathfrak{so}_{2n-1}}{4} \, \,
\cdots 
\overset{\mathfrak{so}_{9}}{4} \, \,
1 \, \,
{\overset{\mathfrak{su}_{3}}{3}} \, \,
1 \, \,
\lbrack \mathfrak{e}_{6}\rbrack  \, \,
\end{align}

\begin{align}
\label{appgen2}
\mathcal{K}^{(1,2_{1},2_{1})}_{N}(\mathfrak{so}_{2n}): \qquad \lbrack \mathfrak{e}_{6}^{(2)}\rbrack  \, \, 
1 \, \,
{\overset{\mathfrak{su}_{3}^{(2)}}{3}} \, \,
1 \, \,
\overset{\mathfrak{so}_{9}}{4} \, \,
\cdots \overset{\mathfrak{so}_{2n-1}}{4} \, \,
\underbrace{
{\overset{\mathfrak{sp}_{n-4}}{1}} \, \,
\overset{\mathfrak{so}_{2n}}{4} \, \,}_{\times (N)} \, \,
 {\overset{\mathfrak{sp}_{n-4}}{1}} \, \,
 \overset{\mathfrak{so}_{2n-1}}{4} \, \,
\cdots 
\overset{\mathfrak{so}_{9}}{4} \, \,
1 \, \,
{\overset{\mathfrak{su}_{3}^{(2)}}{3}} \, \,
1 \, \,
\lbrack \mathfrak{e}_{6}^{(2)}\rbrack  \, \,
\end{align}

\begin{align}
\label{appgen3}
\mathcal{K}^{(1,2_{2},2_{2})}_{N}(\mathfrak{so}_{2n}): \qquad \lbrack \mathfrak{e}_{6}\rbrack  \, \, 
1 \, \,
{\overset{\mathfrak{su}_{3}}{3}} \, \,
1 \, \,
\overset{\mathfrak{so}_{9}}{4} \, \,
\cdots \overset{\mathfrak{so}_{2n-1}}{4} \, \,
\underbrace{
{\overset{\mathfrak{sp}_{n-4}}{1}} \, \,
\overset{\mathfrak{so}_{2n}^{(2)}}{4} \, \,}_{\times (N)} \, \,
 {\overset{\mathfrak{sp}_{n-4}}{1}} \, \,
 \overset{\mathfrak{so}_{2n-1}}{4} \, \,
\cdots 
\overset{\mathfrak{so}_{9}}{4} \, \,
1 \, \,
{\overset{\mathfrak{su}_{3}}{3}} \, \,
1 \, \,
\lbrack \mathfrak{e}_{6}\rbrack  \, \,
\end{align}

\begin{align}
\label{appgen4}
\mathcal{K}^{(1,2_{3},1)}_{N}(\mathfrak{so}_{2n}): \qquad \lbrack \mathfrak{e}_{6}^{(2)}\rbrack  \, \, 
1 \, \,
{\overset{\mathfrak{su}_{3}^{(2)}}{3}} \, \,
1 \, \,
\overset{\mathfrak{so}_{9}}{4} \, \,
\cdots \overset{\mathfrak{so}_{2n-1}}{4} \, \,
\underbrace{
{\overset{\mathfrak{sp}_{n-4}}{1}} \, \,
\overset{\mathfrak{so}_{2n}^{(2)}}{4} \, \,}_{\times (N)} \, \,
 {\overset{\mathfrak{sp}_{n-4}}{1}} \, \,
 \overset{\mathfrak{so}_{2n-1}}{4} \, \,
\cdots 
\overset{\mathfrak{so}_{9}}{4} \, \,
1 \, \,
{\overset{\mathfrak{su}_{3}}{3}} \, \,
1 \, \,
\lbrack \mathfrak{e}_{6}\rbrack  \, \,
\end{align}

\begin{align}
\label{appgen5}
\mathcal{K}^{(1,2_{3},2_{3})}_{N}(\mathfrak{so}_{2n}): \qquad \lbrack \mathfrak{e}_{6}^{(2)}\rbrack  \, \, 
1 \, \,
{\overset{\mathfrak{su}_{3}^{(2)}}{3}} \, \,
1 \, \,
\overset{\mathfrak{so}_{9}}{4} \, \,
\cdots \overset{\mathfrak{so}_{2n-1}}{4} \, \,
\underbrace{
{\overset{\mathfrak{sp}_{n-4}}{1}} \, \,
\overset{\mathfrak{so}_{2n}^{(2)}}{4} \, \,}_{\times (N)} \, \,
 {\overset{\mathfrak{sp}_{n-4}}{1}} \, \,
 \overset{\mathfrak{so}_{2n-1}}{4} \, \,
\cdots 
\overset{\mathfrak{so}_{9}}{4} \, \,
1 \, \,
{\overset{\mathfrak{su}_{3}^{(2)}}{3}} \, \,
1 \, \,
\lbrack \mathfrak{e}_{6}^{(2)}\rbrack  \, \,
\end{align}
We focus on the duality chain for one of model~\eqref{appgen4}. Since this is a $ \mathbbm{Z}_{2}$-twisted theory, we expect to find two distinct continuous families of T-duals for $N=2M$ and $N=2M-1$. Focusing on the latter, the untwisted $\mathfrak{e}_{8} \times \mathfrak{e}_{8}$ dual theory, $ \mathcal{K}^{(1,1,1)}_{M-2}(\mathfrak{so}_{4n-4})$ with $n>6$ is
\begin{align}
\label{weuntw}
\lbrack \mathfrak{sp}_{4}\rbrack  \, \, 
\myoverset{\overset{\mathfrak{sp}_{n-5}}{1}}
{\myunderset{\underset{\mathfrak{sp}_{n-5}}{1}}
{\overset{\mathfrak{so}_{4n-4}}{4}}} \, \, 
\underbrace{
{\overset{\mathfrak{sp}_{2n-6}}{1}} \, \,
{\overset{\mathfrak{so}_{4n-4}}{4}} \, \,}_{\times (M-2)} 
{\overset{\mathfrak{sp}_{2n-6}}{1}} \, \,
{\overset{\mathfrak{so}_{4n-5}}{4}} \, \,
{\overset{\mathfrak{sp}_{2n-7}}{1}} \, \,
{\overset{\mathfrak{so}_{4n-8}}{4}} \, \,
{\overset{\mathfrak{sp}_{2n-9}}{1}} \, \,
{\overset{\mathfrak{so}_{4n-12}}{4}} \, \,
\cdots 
{\overset{\mathfrak{sp}_{3}}{1}} \, \,
{\overset{\mathfrak{so}_{12}}{4}} \, \,
{\overset{\mathfrak{sp}_{1}}{1}} \, \,
{\overset{\mathfrak{g}_{2}}{3}} \, \, 
1 \, \,
\lbrack \mathfrak{f}_{4}\rbrack  \,. 
\end{align}
One can then find the following class of twisted $\mathfrak{so}_{32}$ theories given by $\mathcal{\phantom{}_+\Tilde{K}}^{(2_{1})}_{M-2}( \mathfrak{so}_{4n-4})$,
\begin{align}
\label{t8so}
\lbrack \mathfrak{so}_{6}^{(2)}\rbrack  \, \, 
{\overset{\mathfrak{sp}_{M-2}}{1}} \, \,
\myoverset{\overset{\mathfrak{sp}_{M-3}}{1^{*}}}
{\overset{\mathfrak{so}_{4M+2}^{(2)}}{4^{*}}} \, \,
{\overset{\mathfrak{sp}_{2M-1}}{1}} \, \,
{\overset{\mathfrak{so}_{4M+8}}{4}} \, \,
\cdots
{\overset{\mathfrak{so}_{4n+4M-12}}{4}} \, \,
{\overset{\mathfrak{sp}_{2n+2M-9}}{1}} \, \,
\myoverset{\overset{\mathfrak{sp}_{n+M-6}}{1^{*}}}
{\overset{\mathfrak{so}_{4n+4M-10}^{(2)}}{4^{*}}} \, \,
{\overset{\mathfrak{sp}_{n+M-3}}{1}} \, \,
\lbrack \mathfrak{so}_{14}^{(2)}\rbrack  \, ,
 \end{align} 
for $M>2$. The 6D uplift of this theory, $\mathcal{\phantom{}_+\Tilde{K}}^{(1)}_{M-2}(\mathfrak{so}_{4n-4})$, has five (distinct) 5D twisted theories, similar to \eqref{ex4t2so}. One of them is \eqref{t8so}, as can be seen from \eqref{appsogen1}-\eqref{appsogen7}. Theory~\eqref{t8so} is the only one that has the correct matching data for the duality, which reads
\begin{align}
\text{rk}(\ff)&=8\, , \quad  \text{dim(CB)}=2n^{2}+4nM-7n-6M-3 \, ,\nonumber\\
\kappa_{R}&=4n^{2}+8nM-20n-16M+18\, .
\end{align}
Furthermore, we observe that both 6D uplifts $\mathcal{K}^{(1,1,1)}_{N}(\mathfrak{so}_{2n})$ and $\mathcal{\phantom{}_+\Tilde{K}}^{(1)}_{M-2}(\mathfrak{so}_{4n-4})$ have five twisted 5D theories.

\clearpage
\section{Tensor-twisted \texorpdfstring{$\fso_{32}$}{so} theories and their duals}
\label{app:so32BaseTwistDual}
\begin{table}[h!]
\begin{tabular}{|c|c|c|c|}
\hline
 \multicolumn{2}{|c|}{$\mathfrak{so}_{32}$ rank $M$ twisted theory of type $\mathfrak{g}$ } & $\mathfrak{e}_{8} \times \mathfrak{e}_{8}$ T-dual (rank, $\mathfrak{g}^{(n)}$, base twist)& $\kappa_{P}$\\ \hline
 $ \mathfrak{so}_{8N}$ & 
 $\,  {\overset{\fsp_{M..}}{{\tikzmarknode[black!70!black]{c}{1}}}} \,\overset{ \displaystyle {{\overset{\fsp_{M}..}{{\tikzmarknode[black!70!black]{o}{1}}}} } }{{ \overset{\fso_{4M..}}{{\tikzmarknode[black!70!black]{a}{4}}}} }\,
 {{\tikzmarknode[black!70!black]{l}{...}}}
 \overset{\fso_{4M..}}{4} {{\tikzmarknode[black!70!black]{m}{...}}}\overset{ \displaystyle {{\overset{\fsp_{M}..}{{\tikzmarknode[black!70!black]{u}{1}}}} } }{{\overset{\fso_{4M..}}{{\tikzmarknode[black!70!black]{b}{4}}}} }
 {{\overset{\fsp_{M..}}{{\tikzmarknode[black!70!black]{d}{1}}}} }
 \,$ \,
\begin{tikzpicture}[remember picture,overlay]
\draw[red,<->]([yshift=0.000009ex]a.south) to[bend right=15]node[below]{\scriptsize} ([yshift=0.000009ex]b.south);
\draw[red,<->]([yshift=0.000009ex]l.south) to[bend right=15]node[below]{\scriptsize} ([yshift=0.000009ex]m.south);
\draw[red,<->]([yshift=0.000009ex]c.south) to[bend right=15]node[below]{\scriptsize} ([yshift=0.000009ex]d.south);
\draw[red,<->]([yshift=0.000009ex]o.north) to[bend left=15]node[below]{\scriptsize} ([yshift=0.000009ex]u.north);
\end{tikzpicture}
  
 & (2$M$, $\mathfrak{so}_{4N+2}^{(1)}$, $ \mathbbm{Z}_{2}$) & 2 \\ [0.5cm]
   $\mathfrak{so}_{8N+4}$ & 
 $\,  {\overset{\fsp_{M..}}{{\tikzmarknode[black!70!black]{c}{1}}}} \,\overset{ \displaystyle {{\overset{\fsp_{M}..}{{\tikzmarknode[black!70!black]{o}{1}}}} } }{{ \overset{\fso_{4M..}}{{\tikzmarknode[black!70!black]{a}{4}}}} }\,
 {{\tikzmarknode[black!70!black]{l}{...}}}
 \overset{\fsp_{2M..}}{1} {{\tikzmarknode[black!70!black]{m}{...}}}\overset{ \displaystyle {{\overset{\fsp_{M}..}{{\tikzmarknode[black!70!black]{u}{1}}}} } }{{\overset{\fso_{4M..}}{{\tikzmarknode[black!70!black]{b}{4}}}} }
 {{\overset{\fsp_{M..}}{{\tikzmarknode[black!70!black]{d}{1}}}} }
 \,$ \,
\begin{tikzpicture}[remember picture,overlay]
\draw[red,<->]([yshift=0.000009ex]a.south) to[bend right=15]node[below]{\scriptsize} ([yshift=0.000009ex]b.south);
\draw[red,<->]([yshift=0.000009ex]l.south) to[bend right=15]node[below]{\scriptsize} ([yshift=0.000009ex]m.south);
\draw[red,<->]([yshift=0.000009ex]c.south) to[bend right=15]node[below]{\scriptsize} ([yshift=0.000009ex]d.south);
\draw[red,<->]([yshift=0.000009ex]o.north) to[bend left=15]node[below]{\scriptsize} ([yshift=0.000009ex]u.north);
\end{tikzpicture} &(2$M$, $\mathfrak{so}_{4N+4}^{(1)}$, $ \mathbbm{Z}_{2}$) & 1 \\ [0.5cm]

$ \mathfrak{so}_{4N+2}$ & 
 $ \, \, {\overset{\fsp_{M..}}{{\tikzmarknode[black!70!black]{c}{1}}}} \, \, \overset{ \displaystyle {\overset{\fsp_{ M..}}{{\tikzmarknode[black!70!black]{d}{1}}}}}{
\overset{\fso_{4M..}}{4} 
} ... \overset{\fsp_{2M }}{1} \, \overset{\fsu_{2M }}{2} \, $ \,
\begin{tikzpicture}[remember picture,overlay]
\draw[red,<->]([yshift=0.000009ex]c.north) to[bend left]node[below]{\scriptsize} ([yshift=0.000009ex]d.north);
\end{tikzpicture}

 & ($M$, $\mathfrak{so}_{4N+2}^{(2)}$, $ \mathbbm{Z}_{1}$) & 2 \\ [0.5cm]

 $\mathfrak{so}_{4N}$ & 
 $\,  {\overset{\fsp_{M..}}{{\tikzmarknode[black!70!black]{c}{1}}}} \,\overset{ \displaystyle {{\overset{\fsp_{M}..}{{\tikzmarknode[black!70!black]{o}{1}}}} } } {\overset{\fso_{4M..}}{4}} \,
 ...
 \overset{\fsp_{2M..}}{1} ... \overset{ \displaystyle {{\overset{\fsp_{M}..}{{\tikzmarknode[black!70!black]{u}{1}}}} } }{\overset{\fso_{4M..}}{4}}
 {{\overset{\fsp_{M..}}{{\tikzmarknode[black!70!black]{d}{1}}}} }
 \,$ \,
\begin{tikzpicture}[remember picture,overlay]
\draw[red,<->]([yshift=0.000009ex]c.north) to[bend left]node[below]{\scriptsize} ([yshift=0.000009ex]o.north);
\draw[red,<->]([yshift=0.000009ex]d.north) to[bend right]node[below]{\scriptsize} ([yshift=0.000009ex]u.north);
\end{tikzpicture} &(2$M$, $\mathfrak{su}_{4N-4}^{(1)}$, $ \mathbbm{Z}_{2}$) & 2 \\ [0.5cm]

$\mathfrak{so}_{2N}$ &
$ \overset{ \displaystyle {{\overset{ \fsu_{2M..}}{{\tikzmarknode[black!70!black]{c}{1}}}} } } { \underset{ \displaystyle \, {{\overset{\fsu_{2M..}}{{\tikzmarknode[black!70!black]{d}{1}}}} } }{\, \, \overset{\fsu_{4M..}}{2}}} \, \overset{\fsu_{4M..}}{2} \, \overset{\fsu_{4M..}}{ 2} \hspace{-0.1cm}... \hspace{-0.1cm}\overset{\fsu_{4M+8}}{ 2} \,\overset{\fsu_{4M}/\fsp_{2M}}{1} $ \,
\begin{tikzpicture}[remember picture,overlay]
\draw[red,<->]([yshift=0.000009ex]c.north) to[bend right=50]node[below]{\scriptsize} ([yshift=0.000009ex]d.north);
\end{tikzpicture} 
 & (2$M$, $\mathfrak{su}_{2N-4}^{(1)}$, $ \mathbbm{Z}_{2}$) & 1 \\ [0.7cm]

$ \mathfrak{so}_{8N}$ & 
 $\,  {\overset{\fsp_{M..}}{{\tikzmarknode[black!70!black]{c}{1}}}} \,\overset{ \displaystyle {{\overset{\fsp_{M}..}{{\tikzmarknode[black!70!black]{o}{1}}}} } }{{ \overset{\fso_{4M..}}{{\tikzmarknode[black!70!black]{a}{4}}}} }\,
 {{\tikzmarknode[black!70!black]{l}{...}}}
 \overset{\fso_{4M..}}{4} {{\tikzmarknode[black!70!black]{m}{...}}}\overset{ \displaystyle {{\overset{\fsp_{M}..}{{\tikzmarknode[black!70!black]{u}{1}}}} } }{{\overset{\fso_{4M..}}{{\tikzmarknode[black!70!black]{b}{4}}}} }
 {{\overset{\fsp_{M..}}{{\tikzmarknode[black!70!black]{d}{1}}}} }
 \,$ \,
\begin{tikzpicture}[remember picture,overlay]
\draw[red,<->]([yshift=0.000009ex]a.south) to[bend right=15]node[below]{\scriptsize} ([yshift=0.000009ex]b.south);
\draw[red,<->]([yshift=0.000009ex]l.south) to[bend right=15]node[below]{\scriptsize} ([yshift=0.000009ex]m.south);
\draw[red,<->]([yshift=0.000009ex]c.south) to[bend right=15]node[below]{\scriptsize} ([yshift=0.000009ex]d.south);
\draw[red,<->]([yshift=0.000009ex]o.north) to[bend left=15]node[below]{\scriptsize} ([yshift=0.000009ex]u.north);
\draw[green,<->]([yshift=0.000009ex]c.north) to[bend left]node[below]{\scriptsize} ([yshift=0.000009ex]o.north);
\draw[green,<->]([yshift=0.000009ex]u.north) to[bend left]node[below]{\scriptsize} ([yshift=0.000009ex]d.north);
\end{tikzpicture}

 & (2$M$, $\mathfrak{so}_{4N+2}^{(2)}$, $ \mathbbm{Z}_{2}$) & 2 \\ [0.7cm]

 $\mathfrak{so}_{8N+4}$ & 
 $\,  {\overset{\fsp_{M..}}{{\tikzmarknode[black!70!black]{c}{1}}}} \,\overset{ \displaystyle {{\overset{\fsp_{M}..}{{\tikzmarknode[black!70!black]{o}{1}}}} } }{{ \overset{\fso_{4M..}}{{\tikzmarknode[black!70!black]{a}{4}}}} }\,
 {{\tikzmarknode[black!70!black]{l}{...}}}
 \overset{\fsp_{2M..}}{1} {{\tikzmarknode[black!70!black]{m}{...}}}\overset{ \displaystyle {{\overset{\fsp_{M}..}{{\tikzmarknode[black!70!black]{u}{1}}}} } }{{\overset{\fso_{4M..}}{{\tikzmarknode[black!70!black]{b}{4}}}} }
 {{\overset{\fsp_{M..}}{{\tikzmarknode[black!70!black]{d}{1}}}} }
 \,$ \,
\begin{tikzpicture}[remember picture,overlay]
\draw[red,<->]([yshift=0.000009ex]a.south) to[bend right=15]node[below]{\scriptsize} ([yshift=0.000009ex]b.south);
\draw[red,<->]([yshift=0.000009ex]l.south) to[bend right=15]node[below]{\scriptsize} ([yshift=0.000009ex]m.south);
\draw[red,<->]([yshift=0.000009ex]c.south) to[bend right=15]node[below]{\scriptsize} ([yshift=0.000009ex]d.south);
\draw[red,<->]([yshift=0.000009ex]o.north) to[bend left=15]node[below]{\scriptsize} ([yshift=0.000009ex]u.north);
\draw[green,<->]([yshift=0.000009ex]c.north) to[bend left]node[below]{\scriptsize} ([yshift=0.000009ex]o.north);
\draw[green,<->]([yshift=0.000009ex]u.north) to[bend left]node[below]{\scriptsize} ([yshift=0.000009ex]d.north);
\end{tikzpicture} 
&(2$M$, $\mathfrak{so}_{4N+4}^{(2)}$, $ \mathbbm{Z}_{2}$) & 1 \\ [0.7cm]

$\mathfrak{su}_{2N}$ & 
 $\,  {\overset{\fsp_{2M..}}{{\tikzmarknode[black!70!black]{c}{1}}}} \,{{ \overset{\fsu_{4M..}}{{\tikzmarknode[black!70!black]{a}{2}}}} }\,
 {{\tikzmarknode[black!70!black]{l}{...}}}
 {{\overset{\fsu_{4M..}}{{\tikzmarknode[black!70!black]{g}{2}}}} }
 {{\overset{\fsu_{4M..}}{{\tikzmarknode[black!70!black]{h}{2}}}} }
 {{\tikzmarknode[black!70!black]{m}{...}}}{{\overset{\fsu_{4M..}}{{\tikzmarknode[black!70!black]{b}{2}}}} }
 {{\overset{\fsp_{2M..}}{{\tikzmarknode[black!70!black]{d}{1}}}} }
 \,$ \,
\begin{tikzpicture}[remember picture,overlay]
\draw[red,<->]([yshift=0.000009ex]a.south) to[bend right=15]node[below]{\scriptsize} ([yshift=0.000009ex]b.south);
\draw[red,<->]([yshift=0.000009ex]l.south) to[bend right=15]node[below]{\scriptsize} ([yshift=0.000009ex]m.south);
\draw[red,<->]([yshift=0.000009ex]c.south) to[bend right=15]node[below]{\scriptsize} ([yshift=0.000009ex]d.south);
\draw[red,<->]([yshift=0.000009ex]g.south) to[bend right=15]node[below]{\scriptsize} ([yshift=0.000009ex]h.south);
\end{tikzpicture} & ($2M$, $\mathfrak{su}_{N}$, $ \mathbbm{Z}_{2}$) & 2 \\ [0.7cm]

\multirow{2}{*}{$\mathfrak{su}_{2N}$} & 
 \multirow{2}{*}{$\,  {\overset{\fsu_{2M..}}{{\tikzmarknode[black!70!black]{c}{1}}}} \,{{ \overset{\fsu_{2M..}}{{\tikzmarknode[black!70!black]{a}{2}}}} }\,
 {{\tikzmarknode[black!70!black]{l}{...}}}
 \overset{\fsu_{2M..}}{2}
 {{\tikzmarknode[black!70!black]{m}{...}}}{{\overset{\fsu_{2M..}}{{\tikzmarknode[black!70!black]{b}{2}}}} }
 {{\overset{\fsu_{2M..}}{{\tikzmarknode[black!70!black]{d}{1}}}} }
 \,$ \,
\begin{tikzpicture}[remember picture,overlay]
\draw[red,<->]([yshift=0.000009ex]a.south) to[bend right=15]node[below]{\scriptsize} ([yshift=0.000009ex]b.south);
\draw[red,<->]([yshift=0.000009ex]l.south) to[bend right=15]node[below]{\scriptsize} ([yshift=0.000009ex]m.south);
\draw[red,<->]([yshift=0.000009ex]c.south) to[bend right=15]node[below]{\scriptsize} ([yshift=0.000009ex]d.south);
\end{tikzpicture}} & even N: ($M$, $\mathfrak{so}_{N+4}^{(2)}$, $ \mathbbm{Z}_{1}$) & \multirow{2}{*}{2} \\
& & odd N: ($M/2$, $\mathfrak{so}_{2N+4}^{(1)}$, $ \mathbbm{Z}_{1}$) & \\[0.5cm]
$\mathfrak{e}_{7}$ & 
 ${\overset{\mathfrak{sp}_{M..}}{{\tikzmarknode[black!70!black]{c}{1}}}} \, \, {\overset{\mathfrak{so}_{4M..}}{{{\tikzmarknode[black!70!black]{a}{4}}}}} \, \,  {\overset{\mathfrak{sp}_{3M..}}{{{\tikzmarknode[black!70!black]{l}{1}}}}} \, \, 
\myoverset{{\overset{\mathfrak{sp}_{2M..}}{1}}}
{\overset{\mathfrak{so}_{8M..}}{4}} \, \,
{\overset{\mathfrak{sp}_{3M..}}{{{\tikzmarknode[black!70!black]{m}{1}}}}} \, \,
{\overset{\mathfrak{so}_{4M..}}{{{\tikzmarknode[black!70!black]{b}{4}}}}} \, \,
{\overset{\mathfrak{sp}_{M..}}{{{\tikzmarknode[black!70!black]{d}{1}}}}} \, \,$ \,
\begin{tikzpicture}[remember picture,overlay]
\draw[red,<->]([yshift=0.000009ex]a.south) to[bend right=15]node[below]{\scriptsize} ([yshift=0.000009ex]b.south);
\draw[red,<->]([yshift=0.000009ex]l.south) to[bend right=15]node[below]{\scriptsize} ([yshift=0.000009ex]m.south);
\draw[red,<->]([yshift=0.000009ex]c.south) to[bend right=15]node[below]{\scriptsize} ([yshift=0.000009ex]d.south);
\end{tikzpicture} &(2$M$, $\mathfrak{e}_{6}^{(1)}$, $ \mathbbm{Z}_{2}$) & 1 \\[7mm]
\hline
\end{tabular}
\caption{All $ \mathbbm{Z}_{2}$ tensor twists of $\mathfrak{so}_{32}$ theories and their $\mathfrak{e}_{8} \times \mathfrak{e}_{8}$ duals based on $c_2,r_2$, and their $\kappa_{P}$.}
\label{tab:conjduals}
\end{table}

\clearpage
\bibliographystyle{ytphys}
\bibliography{refs}
\end{document}